\documentclass[trackchanges, twocolumn]{aastex701}
\usepackage{enumitem, amsmath, bm, lipsum, caption, multirow, soul}
\setlist[itemize]{noitemsep, topsep=0pt}
\setlist[enumerate]{noitemsep, topsep=3pt}

\newcommand{\todo}{\ifmmode \text{\color{red}\Huge{\(\bullet\)}} \else {\color{red}{\Huge$\bullet$}}\fi}
\newcommand{\finish}{\ifmmode \text{\color{blue}\Huge{\(\bullet\)}} \else {\color{blue}{\Huge$\bullet$}}\fi}

\newcommand{\nod}{207 }

\begin{document}

\title{BEACON: JWST NIRCam Pure-parallel Imaging Survey. IV. \\ A Systematic Search for Galaxy Overdensities and Evidence for Gas Accretion Mode Transition}

\author[0009-0005-1487-7772]{Ryo Albert Sutanto}
\affiliation{Astronomical Institute, Tohoku University, 6-3 Aramaki, Aoba-ku, Sendai 980-8578, Japan}
\email[show]{ryo.sutanto@astr.tohoku.ac.jp}  

\author[0000-0002-8512-1404]{Takahiro Morishita}
\affiliation{IPAC, California Institute of Technology, MC 314-6, 1200 E. California Boulevard, Pasadena, CA 91125, USA}
\affiliation{Astronomical Institute, Tohoku University, 6-3 Aramaki, Aoba-ku, Sendai 980-8578, Japan}
\email[]{}

\author[0000-0002-2993-1576]{Tadayuki Kodama}
\affiliation{Astronomical Institute, Tohoku University, 6-3 Aramaki, Aoba-ku, Sendai 980-8578, Japan}
\email[]{}

\author[0000-0002-5258-8761]{Abdurro'uf}
\affiliation{Department of Astronomy, Indiana University, 727 East Third Street, Bloomington, IN 47405, USA}
\email{fnuabdur@iu.edu}

\author[0000-0002-7908-9284]{Larry D. Bradley}
\affiliation{Space Telescope Science Institute, 3700 San Martin Drive, Baltimore, MD 21218, USA}
\email[]{}

\author[0000-0002-8651-9879]{Andrew J.\ Bunker}
\affiliation{Department of Physics, University of Oxford, Denys Wilkinson Building, Keble Road, Oxford OX1 3RH, UK}
\email{Andy.Bunker@physics.ox.ac.uk}

\author[0000-0003-3691-937X]{Nima Chartab}
\affiliation{IPAC, California Institute of Technology, MC 314-6, 1200 E. California Boulevard, Pasadena, CA 91125, USA}
\email[]{}

\author[0000-0002-0486-5242]{Nuo Chen}
\affiliation{Astronomical Institute, Tohoku University, 6-3 Aramaki, Aoba-ku, Sendai 980-8578, Japan}
\email[]{}

\author[0000-0001-8587-218X]{Matthew J. Hayes}
\affiliation{Stockholm University, Department of Astronomy and Oskar
Klein Centre for Cosmoparticle Physics, AlbaNova University Centre,
SE-10691, Stockholm, Sweden}
\email{matthew.hayes@astro.su.se}

\author[0000-0003-3367-3415]{George Helou} 
\affiliation{IPAC, California Institute of Technology, MC 314-6, 1200 E. California Boulevard, Pasadena, CA 91125, USA}
\email{gxh@ipac.caltech.edu}

\author[0009-0009-3404-5673]{Novan Saputra Haryana}
\affiliation{Astronomical Institute, Tohoku University, 6-3 Aramaki, Aoba-ku, Sendai 980-8578, Japan}
\email{novan.haryana@astr.tohoku.ac.jp}

\author[0000-0003-4570-3159]{Nicha Leethochawalit}
\affiliation{National Astronomical Research Institute of Thailand (NARIT), Mae Rim, Chiang Mai, 50180, Thailand}
\email{}

\author[0009-0002-8965-1303]{Zhaoran Liu}
\affiliation{Astronomical Institute, Tohoku University, 6-3 Aramaki, Aoba-ku, Sendai 980-8578, Japan}
\affiliation{MIT Kavli Institute for Astrophysics and Space Research, 70 Vassar Street, Cambridge, MA 02139, USA}
\email{zhaoran.liu@astr.tohoku.ac.jp}

\author[0000-0002-3407-1785]{Charlotte A. Mason}
\affiliation{Cosmic Dawn Center (DAWN), Denmark}
\affiliation{Niels Bohr Institute, University of Copenhagen, Jagtvej 128, DK-2200 Copenhagen N, Denmark}
\email{charlotte.mason@nbi.ku.dk}

\author[0000-0002-9946-4731]{Marc Rafelski}
\affiliation{Space Telescope Science Institute, 3700 San Martin Drive, Baltimore, MD 21218, USA}
\affiliation{Department of Physics and Astronomy, Johns Hopkins University, Baltimore, MD 21218, USA}
\email{mrafelski@stsci.edu}

\author[0000-0001-7016-5220]{Michael J. Rutkowski}
\affiliation{Minnesota State University, Mankato, Department of Physics and Astronomy, 141 Trafton Science Center N, Mankato, MN 56001, USA}
\email{michael.rutkowski@mnsu.edu}

\author[0000-0001-9935-6047]{Massimo Stiavelli}
\affiliation{Space Telescope Science Institute, 3700 San Martin Drive, Baltimore, MD 21218, USA}
\email{mstiavel@stsci.edu}

\author[0009-0009-8116-0316]{Kosuke Takahashi}
\affiliation{Astronomical Institute, Tohoku University, 6-3 Aramaki, Aoba-ku, Sendai 980-8578, Japan}
\email[]{}

\author[0000-0002-7064-5424]{Harry I. Teplitz}
\affiliation{IPAC, California Institute of Technology, MC 314-6, 1200 E. California Boulevard, Pasadena, CA 91125, USA}
\email{hit@ipac.caltech.edu}

\author[orcid=0000-0001-9391-305X]{Michele Trenti}
\affiliation{School of Physics, The University of Melbourne, VIC 3010, Australia}
\email{michele.trenti@unimelb.edu.au}

\author[0000-0002-8460-0390]{Tommaso Treu}
\affiliation{Department of Physics and Astronomy, University of California, Los Angeles, 430 Portola Plaza, Los Angeles, CA 90095, USA}
\email{tt@astro.ucla.edu}

\author[0000-0003-0980-1499]{Benedetta Vulcani}
\affiliation{INAF -- Osservatorio Astronomico di Padova, Vicolo Osservatorio 5, 35122 Padova, Italy}
\email{benedetta.vulcani@inaf.it}

\author[0000-0003-3817-8739]{Yechi Zhang}
\affiliation{IPAC, California Institute of Technology, MC 314-6, 1200 E. California Boulevard, Pasadena, CA 91125, USA}
\email{yechi@ipac.caltech.edu}

\begin{abstract}
We systematically search for galaxy overdensities 
using 20 independent fields with a minimum of six filters (F090W, F115W, F150W, F277W, F356W, and F444W) from BEACON, the JWST Cycle 2 NIRCam pure-parallel imaging survey. We apply an adaptive kernel-density estimation method that incorporates the full photometric redshift probability distribution function of each galaxy to map galaxy overdensities, and identify \nod significant ($>4\sigma$) overdensities at $1.5<z<5$. 
We measure the quenched galaxy fraction, the average specific star formation rate (sSFR), the total dark matter mass, and the local galaxy density of each system. By investigating the correlation among these observables, we find that galaxy quenching proceeds in two paths: ($i$)~Overdensities within more massive halos exhibit higher quenched fractions and lower average sSFRs. This trend weakens at $z\gtrsim2$, consistent with cold gas streams penetrating shock-heated massive halos and sustaining star formation activity at early times. 
($ii$)~We also find a dependence of the same parameters on local densities at $z<2$, where the quenched fraction increases and the sSFR decreases toward higher densities. The environmental trend in sSFR weakens at $z\sim2$--$3$ and shows tentative evidence for a reversal at $z>3$, potentially due to a larger cold gas supply in earlier times. Our study reveals a complex interplay between individual galaxies and large-scale environmental properties, marking the onset of environmental effects on galaxy quenching in massive halos at cosmic noon. 
\end{abstract}


\section{Introduction} \label{sec:intro}

The structure growth in our universe originates from fluctuations in the dark matter distributions, which lead to the formation of stars and galaxies through the interplay of gravitational collapse and hydrodynamical processes \citep{1991ApJ...379...52W, cole2000hierarchical, springel2005simulations}. The availability and regulation of gas in these systems play crucial roles in shaping their physical properties. Dense environments at high redshifts, especially along cosmic filaments and their cores, provide unique conditions through the supply of cold gas at an accelerated pace, resulting in the enhancement of early galaxy growth.

Some of the earliest evidence for environmental effects on galaxy evolution was observed among galaxy members in local clusters \citep{butcher1978evolution, dressler1980galaxy}. These studies revealed clear differences in galaxy colors and morphologies compared to their field counterparts, indicating the influence of dense environments on galaxy evolution. Local clusters are dense and dynamically hot, which can affect the galaxies therein through several processes, such as strangulation/starvation \citep{larson1980evolution}, ram-pressure stripping \citep{gunn1972infall, abadi1999ram}, and tidal stripping \citep{moore1996galaxy, moore1998morphological}.

These environmental effects are expected to differ at higher redshifts, where most galaxy systems are still young and not yet virialized \citep{davis2011virialization, muldrew2015protoclusters}. This evolutionary stage, commonly referred to as the \textit{protocluster} phase, precedes the formation of a hot intracluster medium, which is thought to be responsible for several of the environmental processes observed in mature clusters mentioned above. Recent studies have extended such investigations to higher redshifts ($2<z<6$) and found evidence of environmental effects in some systems, seen as an enhancement of quiescent galaxies \citep{naufal2024revealing, 2025A&A...696A.236P, 2025ApJ...982..153M} and an offset in the mass-metallicity relation \citep{shimakawa2015early}.

A common approach to studying environmental effects is to relate galaxy properties directly to their local environments, rather than solely to the scale of the host systems (i.e., halo masses or size). This requires quantifying the environment using spatial information on the two-dimensional sky plane together with redshift. Several methods have been commonly used to characterize galaxy environments, including fixed apertures, nearest-neighbor (k-NN) estimators, Voronoi tessellation, weighted kernel density estimation (wKDE), and more advanced approaches such as AMICO and PZWav (see \citealt{2015ApJ...805..121D} and \citealt{adam2019euclid} for reviews of these methods). 

Studies using these approaches have shown that galaxy properties exhibit significant correlations with the environment up to $z\sim2$, with higher density regions hosting galaxies that, on average, have lower star formation rates (SFRs) and specific star formation rates (sSFRs), but higher stellar masses ($M_*$) \citep{darvish2016effects, 2020ApJ...890....7C, 2024ApJ...961...39S, 2024ApJ...966...18T, hatamnia2025large, Liu25}. Interestingly, some studies have reported a reversal of these trends, particularly in SFR and sSFR, around $z \sim 1$ \citep{elbaz2007reversal, cooper2008deep2}, while more recent studies have suggested that this transition may occur at even higher redshifts ($z > 2$; \citealt{lemaux2022vimos, taamoli2024cosmos2020}). This effect may be attributed to the higher rates of merger events or more efficient gas supply during this epoch. However, this observed behavior could be affected by cosmic variance or selection effects, hampering us from obtaining a robust consensus. Recent work modeling selection effects in environmental studies shows that an sSFR-environment reversal is not expected at $z\sim2-3$, and if present, occurs at $z>3$ \citep{chartab2025latis}.

In addition, physically interpreting this reversal effect requires a framework describing how gas regulation changes across cosmic time and halo mass. Numerical simulations by \citet{2006MNRAS.368....2D} suggest that gas accretion is linked to halo mass, dividing the accretion mode into three regimes: cold, hot, and cold in hot. At low redshifts, massive systems exhibit shock-heated halos that can efficiently heat incoming cold gas, preventing new star formation. However, at high redshift, at $z>2$, gas accretion can remain cold through dense filamentary streams \citep{dekel2009cold}. The transition epoch of this accretion mode—when cold-mode accretion begins to shut off, and hot-mode accretion becomes dominant, corresponds to the “cosmic noon”, the period during which the comoving cosmic star formation rate density reaches its peak \citep{madau2014cosmic}.

Several observational studies support this scenario. For example, the ratios of Ly$\alpha$ luminosity and star formation rate (SFR) to the expected baryon accretion rate (BAR) show a clear flattening above the cold-stream threshold, while exhibiting a scaling relation in the hot-accretion regime. \citep{daddi2022evidence, sillassen2024noema}.
Wide-field dual narrow-band imaging observations of Ly$\alpha$ and H$\alpha$ emitters in and around young protoclusters at $z=2-2.5$ have revealed the physical association of HI gas in dense environments such as cores and surrounding filaments probably fed by cold-mode accretions, as indicated by poor visibility of Ly$\alpha$ emitters with respect to H$\alpha$ emitters in those dense environments (\citealt{shimakawa2017direct, daikuhara2025association}, M. Funaki et al. 2026, in preparation).
Other studies also point to a transition from the metal-deficient to metal-rich systems around $z\sim2$, compared to field galaxies, as evidence for a change in gas accretion mode \citep{chartab2021mosdef, perez2024enhanced, adachi2025enhanced}. 

Although significant progress has been made, studies of galaxy environments at high redshift are often limited to contiguous survey fields, raising concerns that some reported trends may be influenced by cosmic variance \citep[e.g.,][]{trenti2008cosmicvariance}. This limitation complicates the interpretation of environmental effects and their connection to the underlying physical processes. The pure-parallel imaging mode, enabled for HST \citep{trenti2012overdensities,bradley2012brightest,schmidt2014luminosity,morishita2018bright,morishita2020superborg} and now for JWST \citep{williams2025panoramic,2025ApJ...983..152M}, provides several key advantages for overcoming these limitations by observing many independent pointings distributed across the sky with exquisite sensitivity. The pure-parallel imaging strategy substantially reduces the impact of cosmic variance, while the depth of JWST NIRCam observations enables the detection of lower-mass galaxies out to higher redshift compared to that of HST. In addition, the multi-band coverage of the survey yields high-quality photometric redshift measurements, which are well-suited for constraining galaxy clustering. Together, these characteristics make the mode an ideal approach to a systematic and unbiased search for galaxy overdensities and to investigating how galaxy properties depend on both halo mass and local environment at high redshift.

In this paper, we perform a search for galaxy overdensities at $1.5<z<7$ in 20 NIRCam pointings observed in the JWST Cycle 2 pure-parallel NIRCam imaging survey, Bias-free Extragalactic Analysis for Cosmic Origins with NIRCam \citep[BEACON;][]{2025ApJ...983..152M, zhang2026beacon, kreilgaard2026beacon}. Our study adopts the weighted-adaptive kernel density estimation (wKDE) method to fully utilize our photometric redshift probability distribution function. This paper is organized as follows. In Section~\ref{sec:data}, we summarize the BEACON observations, parameter estimations, and the construction of our final galaxy sample for the overdensity search. Section~\ref{sec:search} introduces the weighted-adaptive KDE method. In Section~\ref{sec:results}, we outline the criteria used to define the final catalog of galaxy overdensities, present the results of our overdensity search, and highlight several prominent systems from the catalog. We also describe the procedures used to estimate the physical properties of the identified systems. In Section~\ref{sec:discussion}, we discuss the properties of galaxies both on the system scale, which links to total dark matter mass, and on the local scale, which links to the overdensity factor. Finally, we summarize our findings in Section~\ref{sec:conclusion}.

Throughout this paper, we adopt the standard cosmology model $\Lambda$CDM with $\Omega_m=0.3$, $\Omega_\Lambda=0.7$, and $H_0=70\text{ km s}^{-1} \text{ Mpc}^{-1}$. Photometric measurements are reported in the AB magnitude system \citep{okegun,fukugita1995galaxy}.

\section{Data and Sample} \label{sec:data}
\subsection{Observation}

We use the JWST photometric datasets from the Cycle 2 NIRCam pure-parallel imaging program, BEACON (GO-3990; PI: T. Morishita; Co-PIs: C. Mason, T. Treu, M. Trenti; \citealt{2025ApJ...983..152M}). The configuration of this survey follows a parallel observing mode, in which multiple JWST science instruments operate simultaneously, with one serving as the primary instrument and others as parallel instruments observing different regions of the focal plane. Therefore, the configuration of the parallel observation on each visit to the field is different, particularly the exposure times and filter configurations. To ensure the uniformity of our analysis and high accuracy on the photometric redshift, we select the 20 fields from the full BEACON DR2 \citep{kreilgaard2026beacon} with a requirement of a minimum of 6 filters, with the shortest filter F090W and the longest filter either F444W/F480M. This ensures that we trace the Balmer break features at $z=1.5-7$. From this collection of fields, we have a total area of $\sim400$\,arcmin$^2$ and achieve $10\sigma$ limiting depths in the F444W band of $m_{\mathrm{F444W}} \approx 27.3$--$28.6$ across the field.

The details of the photometry extraction for our data are detailed in \citet{2025ApJ...983..152M}, and we briefly summarize the procedure here. Sources were identified by their detection image, which is the rms-weighted combination of F277W + F356W + F444W, using \textsc{SourceExtractor}  \citep[][]{1996A&AS..117..393B}. We adopt the aperture flux with $0.\!''16$ radius on the PSF-matched image on F444W, scaled by $f_{auto, F444W}/f_{aper, F444W}$ to maintain the colors while using total fluxes for deriving galaxy parameters. All of the measured fluxes are corrected for the galactic dust reddening value at the corresponding field location taken from the NASA/IPAC Extragalactic Database \citep{1998ApJ...500..525S, 2011ApJ...737..103S}. We use the reddening curve of the Milky Way dust law \citep{cardelli1989relationship}.

\subsection{Photometric Redshift} \label{sec:photoz}
We create a photometric redshift catalog utilizing \textsc{eazy-py} \citep{2008ApJ...686.1503B}, which provides us with direct measurements for the zero-point corrections for each catalog. Photometric redshift fitting is performed using the \textsc{eazy} software by fitting linear combination templates to the data using chi-square minimization. In contrast to previous BEACON studies, we carry out our own photometric redshift SED fitting to obtain the complete photometric redshift probability distribution function (PDF). We choose the template library \texttt{tweak\_fsps\_QSF\_12\_v3.param}, which consists of 12 templates derived from the flexible stellar population synthesis model (FSPS) code \citep{2010ApJ...708...58C} and offers the lowest outlier fraction. We set the redshift range to $0<z<12$, with $0.01(1+z)$ increment. We use the magnitude prior provided by \textsc{eazy}, \texttt{prior\_F160W\_TAO.dat}, using F444W.

To account for the uncertainties of the photometric redshift, we use its full probability distribution function, $P(z)$, for our main analysis below. Using the complete PDF of the redshift, we can define multiple redshift values, such as $z_{peak}$, which represents the redshift with the highest probability in the PDF, and $z_{50}$, which represents the median-posterior redshift and is more representative of the PDF’s central tendency. Throughout this paper, when defining a representative redshift value beyond its PDF, we use $z_{50}$.

\begin{figure}
    \centering
    \includegraphics[width=\linewidth]{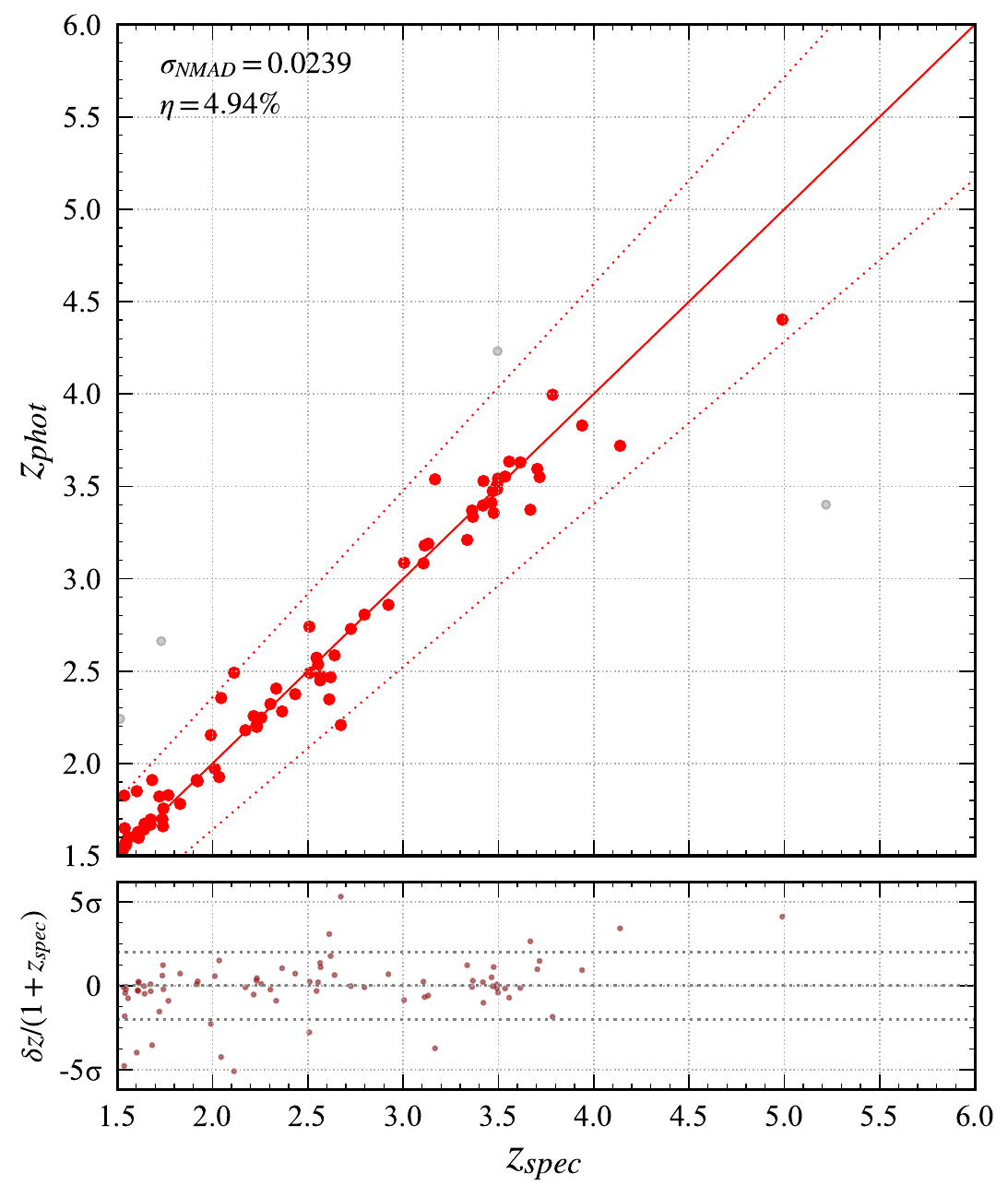}
    \caption{Comparison between \textsc{eazy}-derived photometric redshifts and spectroscopic redshifts compiled from various previous studies. The red dotted line corresponds to the outlier limit between photometric redshift and spectroscopic redshift. Grey dots correspond to the outliers.}
    \label{fig:specz}
\end{figure}

We assess the accuracy of the photometric redshift by cross-matching the galaxy sample from the BEACON fields located in regions overlapping with one of four legacy fields (COSMOS, EGS, GOODS-S, and UDS), where a spectroscopic redshift catalog from various surveys is available, that is, 3DHST \citep{2012ApJS..200...13B, 2014ApJS..214...24S}, COSMOS \citep{2025arXiv250300120K}, GOODS/FORS2 \citep{2008A&A...478...83V}, GOODS/VIMOS \citep{2010A&A...512A..12B}, VANDELS \citep{2021A&A...647A.150G}, CANDELS/HLSP \citep{2011ApJS..197...35G, 2011ApJS..197...36K}, GOODS/JADES \citep{bunker2024jades, d2025jades, curtis2025jades}, and CEERS \citep{finkelstein2023ceers}.  The comparison of our photometric redshift with the spectroscopic redshift is presented in Figure~\ref{fig:specz}. We calculate the median absolute deviation of our photometric redshift fitting by using
\begin{equation}
    \sigma_{\text{NMAD}} = 1.48\times\text{median}\left(\left|\frac{\Delta z-\text{median}(\Delta z)}{1+z_{spec}}\right|\right)
\end{equation}
with $\Delta z=z_{phot} -z_{spec}$. From this measurement, we get $\sigma_{\text{NMAD}}=0.024$, with the number of outliers, defined by $\Delta z/(1+z)>0.15$, reaching below 5\% for galaxies at $z_{phot} > 1.5$. This assures that the photometric measurements of our sample are robust enough to perform an analysis for searching for overdensity. 

\subsection{Spectral Energy Distribution Fitting} \label{sec:sed}
We derive the basic physical parameters of galaxies through a photometric SED fitting analysis using \textsc{CIGALE} \citep[Code Investigating GALaxy Emission;] []{2019A&A...622A.103B}. All available broadband filters are used for each field, with a minimum of six filters common to all fields (F090W, F115W, F150W, F277W, F356W, and F444W). Throughout the fitting process, we fixed the redshift to the median photometric redshift value, $z_{50}$, for each galaxy.

We adopt a delayed exponentially declining star formation history (SFH) model of the form
\begin{equation}
\mathrm{SFR}(t) \propto \frac{t}{\tau^2} \exp{\left(-\frac{t}{\tau}\right)},
\end{equation}
where the onset age of star formation, $t_0$, ranges from $10^{7.6}$ to $10^{10.1}$ years, and the characteristic timescale, $\tau$, ranges from $10$\,Myr to $10$\,Gyr.

The fitting assumes a \citet{2003PASP..115..763C} initial mass function (IMF) and employs simple stellar population (SSP) templates from \citet{2003MNRAS.344.1000B}. The stellar metallicity is varied over the range $Z = 0.0001$–$0.05$, motivated by \citet{bellstedt2025progeny}, for which this choice yields stellar masses and star formation rates with minimal bias.

Nebular emission from \citet{inoue2011rest}, which was generated using \textsc{cloudy} \citep{ferland1998cloudy, ferland20132013}, is included using templates with an ionization parameter of $\log{U} = -2.0$. For dust attenuation, we adopt the modified \citet{2000ApJ...533..682C} law with a nebular color excess of $E(B-V) = 0.0$–$1.5$ and a total-to-selective extinction ratio of $R_V = 4.05$. Rest-frame $U-V$ and $V-J$ colors are also computed to be used for classifying galaxies as quiescent or star-forming using the UVJ diagram.

Note that we verified that the derived parameters for the main sample remain consistent within the uncertainties when using either $z_{50}$ or the peak redshift $z_{\mathrm{peak}}$, indicating that our results are robust against the choice of redshift estimator.

\subsection{Quiescent and Star-forming Galaxies}
To investigate the environmental impact on the galaxy quenching process, we classify a population of quiescent galaxies. We select the population of quiescent galaxies utilizing the modified rest-frame UVJ classification from \citet{2012ApJ...745..179W} following \citet{2015ApJ...811L..12W}, which is defined as
\begin{align*}
    U - V &> 0.8 (V-J) + 0.7 \\
    U-V &> 1.3
\end{align*}
We note that the vertical cut of the UVJ selection is not applied here. By doing so, we allow the inclusion of dusty and quiescent galaxies that would have been excluded by the vertical cut \citep{2014ApJ...788...28V}. 

Additionally, we consider an alternative separation of quiescent and star-forming galaxies based on a specific star formation rate (sSFR) cut following \citet{carnall2023surprising}, defined as sSFR\,$< 0.2/t_H$, where $t_H$ is the age of the Universe at the corresponding redshift. We confirm that the choice of quiescent galaxy selection criterion does not affect the final interpretation of this paper. Therefore, we adopt the UVJ selection throughout our analysis.

\subsection{Galaxy Sample}\label{sec:galsample}
We select our galaxy sample in every field based on the data reduction and photometric measurement criteria:
\begin{enumerate}[noitemsep]
    \item To minimize stellar contamination in our galaxy sample,  we exclude objects with \texttt{CLASS\_STAR} $>0.90$ from the \texttt{SExtractor} output.
    \item Due to different coverage for the short and long channels (caused by the detector chip gaps in the short channel imaging), we only use galaxies covered by the full filter set. For this, we use \texttt{SExtractor} flags, which indicate the photometry of the object to be not truncated or located near the edge of the mosaic, \texttt{FLAG\_SW} $<8$, and \texttt{FLAG\_LW} $<8$.
    \item To maintain uniformity across all fields, we apply a magnitude cut based on the 10$\sigma$ magnitude limit of BEACON\_1526+3560, the shallowest field in our sample, requiring $m_{\rm F444W} < 27.3$\, ABmag.
\end{enumerate}
In addition, we include constraints from photometric redshift and SED fitting measurements:
\begin{enumerate}[noitemsep]
    \item Quality of fit (reduced $\chi^2$)  for the photometric redshift from the \textsc{eazy} output in $0.01<\chi_\nu^2<5$.
    \item Photometric redshift ($z_{50}$) range in $1.25<z_{50}<8$.
    \item Photometric redshift error, $\Delta z/(1+z)<0.4$.
    \item Total integrated photometric redshift PDF at $\Delta z\pm0.5$, $\int_{z_{phot}-0.5}^{z_{phot}+0.5} P(z)dz >0.7$, to constrain the redshift PDF mostly located near the photometric redshift value.

\end{enumerate}
From our 20 fields, we find 8287 galaxies that satisfy all of the above criteria.

\subsection{Mass Completeness}
The clustering of galaxies is known to depend on their stellar mass, where more massive galaxies are more clustered compared to lower mass galaxies. This trend has been demonstrated in previous studies at redshifts below $z \sim 2$ \citep{2008A&A...478..299M, 2011ApJ...728...46W, 2014A&A...568A..24B}. This relationship has also been observed to persist at higher redshifts \citep{2018A&A...612A..42D, 2018MNRAS.481.4885Q, 2018PASJ...70S..11H}. Therefore, it is important to consider the stellar mass distribution of our sample before searching for overdensities. In this study, we estimate the stellar mass completeness limit of our sample to assess whether a mass cut should be applied.

The stellar mass completeness limit is determined following the method described by \citet{2010A&A...523A..13P}. We aim to estimate the minimum stellar mass that can be reliably detected based on the magnitude limit we have defined in the previous section. For each galaxy in our sample, we compute its limiting stellar mass ($M_{\mathrm{lim}, i}$) by scaling its derived stellar mass ($M_{i}$) to our magnitude limit:
\begin{equation}
\log{(M_{\mathrm{lim}, i})} = \log{(M_{i})} + 0.4(m_{\mathrm{\rm F444W}} - m_{\rm F444W, lim}).
\end{equation}

This calculation is performed within redshift bins of width $\Delta z = 0.5$. In each bin, we select the 20\% faintest galaxies. We then calculate the 90\% percentile of their $M_{\mathrm{lim}, i}$ distribution, which is adopted as the stellar mass completeness limit ($M_{\mathrm{lim}}$). This definition corresponds to the stellar mass above which the sample is expected to be 90\% complete at a given redshift. 

\begin{figure}
    \centering
    \includegraphics[width=\linewidth]{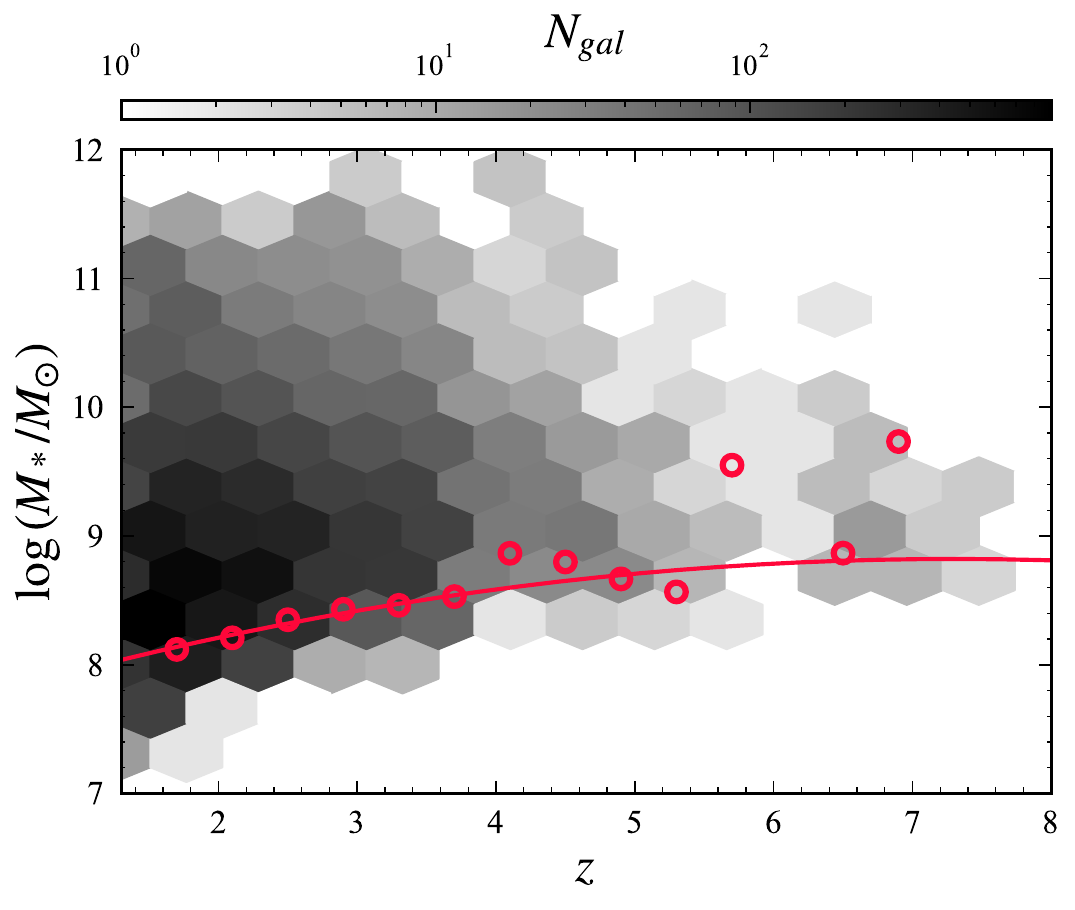}
    \caption{Mass completeness limit for the 20 selected BEACON fields. The 2D hexagon bins show the stellar mass distributions at $0<z<8$. Red open circles show the mass completeness limits calculated in the redshift bins with a step of 0.5, with the red line indicating its quadratic fitted function.}
    \label{fig:completeness}
\end{figure}

The derived mass completeness limit is shown in Figure~\ref{fig:completeness}. However, due to a relatively small sample size, the completeness estimate at $z\sim6$ resulted as an outlier. To mitigate this effect, we determine the stellar mass completeness limit as a function of redshift by interpolating the measurements using a quadratic fit.

Using this approach, we find that the stellar mass completeness limit increases from $\log{(M_*/M_\odot)} = 8.05$ at $z \sim 1.5$ to $\log{(M_*/M_\odot)} = 8.99$ at $z \sim 7$. Based on these results, we adopt a conservative stellar mass cut of $\log{(M_*/M_\odot)} = 9.0$ for the analysis presented below. However, we note that no stellar mass cut is applied, on top of the already applied magnitude cut, when constructing the overdensity maps (Section~\ref{sec:search}) or measuring the overdensity factor (Section~\ref{sec:overview}). Instead, all galaxies are retained to represent the underlying galaxy clustering in our sample.

\section{Systematic Search for Overdensity Candidate} \label{sec:search}
\begin{figure*}
    \centering
    \includegraphics[width=\linewidth]{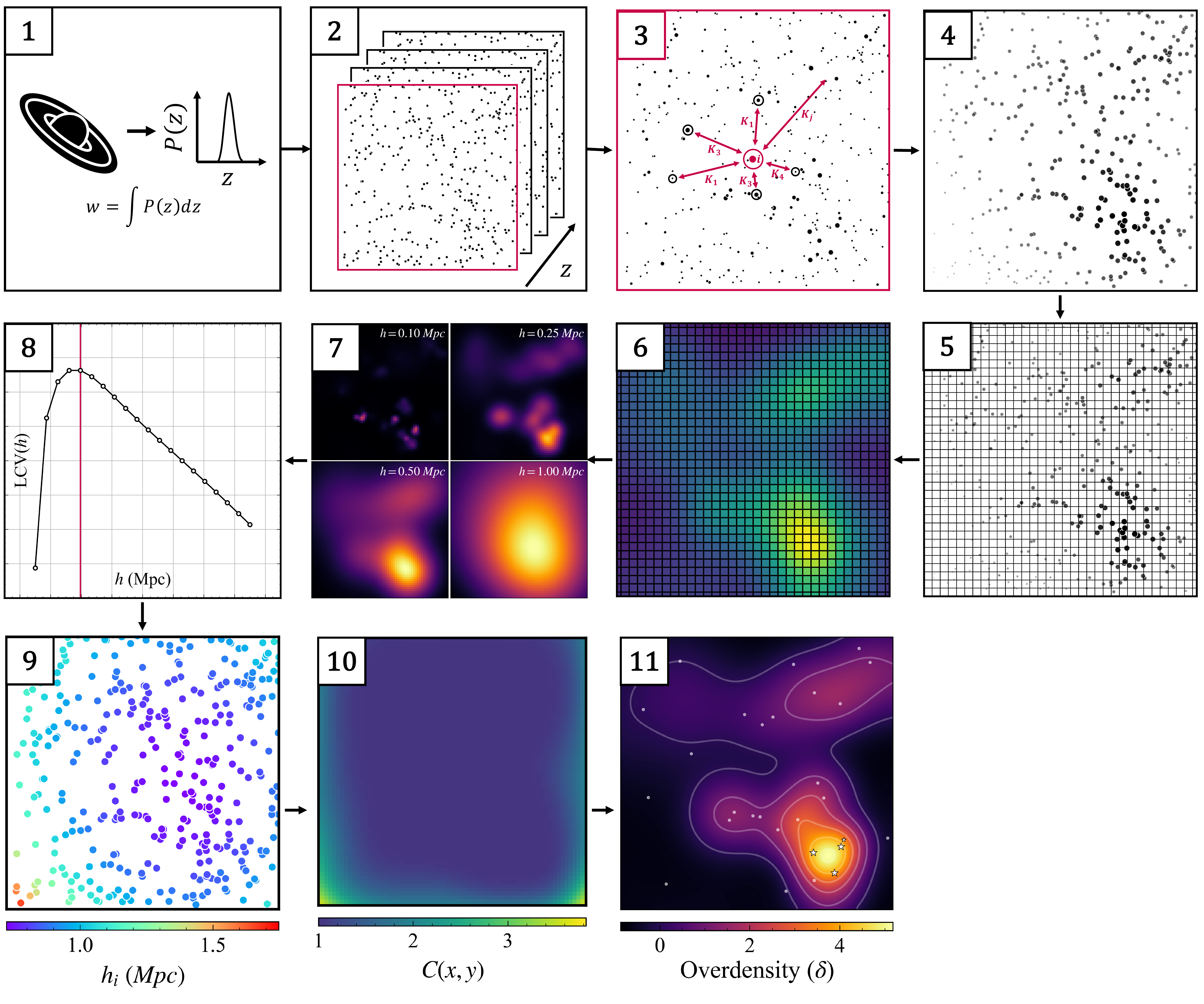}
    \caption{The flowchart of the overdensity map creation, following the direction of arrows. (1) Weighting of each galaxy is calculated by its photometric redshift PDF. (2) Each field is divided into redshift slices following the definition from Section~\ref{sec:zslice}. (3) For each slice (red box), calculate surface density using Equation~\ref{eq:1} around each galaxy (red dot) based on all surrounding galaxies. Different dot sizes represent different weights. (4) The surface density of each galaxy, where a larger size and opacity correspond to a larger local surface density. (5) Dividing the mosaic into a defined pixel grid. (6) Calculation on each pixel grid $(x,y)$ following Equation~\ref{eq:sigmaxy}. (7) Iterative calculation of global bandwidth. (8) Optimal global bandwidth selection based on LCV. (9) Map of local bandwidth of each galaxy calculated with $h_i = h\times\lambda_i$. (10) Edge correction factor map. (11) The final overdensity map with contour step of $\Delta \delta =1$. Each data point corresponds to galaxies that are located inside the redshift slice. Star markers show galaxies that are spatially included in the overdense region.}
    \label{fig:diagram}
\end{figure*}

To identify overdensities within our sample, it is important to quantify the local galaxy densities and investigate how galaxy parameters relate to their local environments. This can be done by reconstructing a galaxy overdensity map, either by creating a comprehensive 3D map or by slicing it into a 2D map by using a redshift slice. A 2D map is suitable for the photometric redshift sample because of its large redshift uncertainties. The framework for this map construction is presented below and illustrated in Figure~\ref{fig:diagram}. We will refer to the numbered blocks in the flowchart as \textit{Step N} throughout this section.

\subsection{Redshift Slices} \label{sec:zslice}

We consider the selection of redshift slices used to create the 2D overdensity map in each field. Previous studies defined their redshift slice width based either on the photometric redshift uncertainty \citep{2015ApJ...805..121D} or on the comoving size ($\Delta\chi$) of galaxy clusters accounting for redshift-space distortion (RSD) effects \citep[e.g.,][]{2020ApJ...890....7C, 2024ApJ...966...18T}.
We explore both approaches. We first evaluate the photo-z–based approach. The typical redshift uncertainty of our sample is estimated from 20 BEACON fields, and we defined this uncertainty as the median absolute deviation of the normalized redshift error,
$\Delta z = (z_{84}-z_{16})/(1+z_{\mathrm{phot}})$, where $z_{16}$ and $z_{84}$ represents the 16th and 84th percentiles of the photometric redshift distribution,
which results in $\sigma_z = 0.07$ across all redshifts. This corresponds to comoving scales of $\Delta\chi\sim140$\,cMpc at $z=1.5$ and $\sim40$ cMpc at $z=5.0$.

Alternatively, we consider a physically motivated approach based on the expected size of massive protoclusters. The corresponding redshift width at a given redshift can be calculated as
\begin{equation}
\Delta z=\frac{\Delta\chi H_0}{c}\sqrt{\Omega_m(1+z)^3+\Omega_\Lambda}.
\end{equation}
We adopt a fixed comoving width $\Delta\chi = 20$ cMpc, which corresponds to the predicted spatial extent of the most massive (Coma-like) protoclusters at $z \sim 5$ \citep{2013ApJ...779..127C}.
However, this width corresponds to a median redshift slice of $\sigma_z \simeq 0.03$, considerably smaller than the typical photometric redshift uncertainty of our sample. To evaluate the impact of slice width, we run our overdensity mapping algorithm using both $\Delta\chi = 20$ cMpc and $\Delta\chi = 90$ cMpc. We find that the choice of redshift slice width does not significantly affect the resulting overdensity maps within our uncertainties (see Appendix~\ref{app:choice}), as it does not alter the spatial distribution of overdense regions nor the derived overdensity values, consistent with the findings of \citet{2024ApJ...961...39S}.
Therefore, to maintain finer redshift resolution in our final maps and catalogs, we adopt a comoving width of $\Delta\chi = 20$ cMpc for the redshift slice definition. We perform the calculation from $z=1.5$ to $z=7.0$, resulting in 214 redshift slices in total. 

For each redshift slice with $z \in [z_a,z_b]$, the weight of each galaxy can be calculated as
\begin{equation}
    w = \int^{z_b}_{z_a} P(z)~dz
\end{equation}
(Step 1 in Figure~\ref{fig:diagram}).
 
\subsection{Weighted-Adaptive Kernel Density Estimation}
We apply the weighted adaptive kernel density estimation (wKDE) introduced by \citet{2015ApJ...805..121D}. This can be done by performing an iterative procedure of adjusting parameters adaptively to calculate the surface density, following the \citet{2020ApJ...890....7C} workflow, with some modifications. Unlike the standard calculation of kernel density estimation (KDE), this method applies the weight of galaxies based on their full PDF with an adaptive algorithm to adjust the creation of the surface density map. The adaptive algorithm takes place in two parts: applied for calculating the smoothing parameter for the map, globally on each redshift slice, and locally at each galaxy location based on its local density.

For each redshift slice $z$ (Step 2), we estimate the local surface density $(\sigma_i)$ for each galaxy $i$, based on the sum of the kernel with the other galaxies $j$ in our sample and their weight $w_j$ calculated by the redshift PDF (Step 3), 
\begin{equation} 
    \sigma_i=\frac{1}{\sum_{j=1, j\neq i}^{N}w_j} \sum_{j=1, j\neq i}^{N} w_j K(\bm{r_i}, \bm{r_j}, h) \label{eq:1}
\end{equation}
where N is the number of galaxies in the sample, $K$ is the kernel function, defined by the position of the galaxy $i$ ($\bm{r_i}$) and the rest $j$ ($\bm{r_j}$), with $h$ as the bandwidth of the kernel function, which provides the smoothing parameter for density maps. The $\sigma_i$ is calculated for every galaxy inside the sample, on every redshift slice (Step 4).

We choose a 2D symmetric Gaussian kernel, considering each NIRCam mosaic is small enough for a circular kernel (e.g., the von Mises kernel) to be used. A 2D symmetric Gaussian kernel is defined as:
\begin{equation}
    K(\bm{r_i}, \bm{r_j}, h)=\frac{1}{2\pi h^2}\exp{\left(-\frac{|\bm{r_i}-\bm{r_j|^2}}{2h^2}\right)} 
\end{equation}
In this case, when calculating the local surface density for each galaxy $\sigma_i$, we first have to define a global kernel bandwidth ($h$) for each redshift slice. It is possible to choose a specific well-established value based on the observation, such as the typical size of the local clusters, as adopted by \citet{2015ApJ...805..121D}, by choosing $h =$ 0.5 Mpc, which corresponds to the typical value of $R_{200}$ for X-ray clusters and groups in the COSMOS field. However, this specific value might be unsuited in our study, due to our lower-mass limit, observational setup, and redshift range. Therefore, we need to choose our own global bandwidth.

It is important to keep the bias-variance tradeoff in the sense that choosing too narrow a bandwidth could give rise an overestimate of variance (undersmoothing), which shows unrelated detailed features (e.g., shot noise) in the density map, while choosing too wide a bandwidth could give rise to a high-bias map (oversmoothing), which creates a too simple map by removing important finer features \citep{2015ApJ...805..121D, 2020ApJ...890....7C}. In this sense, the choice of global bandwidth must depend only on the distribution of the sample across the mosaic. 

Therefore, we employ Likelihood-Cross Variance (LCV) \citep{hall1982cross, 2020ApJ...890....7C} to choose the best global width for each redshift slice. The calculation is done iteratively by grid-searching the global bandwidth value that maximizes the likelihood value, based on our data (Step 7-8). LCV is defined as:
\begin{equation}
    LCV(h) = \frac{1}{N}\sum_{k=1}^N\log{(\sigma_{-k}(\bm{r}))}
\end{equation}
where $N$ is the total number of galaxies in the sample, and $\sigma_
{-k}$ is the surface density for galaxy $k$ excluding itself, which is already calculated by Equation \ref{eq:1}. We obtain typical value of $h=0.28~(1.50)$ cMpc at $z=1.5~(7.0)$. This calculation marks the first adaptive part. 

Within each redshift slice, different regions of the mosaic may have a variety of local densities. Therefore, we cannot directly apply the LCV-chosen global bandwidth to the final calculation, as it will cause the problem of map creation previously discussed. Thus, we also use the second adaptive part by adjusting the global bandwidth $h$ to each galaxy based on its local density (Step 9), defined as local bandwidth $h_i = h\times\lambda_i$, where $\lambda_i$ is proportional to the inverse of local density \citep{abramson1982bandwidth, silverman2018density, 2015ApJ...805..121D},
\begin{equation}
    \lambda_i=\left(\frac{G}{\sigma_i}\right)^{0.5}
\end{equation}
where $G$ is the geometric mean of all the estimated $\sigma_i$ values, defined as,
\begin{equation}
    \log{(G)} = \frac{1}{N} \sum_{i=1}^N \log{(\sigma_i)}
\end{equation}
To construct a complete map of each redshift slice, we can calculate the surface density on each predefined 2D grid, $r=(x,y)$, on the mosaic, either on each pixel or pixel bin, that is group of pixels that are binned together, similar to Equation~\ref{eq:1},
\begin{equation}
    \sigma(x,y)\equiv\sigma(\bm{r})=\frac{1}{\sum_{i=1}^{N}w_i} \sum_{i=1}^{N} w_i K(\bm{r}, \bm{r_i}, h_i) \label{eq:sigmaxy}
\end{equation}
We perform a calculation on the 50 kpc pixel binning (Step 5-6). This corresponds to $2.\!''3$ ($1.\!''2$) at $z=1.5 ~(7.0)$. We adopt this choice by considering that this scale is larger than galaxy sizes but much finer than the global bandwidth and typical cluster scale. 

\begin{figure*}
    \centering
    \includegraphics[width=\linewidth]{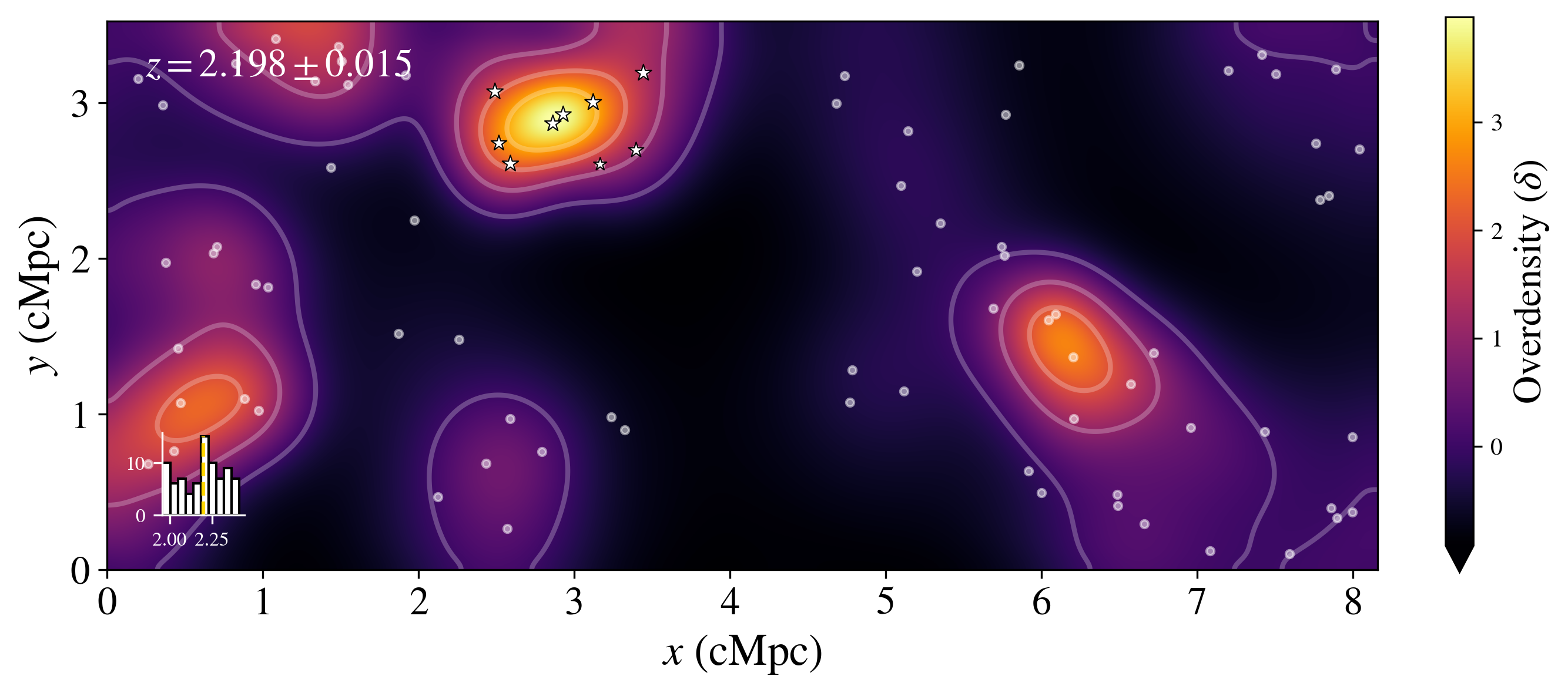}
    \caption{Example of a two-dimensional overdensity map of the BEACON\_0055–3749 field at $z = 2.20$. The map is plotted in the $x$–$y$ plane, converted into comoving coordinates in units of cMpc. White contours indicate one-step increments of the overdensity factor ($\delta$). White dots mark galaxies that have a probability of being included within the map, accounting for their photometric redshift uncertainties. The star symbols denote galaxies located within the $>1\sigma$ overdensity contour, representing the probable members of the overdense region. The inset panel in the bottom-left corner shows the photometric redshift distribution of galaxies within the redshift slice.}
    \label{fig:ovd map}
\end{figure*}

 \subsection{NIRCam Chip Gap Filling}
As the NIRCam instrument contains a $\sim44''$ chip gap separating its two $2'\times2'$ modules, the resulting galaxy distribution is discontinuous across the mosaic. To produce contiguous density maps, we insert artificial galaxies into the chip gap in each redshift slice following \citet{taamoli2024cosmos2020}. The artificial galaxies are generated such that (1) their number density within the chip gap matches the average galaxy number density measured across the NIRCam mosaic in the corresponding redshift slice, and (2) their weights are set to the average galaxy weight in that redshift slice, ensuring that the inclusion of artificial galaxies does not systematically change the density map.

\subsection{Edge Correction}
\citet{2020ApJ...890....7C} and \citet{2024ApJ...966...18T} addressed the problem of calculating the surface density on the edge of the data mosaic using the kernel density estimation algorithm. The kernel function is continuous. On the limited or bounded data, the kernel density estimation encounters a discontinuity and spillover to the region outside the boundary, which causes bias at and near the edges of the data. This effect is referred to as the boundary effect.

The actual survey is always limited by the instrument's field of view, which serves as the boundary to the kernel density estimation. This limitation can cause an underestimation of the surface density at the edge of the mosaic, introduced by the boundary effect. This problem can be easily addressed in a large field by cutting the edge part, because the edge region is small compared to the whole survey area and does not significantly affect the overall map. However, it is crucial to account for the boundary effect in parallel surveys, such as ours, as the random pointings increase the edge-over-area ratio.

There are various ways to perform the edge correction introduced by the boundary effect, namely the reflection method \citep{schuster1985incorporating}, the renormalization method \citep{diggle1985kernel, jones1993simple}, the transformation method \citep{marron1994transformations}, the local polynomial method \citep{jones1993simple, cheng1997boundary}, the pseudo-data method \citep{cowling1996pseudodata}, and the boundary kernel methods \citep{muller1991smooth, chen1999beta, chen2000probability}. 

We adopt the renormalization method, which is the most intuitive and computationally efficient approach to correct for edge effects on the density map, following \citet{2020ApJ...890....7C}, and originally formulated by \cite{jones1993simple}. In this method, the integral of each kernel is normalized to unity, ensuring that the estimated surface density remains unbiased even near the survey boundaries.

\begin{deluxetable*}{cccccccc}
\tablecaption{Catalog of Galaxy Overdensities \label{tab:ovdcat}}
\tablehead{
\colhead{\parbox[c]{2cm}{\centering ID\\[3pt](1)}} &
\colhead{\parbox[c]{2cm}{\centering RA$_{\rm ovd}$ (deg)\\[3pt](2)}} &
\colhead{\parbox[c]{2cm}{\centering Dec$_{\rm ovd}$ (deg)\\[3pt](3)}} &
\colhead{\parbox[c]{1cm}{\centering $z_{\rm ovd}$\\[3pt](4)}} &
\colhead{\parbox[c]{1cm}{\centering $\sigma_z$\\[3pt](5)}} &
\colhead{\parbox[c]{1cm}{\centering $\delta_{\rm max}$\\[3pt](6)}} &
\colhead{\parbox[c]{1cm}{\centering $N_{\rm gal}$\\[3pt](7)}} &
\colhead{\parbox[c]{2cm}{\centering $\log(M_{\rm h}/M_\odot)$\\[3pt](8)}}
}
\startdata
beacon\_0014-3025\_z1p78 & 3.5433 & -30.4037 & 1.78 & 0.08 & 14.82 & 14 &
$12.42^{+0.11}_{-0.14}$ \\
beacon\_0014-3025\_z2p31 & 3.5609 & -30.4170 & 2.31 & 0.13 & 22.08 & 10 &
$12.55^{+0.12}_{-0.16}$ \\
beacon\_0014-3025\_z2p44 & 3.5679 & -30.3930 & 2.44 & 0.11 & 5.01 & 19 &
$12.91^{+0.08}_{-0.14}$ \\
beacon\_0014-3025\_z2p50 & 3.5598 & -30.4141 & 2.50 & 0.13 & 6.35 & 12 &
$12.53^{+0.12}_{-0.11}$ \\
beacon\_0014-3025\_z2p61 & 3.5496 & -30.4048 & 2.61 & 0.09 & 8.07 & 9 &
$12.23^{+0.03}_{-0.05} $\\
beacon\_0014-3025\_z3p38 & 3.6351 & -30.4260 & 3.38 & 0.11 & 13.37 & 11 &
$12.52^{+0.09}_{-0.10} $\\
beacon\_0015-3034\_z1p55 & 3.6942 & -30.5877 & 1.55 & 0.05 & 4.25 & 11 &
$12.86^{+0.13}_{-0.26}$ \\
beacon\_0015-3034\_z1p56 & 3.6802 & -30.5782 & 1.56 & 0.04 & 8.66 & 12 &
$12.08^{+0.06}_{-0.08}$ \\
beacon\_0015-3034\_z1p59 & 3.6577 & -30.5522 & 1.59 & 0.08 & 6.75 & 13 &
$12.64^{+0.16}_{-0.22} $\\
beacon\_0015-3034\_z1p91 & 3.6724 & -30.5510 & 1.91 & 0.07 & 6.55 & 20 &
$12.31^{+0.07}_{-0.06}$ \\
\vdots & \vdots & \vdots & \vdots & \vdots & \vdots & \vdots & \vdots \\
\enddata
\tablecomments{
(1) Overdensity identification number, formatted with BEACON field + photometric redshift; 
(2) Right ascension of the overdensity center; 
(3) Declination of the overdensity center; 
(4) Overdensity redshift determined by the average redshift of galaxy members;
(5) Redshift scatter of potential galaxy member; 
(6) Maximum overdensity factor; 
(7) Total number of potential galaxy members;
(8) Halo mass estimated in Section \ref{sec:halomass}.
}
\tablecomments{Here, we show the first 10 entries of the catalog. The full catalog is published in its entirety in the machine-readable format.
A portion is shown here for guidance regarding its form and content.}
\end{deluxetable*}

Near the edges, the kernel function is truncated by the boundaries, resulting in a non-unity surface integral:
\begin{equation}
    N(x,y) \equiv N(\bm{r}) = \int_S K(\bm{r}, \bm{r_i}, h_i) < 1.
\end{equation}
where $N(x,y)$ is the integrated kernel normalization evaluated at position $(x,y)$ within the valid region of the field $S$. 
Thus, the expectation value of the estimated surface density becomes
\begin{equation}
    \mathbf{E} [\sigma(x,y)] = N(x,y)\ \sigma_\text{True}(x,y)
\end{equation}
Therefore, to remove this bias and recover an unbiased estimate, we can define the correction factor as the inverse of the kernel normalization:
\begin{equation}
    C(x,y) = \frac{1}{N(x,y)}
\end{equation}
and apply it to the surface density calculation:
\begin{equation}
    \sigma_\text{corr}(x,y) = C(x,y)\sigma(x,y)
\end{equation}
The correction factor, $C(x,y)$, ranges from approximately 1 at the center of the density map to about 3.5 near the field boundaries (Step 10). Hence, ignoring the boundary effect can lead to an underestimation of the surface density by up to a factor of 3.5, consistent with the findings of \cite{2020ApJ...890....7C}. 

\section{Galaxy Overdensities from NIRCam Pure-parallel Fields} \label{sec:results}

\subsection{Identifying Overdensities} \label{sec:overview}

\begin{deluxetable*}{cccccccc}
\tablecaption{Overview of the catalog for overdensity galaxy members\label{tab:galcat}}
\tablehead{
\colhead{\parbox[c]{2cm}{\centering Overdensity ID\\[3pt](1)}} &
\colhead{\parbox[c]{1.2cm}{\centering Galaxy ID\\[3pt](2)}} &
\colhead{\parbox[c]{2cm}{\centering RA (deg)\\[3pt](3)}} &
\colhead{\parbox[c]{2cm}{\centering Dec (deg)\\[3pt](4)}} &
\colhead{\parbox[c]{1.2cm}{\centering $z_{\rm phot}$\\[3pt](5)}} &
\colhead{\parbox[c]{1.2cm}{\centering $P_{\rm gal}$\\[3pt](6)}} &
\colhead{\parbox[c]{1.4cm}{\centering $\delta_{\rm gal}$\\[3pt](7)}}
}
\startdata
beacon\_0014-3025\_z1p78 & 3005 & 3.5429 & -30.4028 & 1.81 & 1.00 & 12.69 \\
beacon\_0014-3025\_z1p78 & 8602 & 3.5635 & -30.4033 & 1.72 & 0.76 & 2.00 \\
beacon\_0014-3025\_z1p78 & 8705 & 3.5691 & -30.4069 & 1.76 & 0.87 & 0.97 \\
beacon\_0014-3025\_z1p78 & 8709 & 3.5622 & -30.4018 & 1.77 & 0.93 & 2.20 \\
beacon\_0014-3025\_z1p78 & 9702 & 3.5699 & -30.4047 & 1.74 & 0.99 & 1.14 \\
\vdots & \vdots & \vdots & \vdots & \vdots & \vdots & \vdots \\
beacon\_2346+1256\_z3p13 & 6095 & 356.5978 & 12.9181 & 3.13 & 0.91 & 3.14 \\
beacon\_2346+1256\_z3p13 & 1795 & 356.6038 & 12.9044 & 3.30 & 0.91 & 0.97 \\
beacon\_2346+1256\_z3p13 & 8861 & 356.6006 & 12.9299 & 3.29 & 0.95 & 1.31 \\
beacon\_2346+1256\_z3p13 & 6527 & 356.5974 & 12.9195 & 3.11 & 0.80 & 3.37 \\
beacon\_2346+1256\_z3p13 & 8949 & 356.6092 & 12.9353 & 3.39 & 0.69 & 2.49 \\
\enddata
\tablecomments{
(1) Overdensity identifier number; 
(2) Galaxy identifier number; 
(3) Right ascension of the galaxy; 
(4) Declination of the galaxy; 
(5) Photometric redshift of the galaxy;
(6) Membership probability of the galaxy as described in point 3 in Section~\ref{sec:overview}; 
(7) Overdensity factor of the galaxy.
}
\tablecomments{We only show the first and last 5 entries of the catalog. The full catalog is published in its entirety in the machine-readable format.
A portion is shown here for guidance regarding its form and content.}
\end{deluxetable*}

We calculate the overdensity factor ($\delta$) at each location $r=(x,y)$ by defining,
\begin{equation}
    \delta(x,y) = \frac{\sigma(x,y)}{\overline{{\sigma}}}-1
\end{equation}
where $\overline{\sigma}$ is the median value of $\sigma$ at each redshift slice (Step 11). We discuss how we determine the median sigma for our calculation in the Appendix~\ref{app:validity}. We show an example of an overdensity map in Figure~\ref{fig:ovd map}. 

We note that at higher redshifts, the overdensity factor becomes increasingly uncertain due to the limited field of view and the decreased number of detected galaxies, which results in an insufficient reference sample of field galaxies. We therefore emphasize that our method is unsuitable for estimating the overdensity factor in small fields at high redshift (see Appendix~\ref{app:choice} and ~\ref{app:validity}). Instead, estimating the overdensity factor through a comparison with the expectation number from the luminosity function offers more reliable results \citep[e.g.,][]{zhang2026beacon, kreilgaard2026beacon}. 

In addition, the overdensity factor calculated using this method includes photometric redshift uncertainties, which incorporate the photo-$z$ blurring effect. This can cause the overdensity factor to be underestimated relative to its true value, particularly in cases where the overdensities correspond to real physical structures.

From these obtained maps, we select galaxy overdensities by applying the following criteria:
\begin{enumerate}
    \item The peak of the overdensity should be at least $4\sigma$, following \citet{2023ApJ...943..153B}. This threshold is also supported by the simulation study of \citet{2016ApJ...826..114T}, which shows that 76\% of systems satisfying this criterion will evolve into actual galaxy clusters by $z=0$, with halo masses $M_h(z=0) \geq 10^{14} M_\odot$.
    \item Potential galaxy members should be above the magnitude limit (Section~\ref {sec:galsample}) and located within the 1$\sigma$ overdensity contour. 
    \item The probability that a galaxy resides within the overdensity, based on its photometric redshift PDF, is at least 68\% within the $1\sigma$ redshift interval around the center of the overdensity.
    \item The overdensity contains at least 7 potential galaxy members, to remove spurious detection. 
\end{enumerate}
Lastly, we merge duplicated overdensities in the same field, if those are within $\Delta z<0.1$ and the projected distance between the centroids is $<$1\,cMpc. We initially obtain 1432 overdensity candidates with a significant number of galaxy members ($n_{gal}\geq7$), where this number still suffers from the duplication problem. After performing the merging step, we obtain a clean sample of \nod overdensities. We show the example of the first 10 overdensities in Table~\ref{tab:ovdcat}. For each corresponding overdensity, we also show the example of the galaxy member catalog in Table~\ref{tab:galcat}.

\subsection{Estimating the Total Dark Matter Mass of Overdensities} \label{sec:halomass}

\begin{figure*}
    \centering
    \includegraphics[width=\textwidth]{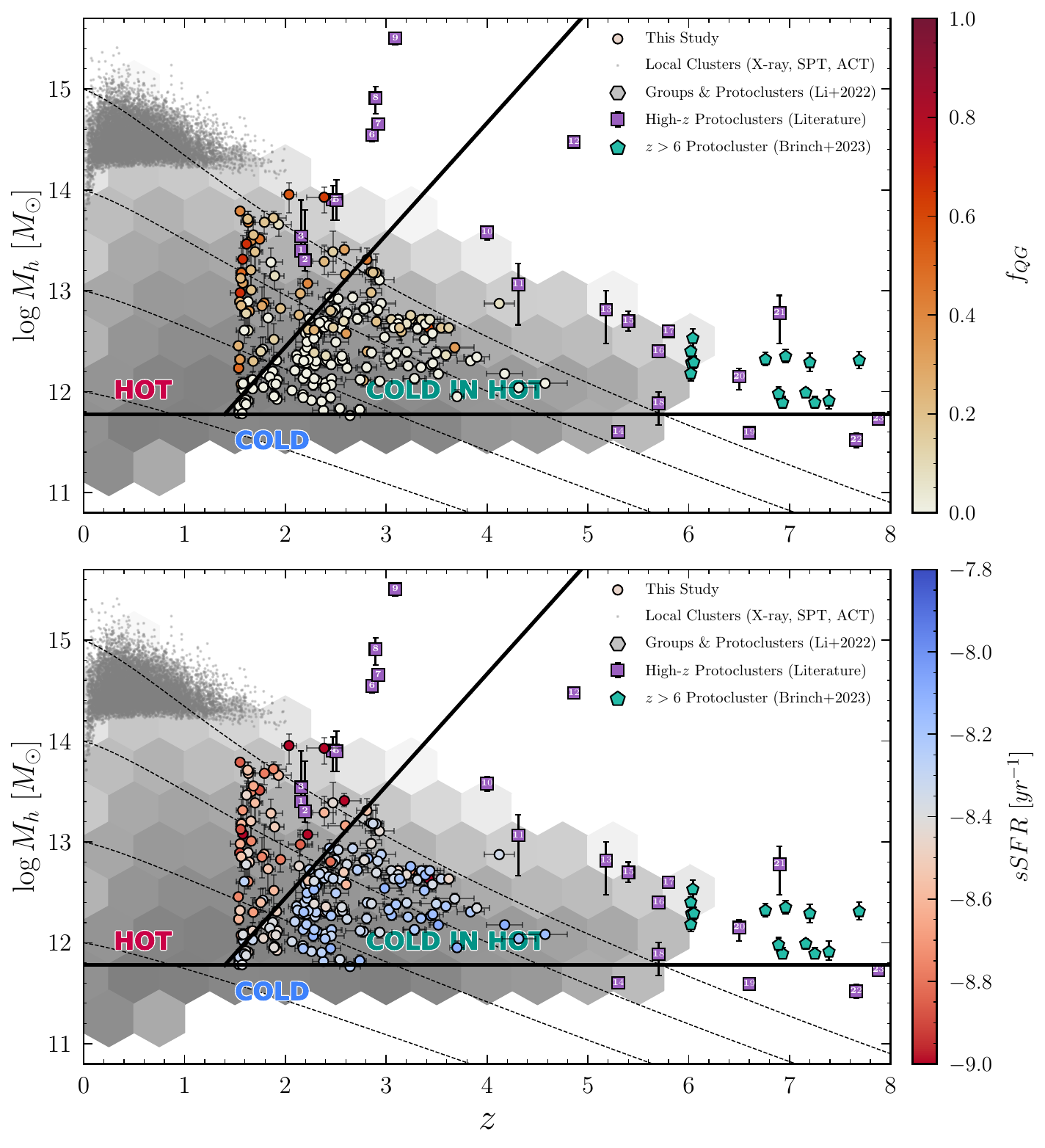}
    \caption{The distribution of total dark matter mass of our overdensities as a function of redshift. The error bar represents 1$\sigma$ error of the total dark matter mass estimation from bootstrapping. Solid circle: data points from this study, color-coded by their measured quiescent galaxy fraction (top) and average specific star formation rate (bottom). Solid lines: dividing lines of the halo accretion modes to cold, hot, and cold in hot modes as predicted by \cite{2006MNRAS.368....2D}. Dashed lines: median halo mass growth history for each present-day halo with masses of $\log{(M_h(z=0))}=12.0, 13.0, 14.0, \text{and } 15.0$, derived by \cite{2013ApJ...770...57B}. Small dot (grey): local clusters detected with Sunyaev-Zel'dovich (SZ) effects with ROSAT All-Sky Survey (X-ray) or Planck Satellite (FIR) \citep{2019A&A...626A...7T}, South Pole Telescope \citep[SPT;][]{2019ApJ...878...55B}, and Atacama Cosmology Telescope \citep[ACT;][]{2025arXiv250721459A}.
    Hex-plots (grey): groups and protoclusters identified by \citet{2022ApJ...933....9L} using halo-based group finder. 
    Pentagon (green): protocluster candidates from \citet{2023ApJ...943..153B} using a similar method to this study. Square (purple): high-redshift protoclusters compiled from various studies. See Section~\ref{sec:halomass} for detailed references.}
    \label{fig:halo mass}. 
\end{figure*}

The total halo mass or total dark matter mass of each identified overdensity can be estimated by using several approaches. The most commonly adopted method relies on the stellar-to-halo mass relation (SHMR), obtained from empirical abundance-matching models \citep{2013ApJ...770...57B, 2019MNRAS.488.3143B, 2025A&A...695A..20S} or halo occupation distribution (HOD) frameworks \citep{2022A&A...664A..61S}. A simpler alternative involves a direct conversion from stellar (or baryonic) mass to halo mass by assuming a fixed baryon-to-dark matter fraction and a one-to-one correspondence between galaxies and halos \citep{2022A&A...667L...3L, 2023ApJ...943..153B, 2025ApJ...988L..19J}. In addition, halo masses may be inferred using empirical relations tied to other galaxy properties, such as the $M_h$–$M_{UV}$ relation \citep{2023MNRAS.521..497M}.

However, the relations derived from the analytical SHMR equation \citep{2013ApJ...770...57B} or the interpolation of the stellar mass-halo mass ratio \citep{2019A&A...622A.103B} do not extend to the massive end at $z>2$. In these relations, the halo mass reaches a peak at a stellar mass of approximately $M_*\sim10^{10.5}$--$10^{11}\, M_\odot$, and begins to decline at higher halo mass. Thus, these relations are not suitable for our stellar mass distribution, which extends up to $\sim 10^{12}M_\odot$. 

To overcome this problem, we adopt the observed HOD-based SHMR derived by \citet{2022A&A...664A..61S}, which includes the distributions of central and satellite galaxies from the COSMOS2020 survey and incorporates ancillary imaging, ensuring a highly complete sample even at the lower-mass end. This relation provides improved constraints at the massive end and extends to redshifts as high as $z=5.5$, making it more suitable for the overdensity systems studied in this work.

Considering uncertainties in stellar masses derived from SED fitting and from the SHMR model, we use a bootstrapping method to estimate the total dark matter mass overdensity. For each bootstrapping realization, we randomly select the mass of each potential galaxy member by assuming a Gaussian posterior centered on the Bayesian stellar mass, with uncertainties derived from \textsc{cigale} as the standard deviation. We then calculate the sum of these randomly selected masses and convert them into the total dark matter mass using uncertainty-perturbed SHMR.

We perform 100 bootstrap realizations for each overdensity and present its final total dark matter mass, $M_h$, as the median of the distribution, along with its $1\sigma$ error derived from the 68\% percentile of the distribution. The final result is presented in Figure~\ref{fig:halo mass}. However, since the galaxies involved in this work are incomplete below the stellar-mass completeness limit, our total dark matter estimates are dominated by the most massive member galaxies, rather than a fully complete total dark matter mass.

For comparison, we also include various protoclusters and galaxy systems reported in previous studies, identified through different methods, compiled in \cite{2016A&ARv..24...14O}. At low redshift, we show galaxy clusters detected via the Sunyaev–Zel’dovich (SZ) effect using X-ray observations from the Planck satellite and the ROSAT All-Sky Survey (RASS), compiled by \citet{2019A&A...626A...7T}. At intermediate redshifts, we compare galaxy groups and protocluster candidates identified using photometric and spectroscopic samples from \citet{2022ApJ...933....9L}, based on the halo-based group finder \citep{2007ApJ...671..153Y}. Our galaxy systems are distributed consistently with these group and protocluster candidates.

We also include high-redshift ($z>6$) protoclusters in \citet{2023ApJ...943..153B}, identified through similar techniques to our study. In addition, we compare with various well-known protoclusters reported in the literature, including:
(1–2) PHz\_G237.01+42.50 \citep{2021MNRAS.503L...1K, 2021A&A...654A.121P},
(3) the Spiderweb protocluster \citep{2023Natur.615..809D},
(4) PCL1002 \citep{2015ApJ...808L..33C},
(5) CLJ1001 \citep{2016ApJ...828...56W},
(6–7) MRC0052–241 and MRC0943–242 \citep{2007A&A...461..823V},
(8) P2Q1 \citep{2014A&A...570A..16C},
(9) SSA22 \citep{topping2018understanding},
(10) DRC protocluster \citep{2018ApJ...856...72O},
(11) SPT2349–56 \citep{2018Natur.556..469M},
(12) SDF\_z47 \citep{2014ApJ...792...15T},
(13) HDF850.1 \citep{2023ApJ...953...53S},
(14) COSMOSAzTEC03 \citep{2011Natur.470..233C},
(15) JADESGDSz5p4 \citep{2024ApJ...962..124H},
(16–17) A2744-ODz5p7 and A2744-ODz5p8 \citep{2025ApJ...982..153M},
(18-19) z57OD and z66OD \citep{2019ApJ...883..142H},
(20) the system reported by \citet{2019ApJ...877...51C},
(21) SPT0311–58 \citep{2024A&A...688A.146A},
(22) SMACS0723–7327 \citep{2022A&A...667L...3L}, and
(23) A2744-ODz7p9 \citep{2023ApJ...947L..24M,morishita25ifu}.

Overall, our galaxy overdensity samples lie above $M_h \sim10^{12}\,M_\odot$, with the most massive mass reaching $\sim 10^{14}~M_\odot$ at $z \simeq 1.5$. Our sample exhibits a right-wedge-like distribution in the halo mass–redshift plane, consistent with the groups and protoclusters identified by \citet{2022ApJ...933....9L}, despite differences in identification methods.

The relatively flat distribution at lower masses can be explained by our selection approach, where we apply a constant overdensity significance threshold ($\sigma_\delta$) that is independent of redshift. However, the same overdensity factor may represent different physical conditions at different epochs. The galaxy overdensity ($\delta_{\mathrm{gal}}$) is proportional to the underlying matter overdensity ($\delta_m$). Under gravitational evolution, these structures undergo spherical collapse and grow over time. Consequently, an identical overdensity factor at higher redshift corresponds to a system that will evolve into a more massive halo at the present day \citep{2013ApJ...779..127C, 2016A&ARv..24...14O}.

At the high-mass end, our sample follows the trend predicted by the median halo mass accretion history from \citet{2013ApJ...770...57B}, consistent with present-day halos of $M_h(z=0) \sim 10^{15}\, M_\odot$.

Finally, the distribution of our galaxy overdensities is above 
the theoretical shock-heating threshold ($M_{\text{shock}}$) and divided by the predicted cold-stream and hot-halo regions defined in \citet{2006MNRAS.368....2D}. We further investigate the differences in galaxy system properties between these two regimes.

\subsection{Notable Examples} \label{sec:candidates}

\begin{figure*}
    \centering
    \renewcommand{\thefigure}{\arabic{figure}a}
    \includegraphics[width=0.9\linewidth]{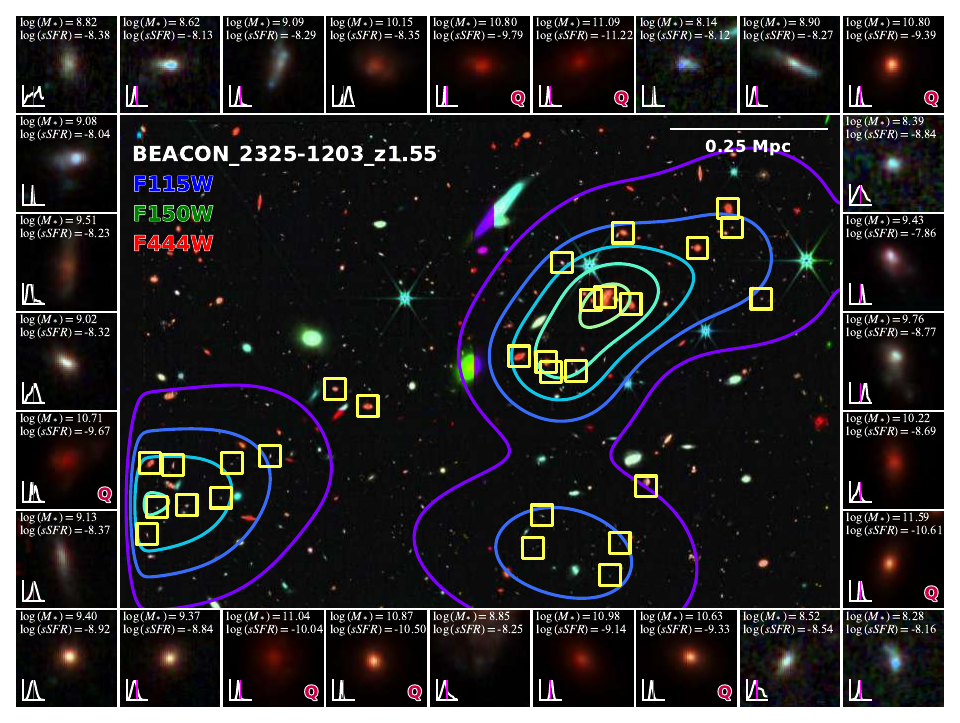}
    \caption{The snapshots of three examples of overdensities and their galaxy member candidates. RGB image is created from a set of filters: F444W (red), F150W (green), and F115W (blue). Each panel shows each overdensity that is described in Section~\ref{sec:candidates}. Yellow circles represent the potential galaxy members in each galaxy system. Each small panel surrounding the main panels shows snapshots of each galaxy member, along with its derived stellar masses and sSFRs. UVJ-selected quiescent galaxies are marked with \textbf{``Q''}. The histogram in the bottom-left corner of each small panel shows the photo-$z$ PDF, with a pink vertical line indicating the redshift of the respective overdensity. Contours on the main panels have different increments: $\delta=1$ step (top two panels) and $\delta=2$ (bottom panel). This figure is a snapshot for the most massive system: BEACON\_2325-1203\_z1.55}
  \label{fig:system1}
\end{figure*}

\begin{figure*}\ContinuedFloat
    \centering
    \renewcommand{\thefigure}{\arabic{figure}b}
    \includegraphics[width=0.9\linewidth]{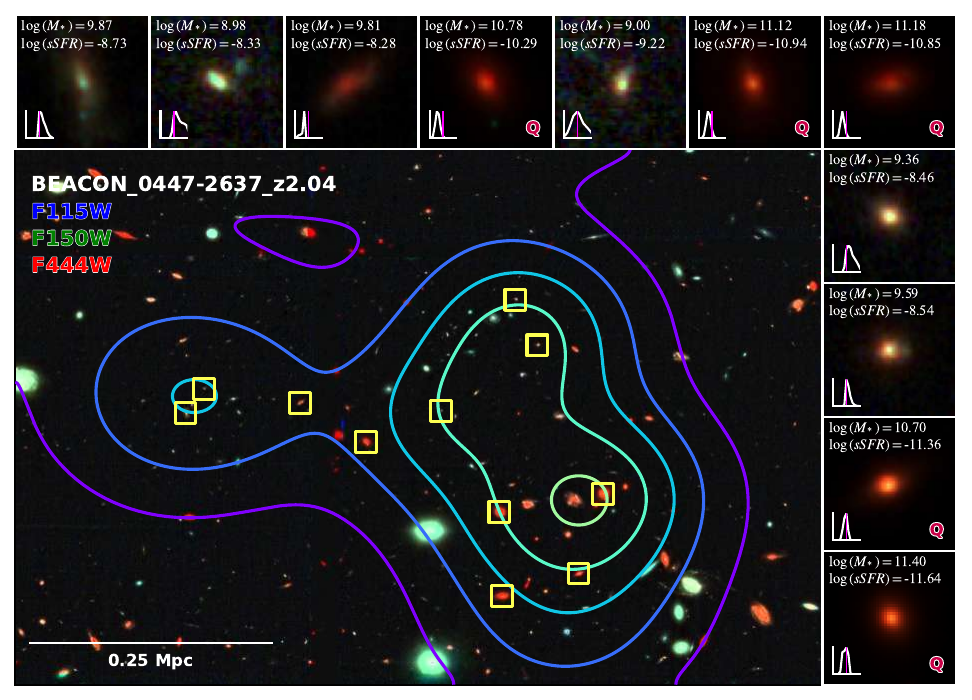} 
    \caption{This figure is a snapshot for the highest quiescent fraction system: BEACON\_0447-2637\_z2.00}
    \label{fig:system2}
\end{figure*}

\begin{figure*}\ContinuedFloat
    \centering
    \renewcommand{\thefigure}{\arabic{figure}c}
    \includegraphics[width=0.9\linewidth]{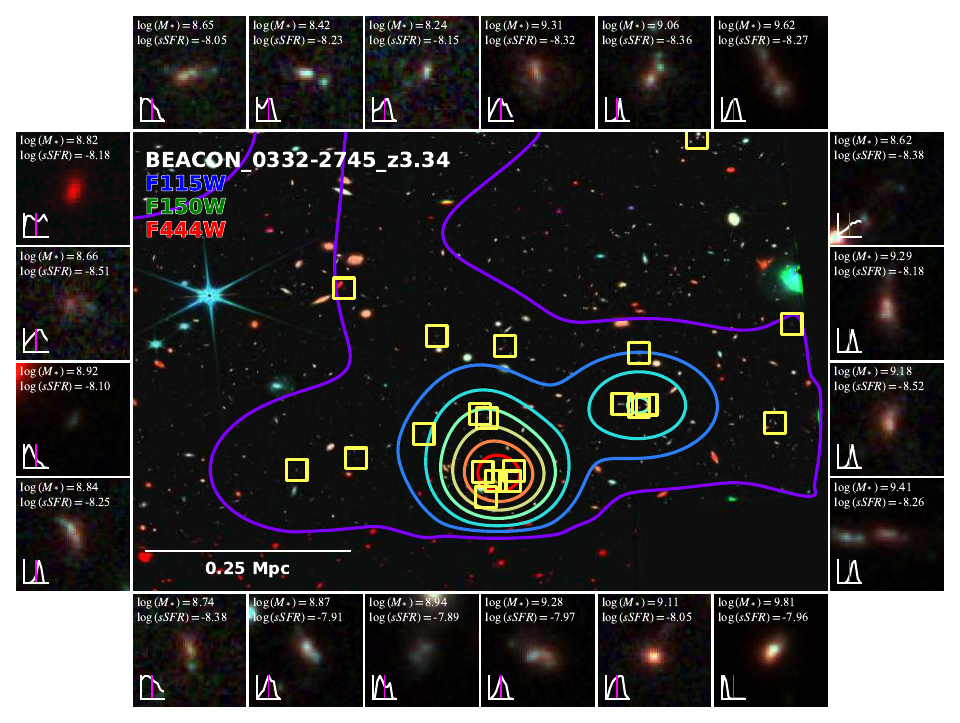} 
    \caption{This figure is a snapshot for the highest overdensity factor system: BEACON\_0332-2745\_z3.32}
    \label{fig:system3}
\end{figure*}

Here, we present a few notable examples of the galaxy overdensities that can be classified as protocluster candidates, defined by their present-day halo mass $M_h(z=0)\geq10^{14}~M_\odot$, assuming these systems evolve following the average halo mass growth trajectories \citep{chiang2013ancient, 2013ApJ...770...57B}. We present three representative examples of galaxy overdensities selected from the subsamples containing the most massive systems, the systems with the highest quiescent fractions, and the systems with the highest overdensity factors.

\subsubsection{Most Massive System: BEACON\_2325-1203\_z1.55}\label{sec:ovdex}

We identify BEACON\_2325-1203\_z1.55 as the most massive overdensity in our sample. The structure lies at $z=1.55$ and exhibits a peak overdensity of $\delta = 7.4$ (corresponding to $10.5\,\sigma$). The density map shows that the 29 potential member galaxies are grouped in three distinct peaks, two of which exhibit elongated morphologies, suggesting the presence of filamentary structures connected to the system (Figure~\ref{fig:system1}). Of these galaxies, 13 lie within the $4\sigma$ contour. 

The system hosts a total stellar mass of $1\times10^{12} M_\odot$, which corresponds to an estimated total dark matter mass of $6.2\times10^{13} M_\odot$. Under classical definitions, such a mass scale is consistent with a galaxy cluster. As shown in Figure~\ref{fig:halo mass}, this can be associated with the local cluster distributions. However, this claim must remain speculative in the absence of actual spectroscopic confirmation or independent evidence from SZ or X-ray observations.

The system includes eight notable massive quiescent galaxies ($M_\star \ge 10^{10.5} M_\odot$), indicating a relatively evolved population already present at this epoch. The overdensity spans a projected comoving extent of $2.5 \times 3.5$ cMpc.

\subsubsection{Highest Quiescent Fraction: 
BEACON\_0447-2637\_z2.00}
We identify BEACON\_0447-2637\_z2.00, located at $z=2.00$, as the overdensity with the highest quiescent fraction of 45\% with 11 potential galaxy members. The density map (Figure~\ref{fig:system2}) shows that the potential member galaxies are concentrated in a compact cluster, with a projected size of $2 \times 2$ cMpc. The system exhibits a peak overdensity of $\delta = 6.6$ ($9.3\sigma$). Of the 11 identified potential galaxy members, 6 are located above the $4\sigma$ contour, including 5 of the massive quiescent galaxies mentioned previously.

The system hosts a total stellar mass of $6.6\times10^{11} M_\odot$, which corresponds to an estimated total dark matter mass of $9.0\times10^{13} M_\odot$. Assuming this overdensity to evolve following the average mass growth, it will become a Coma-type cluster ($M_h\ge10^{15}M_\odot$) at the present day.

The galaxy members span stellar masses from $10^{9} M_\odot$ to $10^{11.5} M_\odot$. Notably, all five galaxies with $M_* \ge 10^{10.5} M_\odot$ in this system are classified as quiescent galaxies, yielding a quiescent fraction of 100\% above this mass limit. To place this result in context, we compare the quiescent fraction in this system with previous studies of field galaxy populations. Using the stellar mass functions (SMFs) of total and quiescent galaxies from \citet{ilbert2013mass}, \citet{muzzin2013evolution}, and \citet{weaver2023cosmos2020}, we derive the field quiescent fraction at $2<z<2.5$ by integrating the SMFs above $M_* > 10^{10.5}M_\odot$. This yields $f_Q(>10^{10.5}) = 23\%$, 24\%, and 21\% for each study, respectively. For \citet{weaver2023cosmos2020}, whose mass completeness extends to lower stellar masses, we additionally compute $f_Q(>10^{9}) = 3.2\%$. This value is broadly consistent with the quiescent fraction measured in our sample, for which we find $f_Q(>10^{9}) = 4.2\%$.

Compared to these field benchmarks, BEACON\_0447-2637\_z2.00 shows a quiescent fraction about five times higher at the massive end, strongly suggesting the presence of environmentally driven quenching, particularly among massive galaxies. Similar results have been reported in previous studies of clusters at comparable redshifts; for example, quiescent fractions among massive galaxies ($>10^{11}M_\odot$ in $z \sim 1.8$ clusters have been found to reach values as high as 100\% \citep{newman2014spectroscopic}. We further compare our result to the protocluster SMF presented by \citet{edward2024stellar}, derived from protoclusters at $2<z<2.5$, which yields a protocluster quiescent fraction of 57\%.

Together, these comparisons indicate that environmental quenching may already be underway by $z \sim 2$, and BEACON\_0447-2637\_z2.00 represents an extreme case among high-redshift overdensities in terms of its quiescent galaxy population. Similar findings at $z \sim 2$–3 have also been reported by \citet{ito2023cosmos2020} and \citet{mcconachie2022spectroscopic}.

\subsubsection{Highest Overdensity Factor: BEACON\_0332-2745\_z3.32}
We identify BEACON\_0332-2745\_z3.32 as the overdensity with the highest overdensity factor of $\delta=25.0$, corresponding to 32\,$\sigma$, which lies at $z=3.32$. The density map (Figure~\ref{fig:system3}) shows that the potential member galaxies are distributed into one compact clustering, with another small clustering on the other side, covering a total projected size of $3\times2.4$ cMpc.

The system hosts a total stellar mass of $3.6\times10^{10} M_\odot$, which corresponds to an estimated total dark matter mass of $1.7\times10^{12} M_\odot$. We identify 20 potential galaxy members, of which 9 lie within the $4\sigma$ contour. All of the potential galaxy members are classified as star-forming galaxies, marking the early phase formation of the galaxy overdensity, a stark contrast to the highly quiescent system BEACON\_0447-2637\_z2.00 at lower redshift.

\subsection{Galaxy Properties with Respect to Their Environment}
\begin{figure*}
    \centering
    \includegraphics[width=\linewidth]{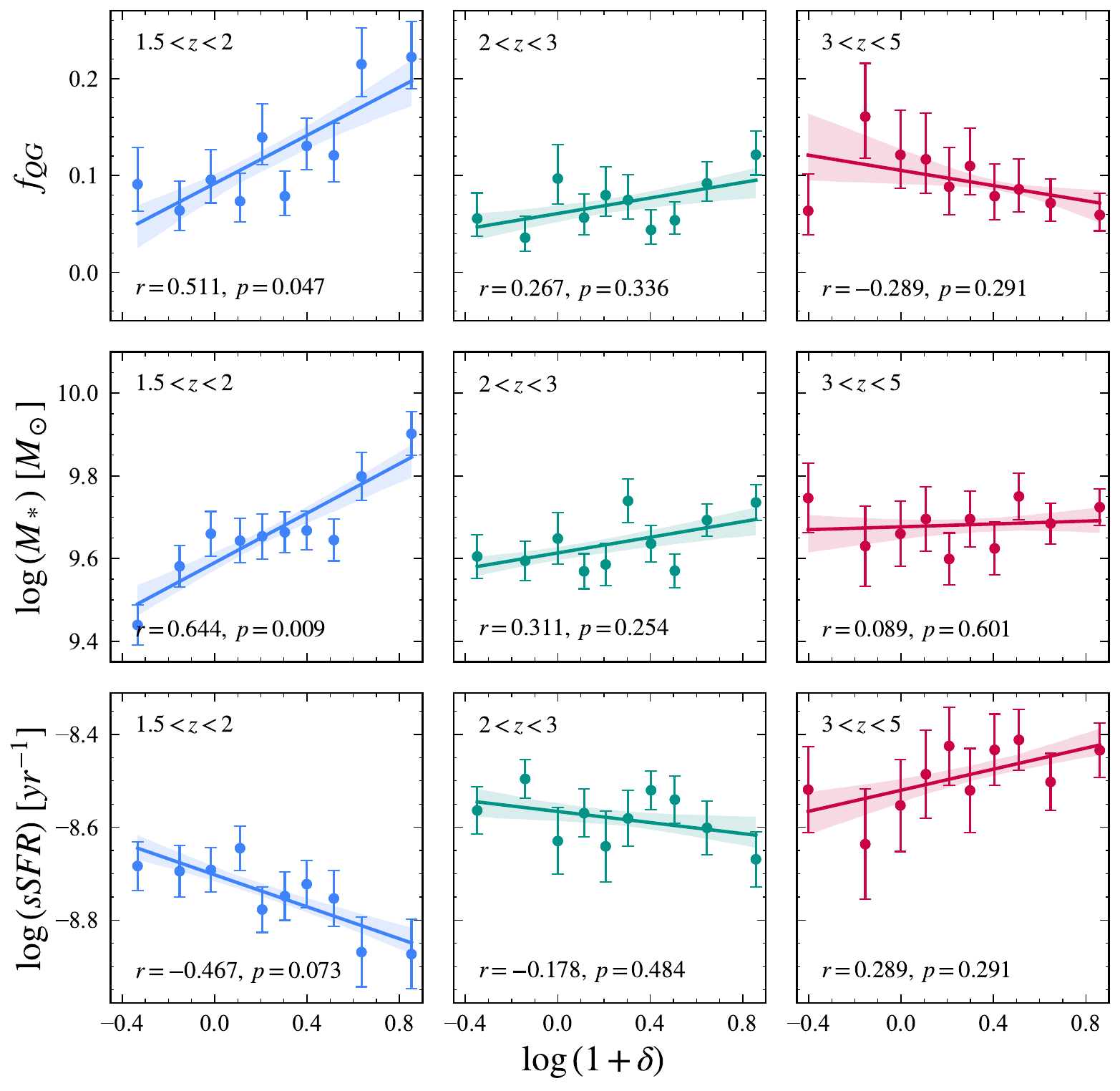}
    \caption{Galaxy properties as a function of local density for the quiescent fractions (top), stellar masses (middle), and sSFRs (bottom), represented by the median value on each bin. On each panel, we show the result of the Kendall rank correlation coefficient ($r$) and its $p$--value.}
    \label{fig:statistics}
\end{figure*}

For each identified overdensity presented in Figure~\ref{fig:halo mass}, we calculate the quiescent galaxy fraction based on the potential galaxy member and apply a color coding to each system. Additionally, we also apply a color coding based on $sSFR$, which is represented by its median in the lower panel of Figure~\ref{fig:halo mass}.

To complement our analysis on how galaxy properties evolve with respect to their global system characteristics, we examine how individual galaxy properties vary with their local environment. Specifically, we investigate the evolution of the quiescent galaxy fractions ($f_{QG}$), stellar masses ($M_*$), and specific star formation rates (sSFRs) as a function of local overdensity ($\delta$), as shown in Figure~\ref{fig:statistics}. To trace their evolution over cosmic time, we divide the sample into redshift bins corresponding to roughly 1 Gyr intervals: $1.5 < z < 2$, $2 < z < 3$, and $3 < z < 5$.

For each redshift bin, the quiescent galaxy fraction is computed within bins of overdensity, considering only galaxies above the stellar mass completeness limit. The associated uncertainties are derived using the binomial confidence interval from \citet{wilson1927probable}. The stellar masses and sSFRs distributions are represented by their median values, with 1$\sigma$ scatter estimated from their standard deviations. In each panel, we also display the best-fit linear relation and its 1$\sigma$ uncertainty.

To assess the strength and significance of these correlations, we compute the Kendall rank correlation coefficient \citep{kendall1938new}. We further account for the measurement uncertainties of each data point through a Monte Carlo error propagation approach following \citet{curran2014monte}.

From Fig.~\ref{fig:statistics} and the Kendall’s $\tau$ values for each parameter, we find that the trends of $f_{QG}$ and $M_*$ steepen toward lower redshift. As a result, these correlations become statistically significant at $z<2$. In contrast, sSFRs show a different behavior along redshift. While it is negatively correlated with local density at $z<2$, which is qualitatively consistent with the other two parameters, it exhibits a reversal toward a positive correlation at higher redshift, although this trend remains statistically weak. We will revisit this in Discussion (Section~\ref{sec:local}).

\section{Discussion} \label{sec:discussion}
We investigate how the properties of galaxies in overdense regions are influenced by two scales: global environment, characterized by the total dark matter mass ($M_h$), and local environment, characterized by the overdensity factor ($\delta$).

\subsection{Properties of Systems with Respect to Their Total Dark Matter Masses}
\begin{figure*}
    \centering
    \includegraphics[width=\linewidth]{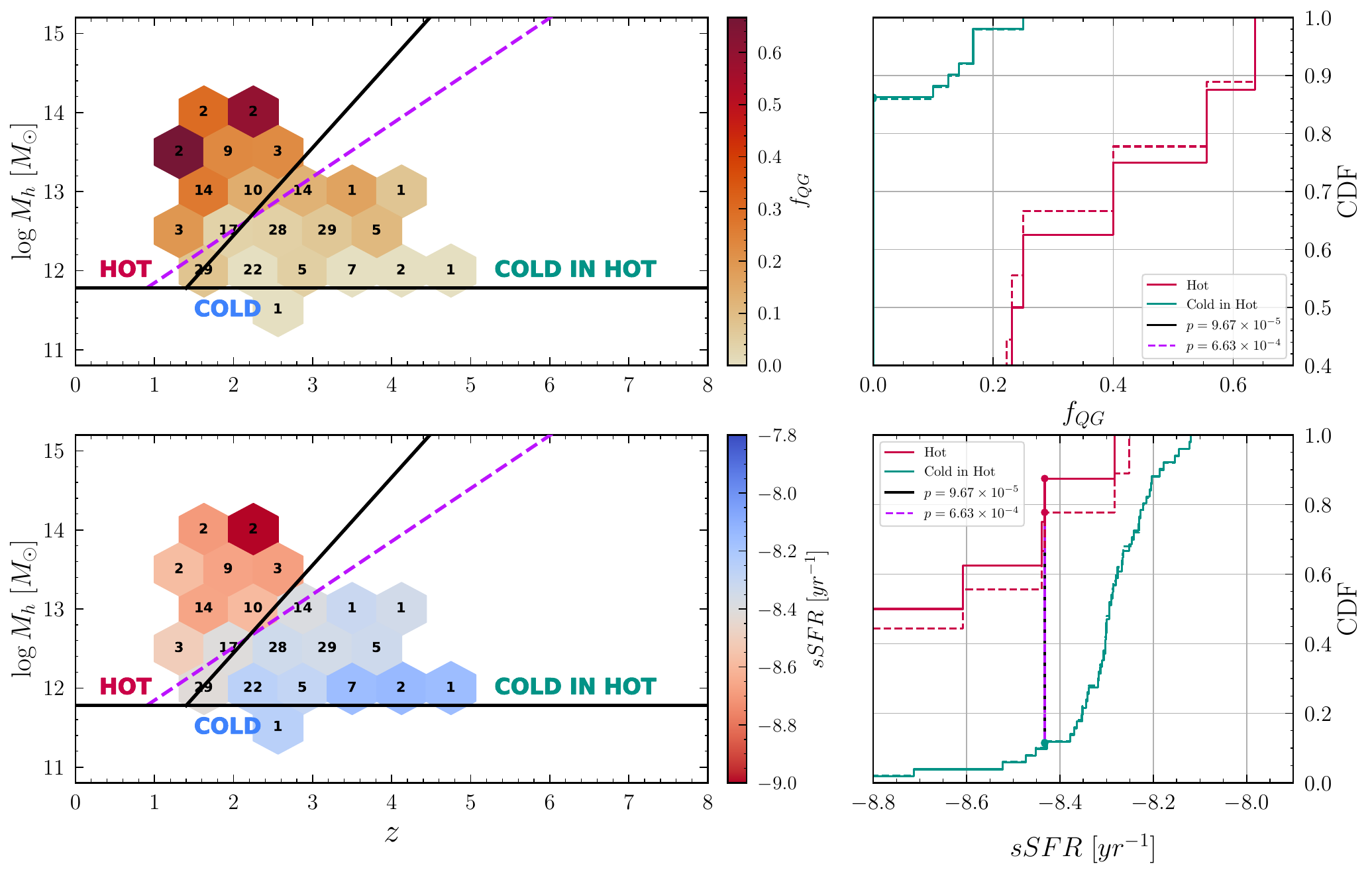}
    \caption{Hexbin plot of the galaxy systems distribution from Figure~\ref{fig:halo mass}. Dividing lines correspond to the definition of the halo accretion modes predicted by \citet{2006MNRAS.368....2D} (black solid line) and \citet{daddi2022bending} (purple dashed line). \textit{Top panel}: Hexbin plot of total dark matter mass with respect to quiescent galaxies fraction. \textit{Bottom panel}: Hexbin plot of total dark matter mass with respect to their $sSFR$. Each bin represents the median of the overdensities inside the bin. The number represents the total number of overdensities within the bin. The right panel shows the Kolmogorov-Smirnov (KS) test results for each parameter distribution on different cold stream transition definitions limited to the system at $1.9<z<2.5$.}
    \label{fig:hexbin}
\end{figure*}

\citet{2006MNRAS.368....2D} proposed that star formation activity in galaxies is regulated by the mode of gas accretion, which depends on the thermodynamic properties of the host halo and eventually gives rise to the observed galaxy bimodality. Halos below a critical shock-heating mass, $M_h < M_{\rm shock} \approx 6 \times 10^{11}M_\odot$, are expected to accrete gas in the cold mode, where infalling gas is not shock-heated because the cooling timescale is shorter than the dynamical timescale. This allows cold gas to efficiently reach the central galaxy. In more massive halos, however, infalling gas is subject to shock-heating to the virial temperature, forming a stable hot halo that suppresses cold inflows and limits the fuel available for star formation.

Narrow streams of cold gas may nevertheless penetrate the hot circumgalactic medium (CGM), continuing to supply star-forming fuel to the central galaxy. This transition is described by the cold-stream model of galaxy formation \citep[also][]{mandelker2020instability}. Cold streams can penetrate hot halos for systems with $M_h < M_{\rm stream}(z)$, where $M_{\rm stream}(z)$ defines the redshift-dependent boundary of the cold-stream regime. Conversely, in halos with $M_h > M_{\rm stream}(z)$, cold streams are expected to be shock-heated nevertheless before reaching the central regions, effectively quenching star formation.

Based on this theoretical framework, the star-forming properties of galaxies in groups and protoclusters are expected to show dependency on their host total dark matter mass as well as redshift. To examine this in our overdensity samples, we construct a hexagonally binned plot (Figure~\ref{fig:hexbin}). First, to qualitatively explore any trends, we color-code each hexagonal point by the quiescent fractions (top) and average specific star formation rates (sSFRs; bottom) of galaxy systems as functions of total dark matter mass and redshift. We overlaid the theoretical boundaries separating the cold, hot, and transitional (cold-in-hot) accretion regimes, as defined by \citet{2006MNRAS.368....2D} and \citet{daddi2022bending}.

Most overdensities identified in our study lie above the critical shock-heating mass. Even so, both quiescent fractions and sSFRs show clear dependencies on total dark matter mass and redshift, consistent with the gas-accretion models of \citet{2006MNRAS.368....2D} and \citet{daddi2022bending}. The gradient traced by the hexbin colors is inclined in the same sense predicted by these models. Along this transition, the quiescent fractions increase while the sSFRs decrease toward lower redshift. This behavior is naturally expected from the following sequence: once a halo becomes fully shock-heated, cold gas accretion is suppressed, halting the inflow of fresh gas needed to sustain star formation. As a result, star formation rates gradually decline, increasing the quiescent fractions and decreasing sSFRs of the member galaxies therein, as observed in our samples.

We quantitatively assess differences in galaxy properties between those in the hot accretion region and those in the cold-in-hot region using a Kolmogorov–Smirnov (K--S) test as shown in the right panel of Figure~\ref{fig:hexbin}. To minimize redshift-driven variation and isolate halo-mass dependence, we restrict the analysis to overdensities at $2.0 < z < 2.5$, which provides sufficient samples in both regimes. The K--S test reveals statistically significant differences ($p<0.05$) between system properties in the hot region and those in the cold-in-hot region, except for the \citet{2006MNRAS.368....2D} boundary when considering the quiescent fractions. These results are consistent with \citet{daddi2022evidence}, which found an improved statistical separation when using the cold-stream boundary of \citet{daddi2022bending}.

This finding suggests that the star-forming properties of overdensities intrinsically follow a bimodal distribution, controlled by the cold-stream threshold, marking the transition at which a halo becomes sufficiently virialized to prevent continuous fresh cold gas accretion from entering the system and eventually suppressing star formation.

\begin{figure*}
    \centering
    \includegraphics[width=.9\linewidth]{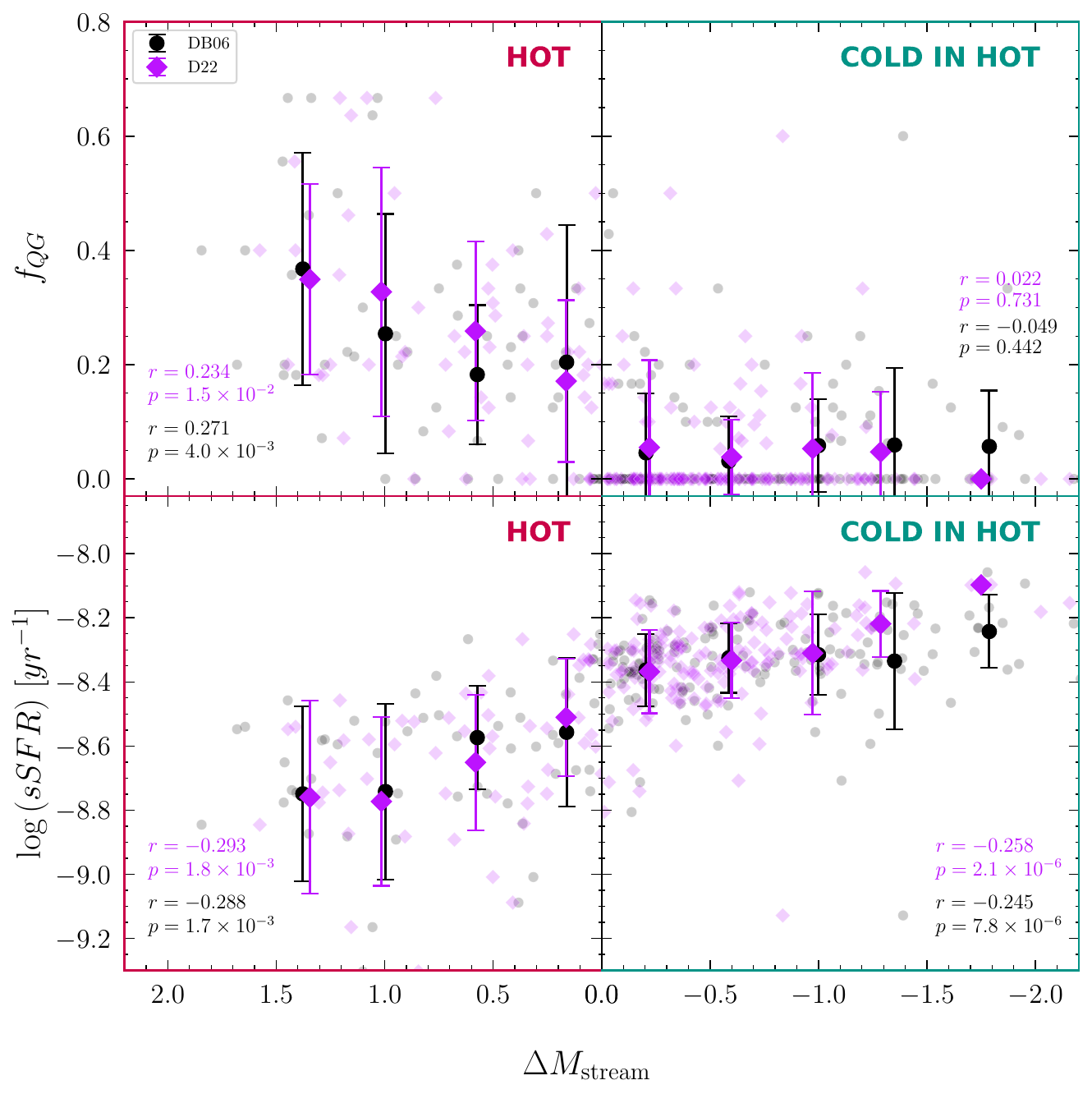}
    \caption{Quiescent fractions (top) and average sSFRs (bottom) with respect to the distance of total dark matter mass to the cold-stream definition ($\Delta M_{\rm stream}$). The half-transparent data point corresponds to each system's properties, while the solid-colored data point and its error bar correspond to the mean and standard deviation (scatter) of each bin. The left panel shows systems in the hot region, and the right panel shows systems in the cold-in-hot region. The DB06 data point (black) is based on the \citet{2006MNRAS.368....2D} cold-streams definition, while D22 (purple) is based on \citet{daddi2022bending}. On each panel, we show the Spearman rank correlation coefficient, $r$, and its $p$--value for both cold-stream definitions.}
    \label{fig:distance}
\end{figure*}

We also examine how these parameters in our sample evolve with respect to the distance from the cold-stream definition, $M_\text{stream}$, as defined by both \citet{2006MNRAS.368....2D} and \citet{daddi2022bending}. We define
\begin{equation}
    \Delta M_{\rm stream} = \log(M_h) - \log(M_{\rm stream}(z)),
\end{equation}
where $\Delta M_{\rm stream}$ is vertical distance of total dark matter mass from the cold-stream mass definition. Figure~\ref{fig:distance} shows the quiescent fractions and average sSFRs of each overdensity as a function of $\Delta M_{\rm stream}$. To quantify these trends, we calculate the Spearman rank correlation coefficient, $r$, and its $p$--value for each panel in Figure~\ref{fig:distance} using both cold-stream definitions. The quiescent fractions show a clear dependence on $\Delta M_{\rm stream}$. It increases as the distance grows larger, particularly for systems in the hot region, with $r\sim0.2$ and $p<0.05$. In contrast, this trend is absent in the cold-in-hot region, with $r\sim0$ and $p>0.05$, which presents a flattening trend, similar to the results from \citet{daddi2022evidence}, although with different tracers. This can happen mainly because most systems contain no quiescent galaxies, resulting in quiescent fractions of zero. This behavior supports the cold-stream scenario, where dense cold gas can still penetrate massive halos in the transitional regime, sustaining star formation. A similarly clear trend is seen in sSFRs, although with a smoother transition. As $\Delta M_{\rm stream}$ increases, the average sSFRs decrease in both hot and cold-in-hot systems, consistent with $r\sim-0.25$, indicating more efficient star-formation suppression as the system becomes more massive. 

This result is consistent with predicted satellite galaxy behavior at high redshift from simulations. \citet{van2017environmental}, using the hydrodynamical EAGLE simulation, reported a clear dependence of gas accretion suppression on the host total dark matter mass at $z \sim 2$: larger host halos correspond to lower gas accretion rates, even at high redshift. This aligns with quenching scenarios in which the main halo suppresses gas accretion onto infalling galaxies, effectively enhancing the efficiency of the galaxy quenching processes.

Therefore, in addition to the bimodality, our finding suggests that star-forming properties are also affected by a system's distance from the cold-stream threshold, indicating that the bimodality does not completely separate the two regimes. Instead, there appears to be a transitional phase as systems move from the cold-in-hot phase to the hot phase.

\subsection{Environmental Effect at the Local Scale} \label{sec:local}

Figure~\ref{fig:statistics} shows the trends between galaxy parameters and their dependence on the overdensity factor, which represents the local-scale environment. Up to $z\approx3$, all parameters show consistent behavior: the quiescent fractions and the median stellar masses increase, while sSFRs decrease, with overdensity. However, Kendall's tau value indicates that these trends are statistically significant only at $1.5<z<2$, as shown by the $p$--values below 0.05, allowing us to reject the null hypothesis that the observed correlations are random.

With caution regarding direct comparisons of overdensity values, we compare these trends with previous studies. The environmental dependence of the quiescent fractions has been reported by \citet{toni2025cosmos} and \citet{shi2024nature}. \citet{toni2025cosmos} found that quiescent fractions increase with increasing group richness (number of galaxies within the system), which itself correlates with local density, with this trend emerging at $z<2$. \citet{shi2024nature} also found similar behavior between the quiescent fractions and the overdensity, and reported no significant environmental dependence beyond $z>2$, consistent with our results.

For galaxy properties, our findings are generally consistent with previous work. The stellar masses increases with overdensity up to $z<3$, consistent with \citet{2024ApJ...966...18T}, \citet{shi2024nature} \citet{hatamnia2025large}, and \citet{chartab2025latis}. Similarly, \citet{mcconachie2025galaxies} found that the majority $(>75\%)$ of their spectroscopically-confirmed massive quiescent galaxies at $3<z<5$ preferentially reside in overdensity peaks. They suggest that the mass build-up of these galaxies reflects a significant contribution from ex-situ star formation via major mergers. This is supported by simulation results from \citet{huvsko2023buildup}, which used the \textsc{GALFORM} semi-analytical model with the Planck-Millennium simulation, indicating that most stellar mass of massive galaxies forms through ex-situ processes, including mergers. Observationally, merger rates also correlate with environment; for example, \citet{shibuya2025galaxy} shows that the relative major merger rate strongly depends on overdensity for galaxies at $2<z<5$. Similar trends have been reported in both local and intermediate-redshift galaxy samples, indicating that local density plays an important role in shaping the galaxy stellar mass function \citep{vulcani2012importance, tomczak2017glimpsing}. Additionally, recent observations of the high-redshift protocluster, including J1001/Hyperion at $z=2.5$ \citep{sun2024jwst, sikorski2026hst}, in Cosmic Vine at $z=3.44$ \citep{jin2024cosmic, sillassen2026cosmic}, and Bigfoot at $z=3.98$ \citep{tanaka2024protocluster, sun2025bigfoot}, also provide further evidence of an excess of massive star-forming galaxies, or enhanced quiescent fractions compared to the field.

For sSFRs, our results agree with previous studies  \citep{elbaz2007reversal, cooper2008deep2, darvish2016effects, 2020ApJ...890....7C}, who found an anti-correlation with overdensity. However, more recent studies \citep{lemaux2022vimos, 2024ApJ...966...18T, shi2024nature, hatamnia2025large} report a reversal in this relation, with sSFRs positively correlating with overdensity, although the transition epoch remains uncertain. \citet{shi2024nature} note that discrepancies may arise from the mass-completeness limit in earlier studies, which can cause mass quenching to dominate observed trends. We too observe the reversal trends in our study from the plot in Figure~\ref{fig:statistics} and Kendall's statistics. However, our limited sample size, particularly at high redshift, reduces the statistical significance of these results. In addition, increasing photometric redshift uncertainties at higher redshifts may further contribute to this effect. This highlights the need for future studies extending to lower stellar masses to confirm whether these potential reversal trends persist. 

Combining these results, we conclude that galaxy properties correlate with local density up to $z<2$. Galaxies residing in higher-density environments tend to be more massive, and a large fraction are either undergoing or have already undergone quenching. These findings suggest that dense environments may facilitate rapid mass assembly and/or accelerate quenching processes through mechanisms such as mass quenching or environmentally driven gas suppression.

Such behavior is consistent with the overconsumption scenario \citep{mcgee2014overconsumption, balogh2016evidence}, in which satellite galaxies in dense regions quench as a result of suppressed cold-gas accretion combined with outflow-driven gas depletion after entering a massive halo, a process often referred to as starvation or strangulation. This mechanism is expected to be more efficient for massive galaxies, in agreement with our findings on the dependence of the quiescent fractions and sSFRs on both total dark matter mass and local density. Furthermore, \citet{van2017environmental} demonstrated that gas accretion rates onto galaxies decrease with increasing local density, consistent with our results that galaxies residing in denser environments experience reduced gas supply.

The influence of the cosmic environment on the gas reservoir in the galaxy has been observed by several previous studies. For example, \citet{wang2018revealing} found that gas content is strongly dependent on the cluster-centric radius of galaxies, with galaxies becoming increasingly gas-poor as they fall toward the cluster center. Additionally, \citet{tadaki2019environmental} found that molecular gas masses are enhanced in lower-mass galaxies ($10.5<\log{(M_*/M_\odot)}<11$) relative to field galaxies. However, more massive galaxies ($log{(M_*/M_\odot)}>11$) exhibit no such enhancement and instead have molecular gas masses comparable to those of field galaxies. This lack of enhanced gas reservoirs suggests that gas accretion is already suppressed in these systems, consistent with expectations from halo quenching \citep{2006MNRAS.368....2D}.

Taken together, these studies further support the observed local environmental effect on stellar masses, sSFRs, and quiescent fractions, which are linked to the regulation of gas supply, with both local density and total dark matter influencing the ability of galaxies to accrete and retain cold gas. 

\section{Conclusion} \label{sec:conclusion}
In this paper, we utilized the 20 deepest fields in the Bias-free Extragalactic Analysis for Cosmic Origins with NIRCam (BEACON) survey to systematically search for galaxy overdensities at $1.5<z<7$, an unprecedented number of independent fields observed to JWST depths. From the identified overdensities and quantified local densities, we investigate the effect of the environment on galaxy properties, especially their quenching mechanism. We summarize our findings as follows:
\begin{enumerate}
    \item We have identified \nod significant ($>4\sigma$) galaxy overdensities at $1.5<z<5$ using photometric-redshift PDF-weighted adaptive kernel density estimation in 20 independent NIRCam fields from BEACON, with peak overdensity factors ($\delta$) ranging from 3 to 30. Using the stellar-to-halo mass relation (SHMR), we determine the total dark matter mass of each system, spanning from $5.9\times10^{11}$ to $1.0\times10^{14}~ M_\odot$
    \item Global-scale trend (bimodality): Total dark matter masses correlate with star-forming properties, particularly quiescent fractions and average sSFRs, in a manner that depends on their location relative to the halo accretion modes. Overdensities in the cold-in-hot regime show low quiescent fractions (mostly zero) and higher average sSFRs, while overdensities in the hot regime show higher quiescent fractions and suppressed average sSFRs.
    \item Global-scale trend ($\Delta M_{\rm stream}$): Overdensities also show a dependence on the vertical distance of total dark matter mass from the cold-stream mass definition, $\Delta M_{\rm stream}$. In the cold-in-hot regime, the quiescent fractions show a flattening trend, while the average sSFRs show a gradual decline as systems move closer to the threshold. In the hot regime, in contrast, the quiescent fractions show a positive correlation with $\Delta M_{\rm stream}$, while sSFRs become increasingly suppressed farther from the cold-stream definition.
    \item Together, these results confirm the existence of the two halo accretion regimes, i.e., a "hot" regime, in which massive halos efficiently suppress star formation at $z\lesssim2$, and a "cold-in-hot" regime at higher redshifts, in which star formation persists despite high total dark matter mass, as narrow cold streams are still able to penetrate the system.
    \item Local-scale trend: We observe significant correlations between local density and galaxy properties at $z<2$, where higher densities correspond to increased quiescent fractions, higher median stellar masses, and suppressed sSFRs. These trends weaken significantly at $z>2$, indicating that environmental quenching mechanisms only become dominant as the cosmic time increases. These findings are consistent with the overconsumption and/or the environmentally driven gas suppression scenarios.
\end{enumerate}
This study has shown that quenching mechanisms can occur on two distinct scales: a global scale, affected by the total mass of the galaxy system, and a local scale, affected by the location of galaxies relative to their surroundings. We also showed that both mechanisms become more effective during and after cosmic noon. Our results are broadly consistent with previous studies and offer additional observational constraints on galaxy evolution scenarios such as cold-stream accretion and overconsumption models.
Spectroscopic follow-up will be required to confirm the overdense structures identified in this work, to robustly assess their physical nature, and to further test the conclusions presented here. Future studies combining pure-parallel observations with the high spatial resolution of JWST will also enable detailed investigations of the connection between galaxy morphology and environment, as well as the occurrence of rare populations, including clumpy galaxies and dusty star-forming galaxies, in various environments.

\begin{acknowledgments}
The kernel density estimation calculations were carried out on the Multi-wavelength Data Analysis System operated by the Astronomy Data Center (ADC), National Astronomical Observatory of Japan. Support for JWST program 3990 was provided by NASA through the Space Telescope Science Institute, which is operated by the Association of Universities for Research in Astronomy, Inc., under NASA contract NAS 5-03127. RAS, TK, NSH, and KT acknowledge financial support from the Japanese Government (Ministry of Education, Culture, Sports, Science and Technology; MEXT), the Graduate Program on Physics for the Universe (GP-PU) at Tohoku University, JSPS KAKENHI Grant Numbers 24H00002 (Specially Promoted Research by T. Kodama et al.), 22K21349 (International Leading Research by S. Miyazaki et al.), and JSPS Core-to-Core Program (JPJSCCA20210003; M. Yoshida et al.). MJH is supported by the Swedish Research Council (Vetenskapsrådet) and is a Fellow of the Knut and Alice Wallenberg Foundation. All of the data presented in this paper were obtained from the Mikulski Archive for Space Telescopes (MAST) at the Space Telescope Science Institute. The specific observations analyzed can be accessed via \dataset[10.17909/swvp-0160]{https://doi.org/10.17909/swvp-0160}.
\end{acknowledgments}

\vspace{5mm}
\facilities{James Webb Space Telescope (JWST)}

\software{\textsc{astropy} \citep{2013A&A...558A..33A, 2018AJ....156..123A, 2022ApJ...935..167A}
          \textsc{eazy-py} \citep{2008ApJ...686.1503B}, 
          \textsc{emcee} \citep{foreman2013emcee},
          \textsc{CIGALE} \citep{2019A&A...622A.103B},          
          \textsc{SExtractor} \citep{1996A&AS..117..393B},
          \textsc{UniverseMachine} \citep{2019MNRAS.488.3143B}
          }

\appendix
\twocolumngrid

\section{Impact of the Filter Coverage on Derived Properties}
\begin{figure}
    \centering
    \includegraphics[width=\linewidth]{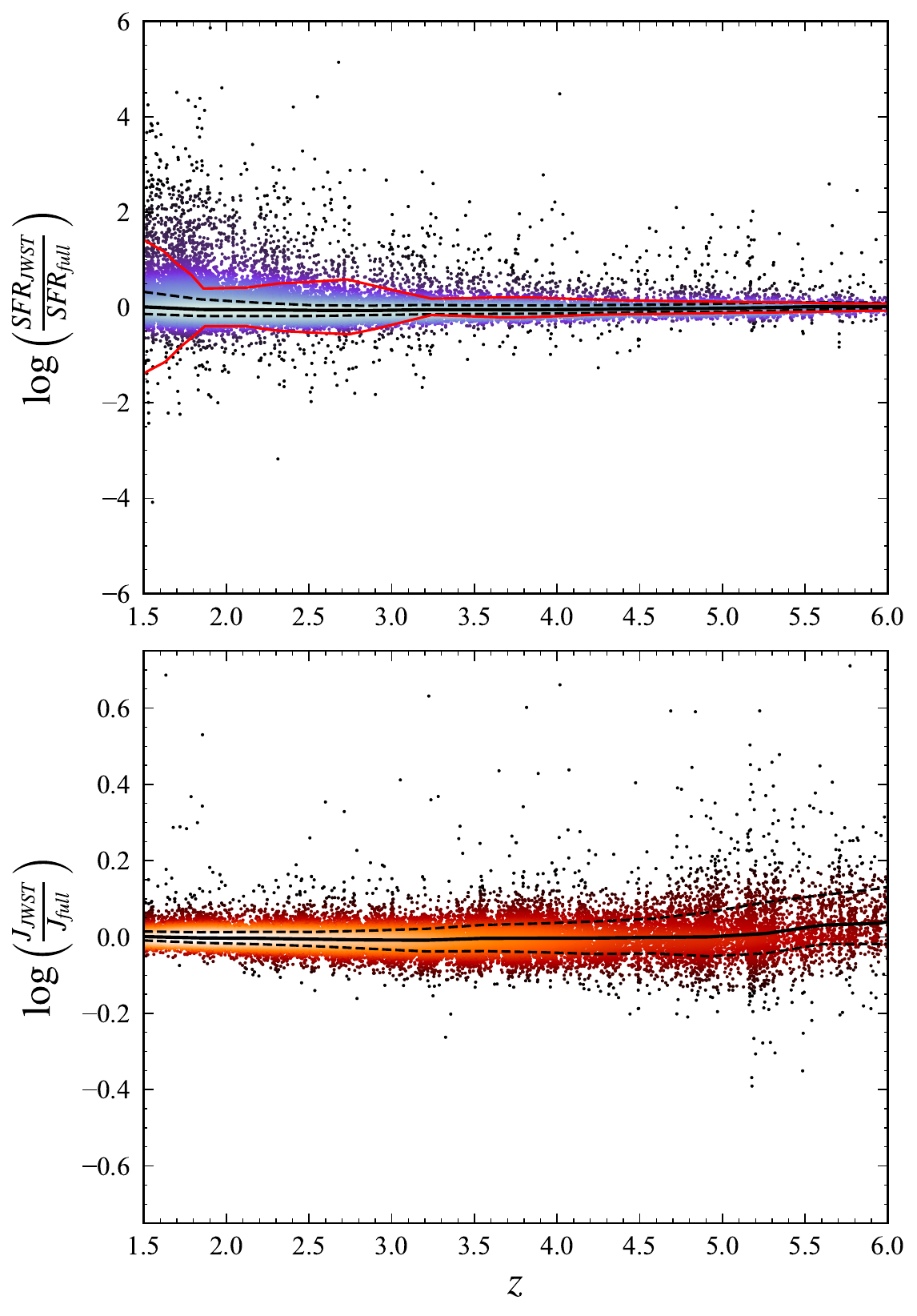}
    \caption{Comparison of SFR (top panel) and J-band rest-frame luminosity (bottom panel) derived from JWST-band only and full band obtained from ULTIMATE-deblending catalog. \textit{Black line}: Median offset (solid) and its lower and upper 1 sigma percentile (dashed). On the top panel, we also show a comparison of the median offset and 1 sigma standard deviation to \citet{sun2026ultimate}. }
    \label{fig:ratio}
\end{figure}

\begin{figure}
    \centering
    \includegraphics[width=\linewidth]{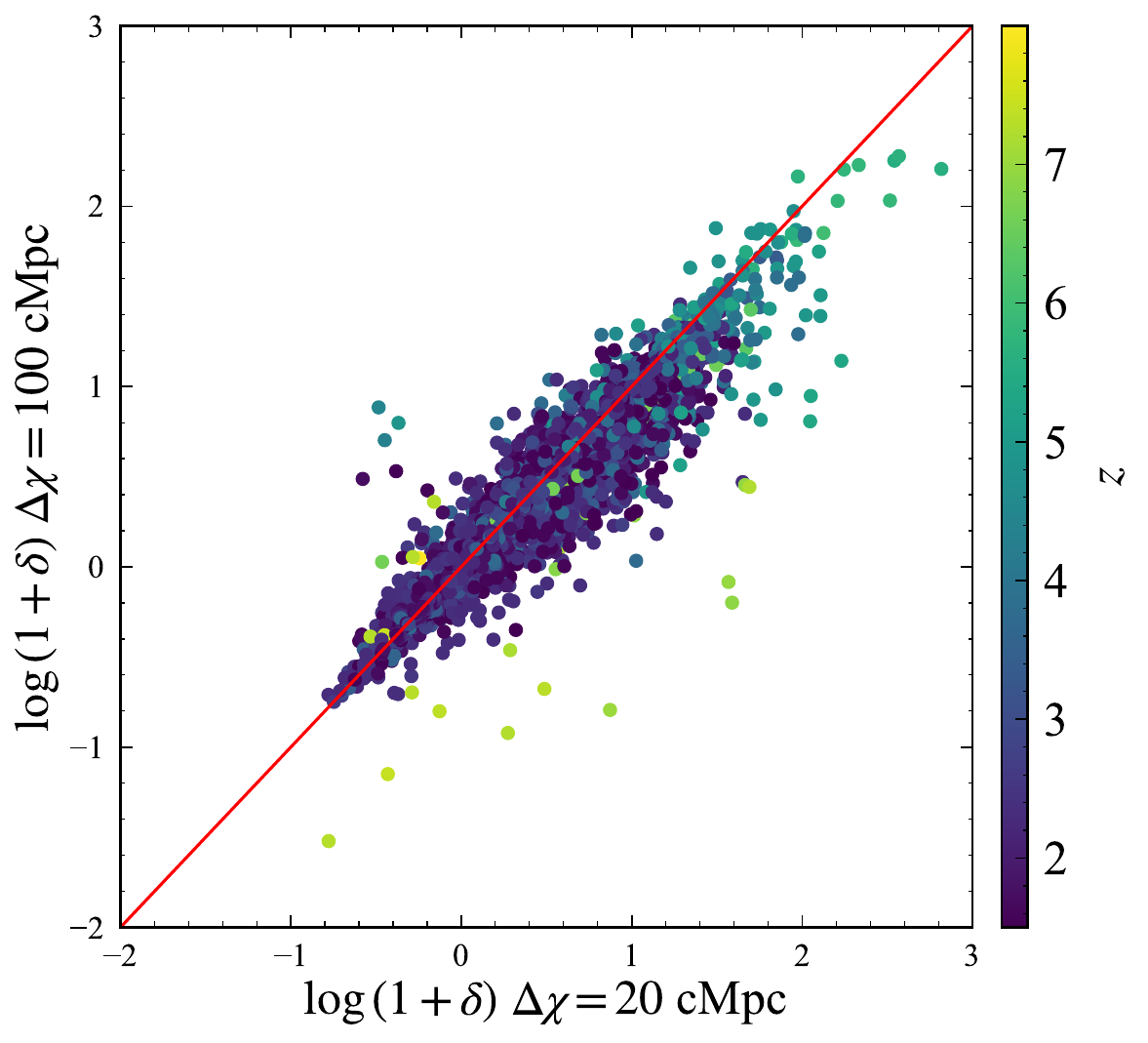}
    \caption{Comparison of the measured individual galaxy overdensity factor $\delta$ adopting two different redshift slices: 20 cMpc (x-axis) and 100 cMpc (y-axis). Each data point represents an individual galaxy, which is color-coded by its redshift. The red line is a 1:1 relation between the two overdensity factors. }
    \label{fig:delta_comparison}
\end{figure}
With the pure-parallel observing strategy of our field, our dataset is primarily limited to the JWST NIRCam filters. Consequently, the physical properties derived from SED fitting are constrained by the rest-frame wavelength coverage provided by these filters. This limitation is particularly important at $z<2$, where the rest-frame UV is only marginally sampled by the F090W filter, potentially introducing biases in the derived SFRs and sSFRs. In addition, the NIRCam filter set cannot fully recover the rest-frame $J$-band at $z>3$, which is used for our quiescent galaxy classification.

In this section, we investigate how the derived galaxy properties differ when using the full wavelength coverage, spanning the rest-frame UV to FIR, compared to using only the JWST bands, following the methodology of \citet{sun2026ultimate}. We utilize the photometric catalogs from the ULTIMATE-deblending project, which provide photometry in 50 bands ranging from CFHT/$U$ to JWST/MIRI F1800W in the PRIMER-UDS and PRIMER-COSMOS fields. We perform SED fitting using the same procedure described in Section~\ref{sec:sed} for two sets of photometry: (1) the ``full'' band set, which includes all available bands in the catalog, and (2) the JWST-only band set, which follows the BEACON filter coverage. To ensure a fair comparison in this study, we apply the same galaxy selection criteria described in Section~\ref{sec:galsample} to both samples. The comparison of the derived galaxy properties as a function of redshift is presented in Figure~\ref{fig:ratio}.

For SFR, the JWST-only estimates exhibit a negligible median offset compared to those from full-band photometry across all redshifts. However, we do not find a significant increase at $z<2$, where rest-frame UV coverage is more limited. This suggests that the absence of ancillary UV data does not introduce significant systematic biases in our SFR measurements.

Turning to the rest-frame $J$-band luminosity, we find that the scatter between the JWST-only and full-band estimates gradually increases with redshift, particularly beyond $z>3$, where NIRCam filters no longer fully sample the rest-frame $J$-band. Nevertheless, we do not observe any reduction in performance. For $z>5$, the JWST-only measurements exhibit a modest positive median offset that remains approximately constant with redshift.

Taken together, these results demonstrate that the discrepancies between the JWST-only and full-band measurements are minor compared to the intrinsic uncertainties of the SED fitting. We therefore conclude that the galaxy properties derived from the BEACON filter set, such as SFRs and quiescent-galaxy classifications, are robust enough for the analyses presented in this work.

\section{Choice of Redshift Slices Bandwidth}\label{app:choice}
We investigate how the choice of redshift-slice bandwidth affects the measurement of galaxy overdensity factors and the identification of galaxy overdensities. Specifically, we compare results obtained using $\Delta\chi = 20$ cMpc and $\Delta\chi = 100$ cMpc. Figure~\ref{fig:delta_comparison} shows the individual galaxy overdensity factors measured with the two bandwidths. Most data points lie close to the 1:1 relation, indicating strong agreement between the two choices. This demonstrates that the measured overdensity factors are largely insensitive to the adopted redshift-slice width. The agreement becomes weaker at $z \gtrsim 5$, consistent with our findings in Figure~\ref{fig:simul} (Appendix~\ref{app:validity}).

We also compare the number of identified overdensities obtained with the two bandwidths. Adopting a 100 cMpc redshift slice changes the total number of detected overdensities by less than 10\%, with the difference primarily arising from the inclusion of additional interloping structures within the larger redshift interval. Overall, these results indicate that the choice of redshift-slice bandwidth has only a minor impact on overdensity measurements and does not affect the main conclusions of this study.

\section{Validity of the Absolute Overdensity Factor} \label{app:validity}
\begin{figure}
    \centering
    \includegraphics[width=\linewidth]{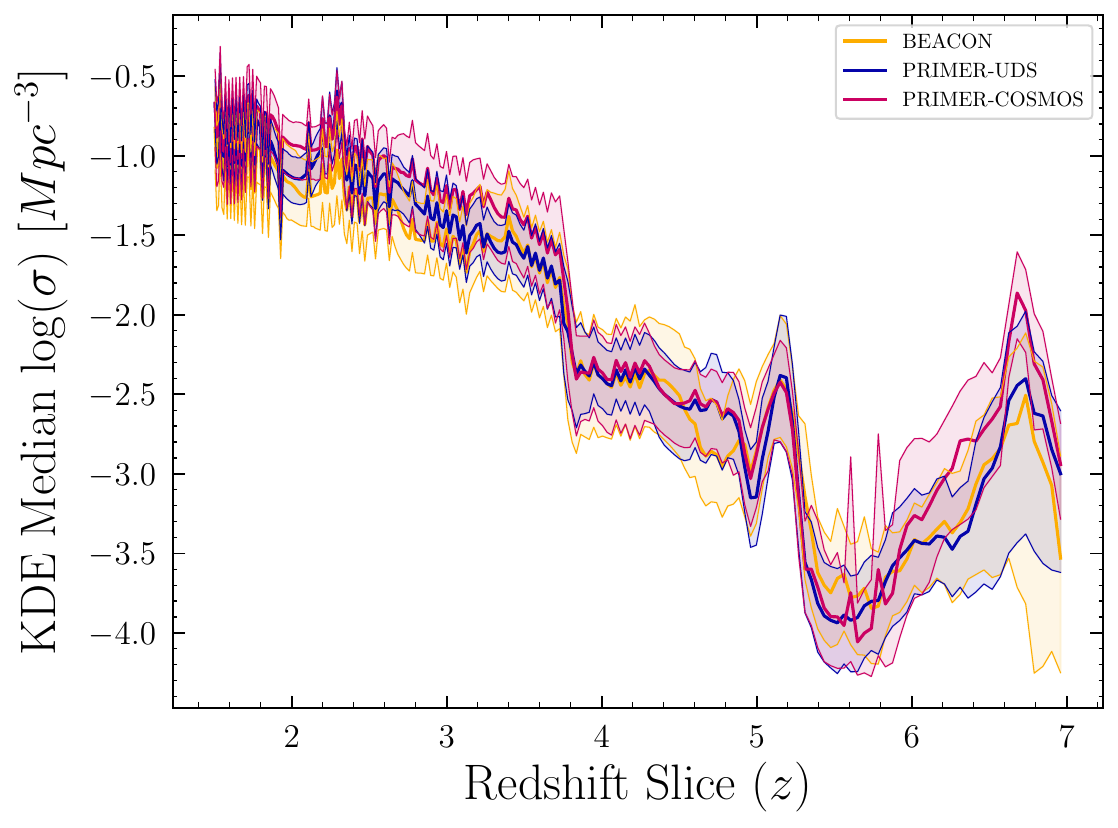}
    \caption{Comparison of measured median of surface density by kernel density estimation from the BEACON, PRIMER-UDS, and PRIMER-COSMOS fields. Each field is represented by different colors, where the solid lines represent the median of the distribution on each redshift and the shaded region represents the range given by the 16th and 84th percentiles.}
    \label{fig:simul}
\end{figure}

The surface density maps derived from the adaptive kernel density estimation are presented in arbitrary, non-normalized units by definition. As described in Section~\ref{sec:overview}, the overdensity factor at each position is calculated by normalizing the local surface density by the median surface density in each redshift slice. However, this normalization is field-dependent, meaning that the median value varies between the BEACON fields. This variation arises from the relatively small footprint of each field, which could be susceptible to cosmic variance in some cases. Consequently, the median surface density of our fields may not represent a typical field environment. To mitigate this, we have adopted a single median value computed across all BEACON fields within each redshift slice.

To further assess this issue, we simulated our observation and analysis procedure on other JWST fields with comparable depth, namely, the PRIMER survey (JWST-GO-1837; \citealt{2021jwst.prop.1837D}) in the UDS and COSMOS regions. Specifically, we randomly placed a NIRCam mosaic footprint within each PRIMER field and extracted galaxies located inside the footprint. This process was repeated 50 times per field to generate independent realizations. For each realization, we performed photometric redshift fitting using the same configuration described in Section~\ref{sec:photoz}, and constructed overdensity maps following the procedure in Section~\ref{sec:search}. Through this simulation, we aim to evaluate how well a small NIRCam footprint can recover the true large-scale overdensity pattern and to determine the reliability of using such a limited field to represent the actual density contrast. 

In Figure~\ref{fig:simul}, we show the measured median value of the KDE surface density for 20 BEACON fields, 50 realizations from PRIMER-UDS, and 50 realizations from PRIMER-COSMOS across redshift slices, along with their 68\% percentile range. The measured medians of surface density are consistent throughout the fields. For our final catalog and the following analysis, we calibrate the estimated overdensity factor to the median of these three surveys. The variation in Figure~\ref{fig:simul} also proves that the determination of the number density on each redshift slice becomes more unstable, especially at $z>5$, due to the sparsity of the galaxies in this redshift range. 

\section{Different Methods of total dark matter mass Determination}
\begin{figure}
    \centering
    \includegraphics[width=1\linewidth]{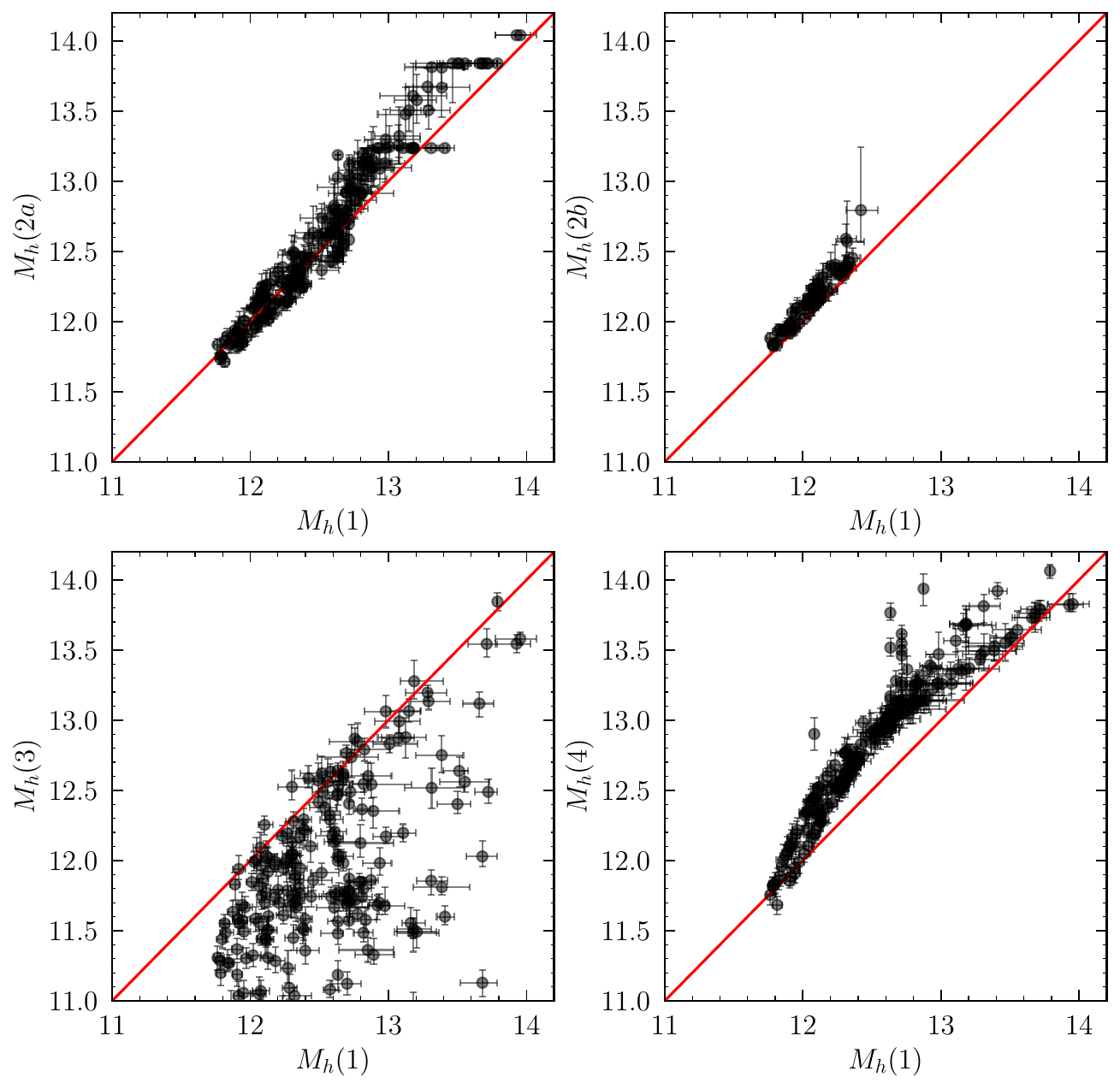}
    \caption{Comparison of total dark matter mass estimates obtained using different methods. The masses derived from each method are compared with our fiducial HOD-based estimates from \citet{2022A&A...664A..61S}. The red line denotes the 1:1 relation, with deviations from this line indicating systematic differences between the methods.}
    \label{fig:halocomparison}
\end{figure}
Motivated by \citet{sillassen2024noema}, who combined multiple halo-mass estimators to reduce the systematic uncertainties associated with any individual method, we also try to determine the total dark matter mass within each overdensity using several independent approaches. The adopted methods are summarized below:
\begin{enumerate}
    \item \textbf{Halo occupation distribution}: We derive the total dark matter mass by summing the total stellar masses of all potential galaxy members within the overdensity contour and converting the resulting stellar mass to halo mass using the HOD-based SHMR calibration from \citet{2022A&A...664A..61S}. This is our fiducial method, which was used for the reported total dark matter mass in this study. 
    \item \textbf{Summed abundance matching}: We estimate the total dark matter mass in the same manner as method (1), but using abundance-matching SHMR. We consider two calibrations: (2a) \citet{2019MNRAS.488.3143B} and (2b) \citet{2025A&A...695A..20S}. However, since the \citet{2019MNRAS.488.3143B} model is calibrated up to $\log{(M_*/M_\odot)}\lesssim11$, we only apply this model to systems with the total stellar masses lying within this range.  
    \item \textbf{Iterative virial-mass estimate:} We follow the methodology of \citet{sillassen2024noema} and \citet{witten2026too} to estimate the dark matter mass enclosed within the virial radius, $M_h(R<R_{\rm vir})$. We first define the overdensity center from the overdensity contour and adopt an initial radius of $R_{\rm vir}=100$ pkpc, which is substantially larger than the expected virial radii of systems at the redshifts considered in this study. Since this method employs a different definition of halo mass from the total halo-mass estimates described above, we derive a relation between virial mass and total stellar mass using the TNG300-1 \citep{nelson2019illustristng} and TNG-Cluster \citep{nelson2024introducing} simulations, following \citet{witten2025not}. For each FoF halo, we extract the group virial mass (\texttt{Group\_M\_Crit200}) and its virial radius (\texttt{Group\_R\_Crit200}). We then calculate the total stellar mass of subhalos that are located within halos virial radius. Here, the stellar mass of subhalos is defined as the total stellar mass within twice the stellar half mass radius (\texttt{SubhaloMassInRadStellar}). The stellar mass of each subhalo is defined as the stellar mass contained within twice the stellar half-mass radius (\texttt{SubhaloMassInRadStellar}). We parameterize the relation between total stellar mass and virial mass using the stellar-to-halo mass relation (SHMR) form introduced by \citet{moster2010constraints},
    \begin{equation} 
        \frac{M_*}{M_{vir}} = 2N\left[\left(\frac{M_{vir}}{M_1}\right)^{-\beta}+\left(\frac{M_{vir}}{M_1}\right)^\gamma\right]^{-1}, 
    \end{equation}
    where $N$ is the normalization of the stellar-to-halo mass ratio, $M_1$ is the characteristic mass, and $\beta$ and $\gamma$ are the low- and high-mass ends slopes, respectively. The four free parameters are fitted to the running median relation using the MCMC sampler \textsc{emcee} \citep{foreman2013emcee}, following the methodology of \citet{engler2021distinct} and \citet{girelli2020stellar}.
    The total stellar mass of candidate member galaxies within the adopted radius is converted to virial mass using this SHMR. We further correct for the contribution of galaxies below our stellar-mass completeness limit, $\log(M_*/M_\odot)<9$, following \citet{weaver2023cosmos2020}. The resulting halo mass is then used to compute an updated virial radius and a new stellar-mass-weighted centroid. This procedure is repeated iteratively until the halo mass converges.
    \item \textbf{Constant baryon-fraction conversion:} We estimate the total dark matter mass by assuming a constant stellar-to-dark matter fraction. We adopt a 1\% conversion factor, which is a typical stellar mass to halo mass ratio for galaxies at our redshift regime \citep{2013ApJ...770...57B, 2019MNRAS.488.3143B}.
\end{enumerate}
Figure~\ref{fig:halocomparison} compares the estimated total dark matter masses derived using the different methods. The HOD-based method of \citet{2022A&A...664A..61S} is broadly consistent with the other approaches, particularly methods (2a), (2b), and (4). Method (4) tends to yield systematically higher total dark matter masses, likely because it adopts a constant stellar-to-halo mass conversion factor, which is expected to overestimate halo masses in the intermediate-mass regime.

We also find a substantial discrepancy between the HOD-based estimates and the virial masses derived using method (3). This difference is expected, as method (3) measures only the mass enclosed within the core region of each overdensity and therefore accounts for only a fraction of the member galaxies. Consequently, the resulting virial masses represent lower limits to the total dark matter masses of the overdensity systems.

Overall, the different mass-estimation methods either produce consistent results or introduce systematic shifts in the inferred total dark matter masses. Since these shifts do not change the relative distribution of masses among the overdensity systems, the main conclusions of this study remain robust against the choice of mass-estimation method.

\bibliography{references}

@ARTICLE{morishita25ifu,
       author = {{Morishita}, Takahiro and {Stiavelli}, Massimo and {Vanzella}, Eros and {Bergamini}, Pietro and {Boyett}, Kristan and {Chiaberge}, Marco and {Grillo}, Claudio and {Leethochawalit}, Nicha and {Messa}, Matteo and {Roberts-Borsani}, Guido and {Rosati}, Piero and {Shajib}, Anowar J.},
        title = "{Metallicity Scatter Originating from Subkiloparsec Starbursting Clumps in the Core of a Protocluster at z = 7.88}",
      journal = {\apj},
         year = 2025,
        month = may,
       volume = {985},
       number = {1},
          eid = {83},
        pages = {83},
          doi = {10.3847/1538-4357/adc4c3},
archivePrefix = {arXiv},
       eprint = {2501.11879},
 primaryClass = {astro-ph.GA},
       adsurl = {https://ui.adsabs.harvard.edu/abs/2025ApJ...985...83M}
}

@ARTICLE{trenti2008cosmicvariance,
       author = {{Trenti}, M. and {Stiavelli}, M.},
        title = "{Cosmic Variance and Its Effect on the Luminosity Function Determination in Deep High-z Surveys}",
      journal = {\apj},
         year = 2008,
        month = apr,
       volume = {676},
       number = {2},
        pages = {767-780},
          doi = {10.1086/528674},
archivePrefix = {arXiv},
       eprint = {0712.0398},
 primaryClass = {astro-ph},
       adsurl = {https://ui.adsabs.harvard.edu/abs/2008ApJ...676..767T}
}

@ARTICLE{2025ApJ...983..152M,
       author = {{Morishita}, Takahiro and {Mason}, Charlotte A. and {Kreilgaard}, Kimi C. and {Trenti}, Michele and {Treu}, Tommaso and {Vulcani}, Benedetta and {Zhang}, Yechi and {Abdurro'uf} and {Alavi}, Anahita and {Atek}, Hakim and {Bah{\'e}}, Yannick and {Brada{\v{c}}}, Maru{\v{s}}a and {Bradley}, Larry D. and {Bunker}, Andrew J. and {Coe}, Dan and {Colbert}, James and {Gelli}, Viola and {Hayes}, Matthew J. and {Jones}, Tucker and {Kodama}, Tadayuki and {Leethochawalit}, Nicha and {Liu}, Zhaoran and {Malkan}, Matthew A. and {Mehta}, Vihang and {Metha}, Benjamin and {Newman}, Andrew B. and {Rafelski}, Marc and {Roberts-Borsani}, Guido and {Rutkowski}, Michael J. and {Scarlata}, Claudia and {Stiavelli}, Massimo and {Sutanto}, Ryo A. and {Takahashi}, Kosuke and {Teplitz}, Harry I. and {Wang}, Xin},
        title = "{BEACON: JWST NIRCam Pure-parallel Imaging Survey. I. Survey Design and Initial Results}",
      journal = {\apj},
         year = 2025,
        month = apr,
       volume = {983},
       number = {2},
          eid = {152},
        pages = {152},
          doi = {10.3847/1538-4357/adbbdc},
archivePrefix = {arXiv},
       eprint = {2412.04211},
 primaryClass = {astro-ph.GA},
       adsurl = {https://ui.adsabs.harvard.edu/abs/2025ApJ...983..152M}
}

@ARTICLE{2025A&A...696A.236P,
       author = {{P{\'e}rez-Mart{\'\i}nez}, J.~M. and {Dannerbauer}, H. and {Emonts}, B.~H.~C. and {Allison}, J.~R. and {Champagne}, J.~B. and {Indermuehle}, B. and {Norris}, R.~P. and {Serra}, P. and {Seymour}, N. and {Thomson}, A.~P. and {Casey}, C.~M. and {Chen}, Z. and {Daikuhara}, K. and {De Breuck}, C. and {D'Eugenio}, C. and {Drouart}, G. and {Hatch}, N. and {Jin}, S. and {Kodama}, T. and {Koyama}, Y. and {Lehnert}, M.~D. and {Macgregor}, P. and {Miley}, G. and {Naufal}, A. and {R{\"o}ttgering}, H. and {S{\'a}nchez-Portal}, M. and {Shimakawa}, R. and {Zhang}, Y. and {Ziegler}, B.},
        title = "{COALAS: III. The ATCA CO(1{\textendash}0) look at the growth and death of H{\ensuremath{\alpha}} emitters in the Spiderweb protocluster at z = 2.16}",
      journal = {\aap},
         year = 2025,
        month = apr,
       volume = {696},
          eid = {A236},
        pages = {A236},
          doi = {10.1051/0004-6361/202450785},
archivePrefix = {arXiv},
       eprint = {2411.12138},
 primaryClass = {astro-ph.GA},
       adsurl = {https://ui.adsabs.harvard.edu/abs/2025A&A...696A.236P}
}

@ARTICLE{Liu25,
       author = {{Liu}, Zhaoran and {Kodama}, Tadayuki and {Lemaux}, Brian C. and {Kubo}, Mariko and {P{\'e}rez-Mart{\'\i}nez}, Jose Manuel and {Koyama}, Yusei and {Tanaka}, Ichi and {Daikuhara}, Kazuki and {Gal}, Roy R. and {Hung}, Denise and {Konishi}, Masahiro and {Kushibiki}, Kosuke and {Laishram}, Ronaldo and {Lubin}, Lori M. and {Motohara}, Kentaro and {Takahashi}, Hidenori},
        title = "{Environmental Regulation of Dust and Star Formation Unveiled by Subaru Dual Narrowband Imaging: Degree-scale Balmer Decrement Mapping Across a z = 0.9 Supercluster}",
      journal = {\apj},
         year = 2025,
        month = dec,
       volume = {995},
       number = {1},
          eid = {81},
        pages = {81},
          doi = {10.3847/1538-4357/ae0c02},
archivePrefix = {arXiv},
       eprint = {2510.10386},
 primaryClass = {astro-ph.GA},
       adsurl = {https://ui.adsabs.harvard.edu/abs/2025ApJ...995...81L}
}

@ARTICLE{2015ApJ...805..121D,
       author = {{Darvish}, Behnam and {Mobasher}, Bahram and {Sobral}, David and {Scoville}, Nicholas and {Aragon-Calvo}, Miguel},
        title = "{A Comparative Study of Density Field Estimation for Galaxies: New Insights into the Evolution of Galaxies with Environment in COSMOS out to z{\ensuremath{\sim}}3}",
      journal = {\apj},
         year = 2015,
        month = jun,
       volume = {805},
       number = {2},
          eid = {121},
        pages = {121},
          doi = {10.1088/0004-637X/805/2/121},
archivePrefix = {arXiv},
       eprint = {1503.07879},
 primaryClass = {astro-ph.GA},
       adsurl = {https://ui.adsabs.harvard.edu/abs/2015ApJ...805..121D}
}

@ARTICLE{2020ApJ...890....7C,
       author = {{Chartab}, Nima and {Mobasher}, Bahram and {Darvish}, Behnam and {Finkelstein}, Steve and {Guo}, Yicheng and {Kodra}, Dritan and {Lee}, Kyoung-Soo and {Newman}, Jeffrey A. and {Pacifici}, Camilla and {Papovich}, Casey and {Sattari}, Zahra and {Shahidi}, Abtin and {Dickinson}, Mark E. and {Faber}, Sandra M. and {Ferguson}, Henry C. and {Giavalisco}, Mauro and {Jafariyazani}, Marziye},
        title = "{Large-scale Structures in the CANDELS Fields: The Role of the Environment in Star Formation Activity}",
      journal = {\apj},
         year = 2020,
        month = feb,
       volume = {890},
       number = {1},
          eid = {7},
        pages = {7},
          doi = {10.3847/1538-4357/ab61fd},
archivePrefix = {arXiv},
       eprint = {1912.04890},
 primaryClass = {astro-ph.GA},
       adsurl = {https://ui.adsabs.harvard.edu/abs/2020ApJ...890....7C}
}

@ARTICLE{2023ApJ...943..153B,
       author = {{Brinch}, Malte and {Greve}, Thomas R. and {Weaver}, John R. and {Brammer}, Gabriel and {Ilbert}, Olivier and {Shuntov}, Marko and {Jin}, Shuowen and {Liu}, Daizhong and {Gim{\'e}nez-Arteaga}, Clara and {Casey}, Caitlin M. and {Davidson}, Iary and {Fujimoto}, Seiji and {Koekemoer}, Anton M. and {Kokorev}, Vasily and {Magdis}, Georgios and {McCracken}, H.~J. and {McPartland}, Conor J.~R. and {Mobasher}, Bahram and {Sanders}, David B. and {Toft}, Sune and {Valentino}, Francesco and {Zamorani}, Giovanni and {Zavala}, Jorge and {Cosmos Team}},
        title = "{COSMOS2020: Identification of High-z Protocluster Candidates in COSMOS}",
      journal = {\apj},
         year = 2023,
        month = feb,
       volume = {943},
       number = {2},
          eid = {153},
        pages = {153},
          doi = {10.3847/1538-4357/ac9d96},
archivePrefix = {arXiv},
       eprint = {2210.17334},
 primaryClass = {astro-ph.GA},
       adsurl = {https://ui.adsabs.harvard.edu/abs/2023ApJ...943..153B}
}

@ARTICLE{2024ApJ...966...18T,
       author = {{Taamoli}, Sina and {Mobasher}, Bahram and {Chartab}, Nima and {Darvish}, Behnam and {Weaver}, John R. and {Hemmati}, Shoubaneh and {Casey}, Caitlin M. and {Sattari}, Zahra and {Brammer}, Gabriel and {Capak}, Peter L. and {Ilbert}, Olivier and {Kartaltepe}, Jeyhan S. and {McCracken}, Henry J. and {Moneti}, Andrea and {Sanders}, David B. and {Scoville}, Nicholas and {Steinhardt}, Charles L. and {Toft}, Sune},
        title = "{Large-scale Structures in COSMOS2020: Evolution of Star Formation Activity in Different Environments at 0.4 < z < 4}",
      journal = {\apj},
         year = 2024,
        month = may,
       volume = {966},
       number = {1},
          eid = {18},
        pages = {18},
          doi = {10.3847/1538-4357/ad32c5},
archivePrefix = {arXiv},
       eprint = {2312.10222},
 primaryClass = {astro-ph.GA},
       adsurl = {https://ui.adsabs.harvard.edu/abs/2024ApJ...966...18T}
}

@article{hall1982cross,
  title={Cross-validation in density estimation},
  author={Hall, Peter},
  journal={Biometrika},
  volume={69},
  number={2},
  pages={383--390},
  year={1982},
  publisher={Oxford University Press}
}

@article{abramson1982bandwidth,
  title={On bandwidth variation in kernel estimates-a square root law},
  author={Abramson, Ian S},
  journal={The annals of Statistics},
  pages={1217--1223},
  year={1982},
  publisher={JSTOR}
}

@book{silverman2018density,
  title={Density estimation for statistics and data analysis},
  author={Silverman, Bernard W},
  year={2018},
  publisher={Routledge}
}

@ARTICLE{2010ApJ...708...58C,
       author = {{Conroy}, Charlie and {White}, Martin and {Gunn}, James E.},
        title = "{The Propagation of Uncertainties in Stellar Population Synthesis Modeling. II. The Challenge of Comparing Galaxy Evolution Models to Observations}",
      journal = {\apj},
         year = 2010,
        month = jan,
       volume = {708},
       number = {1},
        pages = {58-70},
          doi = {10.1088/0004-637X/708/1/58},
archivePrefix = {arXiv},
       eprint = {0904.0002},
 primaryClass = {astro-ph.CO},
       adsurl = {https://ui.adsabs.harvard.edu/abs/2010ApJ...708...58C}
}

@ARTICLE{2008ApJ...686.1503B,
       author = {{Brammer}, Gabriel B. and {van Dokkum}, Pieter G. and {Coppi}, Paolo},
        title = "{EAZY: A Fast, Public Photometric Redshift Code}",
      journal = {\apj},
         year = 2008,
        month = oct,
       volume = {686},
       number = {2},
        pages = {1503-1513},
          doi = {10.1086/591786},
archivePrefix = {arXiv},
       eprint = {0807.1533},
 primaryClass = {astro-ph},
       adsurl = {https://ui.adsabs.harvard.edu/abs/2008ApJ...686.1503B}
}

@ARTICLE{1996A&AS..117..393B,
       author = {{Bertin}, E. and {Arnouts}, S.},
        title = "{SExtractor: Software for source extraction.}",
      journal = {\aaps},
         year = 1996,
        month = jun,
       volume = {117},
        pages = {393-404},
          doi = {10.1051/aas:1996164},
       adsurl = {https://ui.adsabs.harvard.edu/abs/1996A&AS..117..393B}
}

@ARTICLE{1998ApJ...500..525S,
       author = {{Schlegel}, David J. and {Finkbeiner}, Douglas P. and {Davis}, Marc},
        title = "{Maps of Dust Infrared Emission for Use in Estimation of Reddening and Cosmic Microwave Background Radiation Foregrounds}",
      journal = {\apj},
         year = 1998,
        month = jun,
       volume = {500},
       number = {2},
        pages = {525-553},
          doi = {10.1086/305772},
archivePrefix = {arXiv},
       eprint = {astro-ph/9710327},
 primaryClass = {astro-ph},
       adsurl = {https://ui.adsabs.harvard.edu/abs/1998ApJ...500..525S}
}

@ARTICLE{2011ApJ...737..103S,
       author = {{Schlafly}, Edward F. and {Finkbeiner}, Douglas P.},
        title = "{Measuring Reddening with Sloan Digital Sky Survey Stellar Spectra and Recalibrating SFD}",
      journal = {\apj},
         year = 2011,
        month = aug,
       volume = {737},
       number = {2},
          eid = {103},
        pages = {103},
          doi = {10.1088/0004-637X/737/2/103},
archivePrefix = {arXiv},
       eprint = {1012.4804},
 primaryClass = {astro-ph.GA},
       adsurl = {https://ui.adsabs.harvard.edu/abs/2011ApJ...737..103S}
}

@ARTICLE{2019A&A...622A.103B,
       author = {{Boquien}, M. and {Burgarella}, D. and {Roehlly}, Y. and {Buat}, V. and {Ciesla}, L. and {Corre}, D. and {Inoue}, A.~K. and {Salas}, H.},
        title = "{CIGALE: a python Code Investigating GALaxy Emission}",
      journal = {\aap},
         year = 2019,
        month = feb,
       volume = {622},
          eid = {A103},
        pages = {A103},
          doi = {10.1051/0004-6361/201834156},
archivePrefix = {arXiv},
       eprint = {1811.03094},
 primaryClass = {astro-ph.GA},
       adsurl = {https://ui.adsabs.harvard.edu/abs/2019A&A...622A.103B}
}

@ARTICLE{2013A&A...558A..33A,
       author = {{Astropy Collaboration} and {Robitaille}, Thomas P. and {Tollerud}, Erik J. and {Greenfield}, Perry and {Droettboom}, Michael and {Bray}, Erik and {Aldcroft}, Tom and {Davis}, Matt and {Ginsburg}, Adam and {Price-Whelan}, Adrian M. and {Kerzendorf}, Wolfgang E. and {Conley}, Alexander and {Crighton}, Neil and {Barbary}, Kyle and {Muna}, Demitri and {Ferguson}, Henry and {Grollier}, Fr{\'e}d{\'e}ric and {Parikh}, Madhura M. and {Nair}, Prasanth H. and {Unther}, Hans M. and {Deil}, Christoph and {Woillez}, Julien and {Conseil}, Simon and {Kramer}, Roban and {Turner}, James E.~H. and {Singer}, Leo and {Fox}, Ryan and {Weaver}, Benjamin A. and {Zabalza}, Victor and {Edwards}, Zachary I. and {Azalee Bostroem}, K. and {Burke}, D.~J. and {Casey}, Andrew R. and {Crawford}, Steven M. and {Dencheva}, Nadia and {Ely}, Justin and {Jenness}, Tim and {Labrie}, Kathleen and {Lim}, Pey Lian and {Pierfederici}, Francesco and {Pontzen}, Andrew and {Ptak}, Andy and {Refsdal}, Brian and {Servillat}, Mathieu and {Streicher}, Ole},
        title = "{Astropy: A community Python package for astronomy}",
      journal = {\aap},
         year = 2013,
        month = oct,
       volume = {558},
          eid = {A33},
        pages = {A33},
          doi = {10.1051/0004-6361/201322068},
archivePrefix = {arXiv},
       eprint = {1307.6212},
 primaryClass = {astro-ph.IM},
       adsurl = {https://ui.adsabs.harvard.edu/abs/2013A&A...558A..33A}
}

@ARTICLE{2018AJ....156..123A,
       author = {{Astropy Collaboration} and {Price-Whelan}, A.~M. and {Sip{\H{o}}cz}, B.~M. and {G{\"u}nther}, H.~M. and {Lim}, P.~L. and {Crawford}, S.~M. and {Conseil}, S. and {Shupe}, D.~L. and {Craig}, M.~W. and {Dencheva}, N. and {Ginsburg}, A. and {VanderPlas}, J.~T. and {Bradley}, L.~D. and {P{\'e}rez-Su{\'a}rez}, D. and {de Val-Borro}, M. and {Aldcroft}, T.~L. and {Cruz}, K.~L. and {Robitaille}, T.~P. and {Tollerud}, E.~J. and {Ardelean}, C. and {Babej}, T. and {Bach}, Y.~P. and {Bachetti}, M. and {Bakanov}, A.~V. and {Bamford}, S.~P. and {Barentsen}, G. and {Barmby}, P. and {Baumbach}, A. and {Berry}, K.~L. and {Biscani}, F. and {Boquien}, M. and {Bostroem}, K.~A. and {Bouma}, L.~G. and {Brammer}, G.~B. and {Bray}, E.~M. and {Breytenbach}, H. and {Buddelmeijer}, H. and {Burke}, D.~J. and {Calderone}, G. and {Cano Rodr{\'\i}guez}, J.~L. and {Cara}, M. and {Cardoso}, J.~V.~M. and {Cheedella}, S. and {Copin}, Y. and {Corrales}, L. and {Crichton}, D. and {D'Avella}, D. and {Deil}, C. and {Depagne}, {\'E}. and {Dietrich}, J.~P. and {Donath}, A. and {Droettboom}, M. and {Earl}, N. and {Erben}, T. and {Fabbro}, S. and {Ferreira}, L.~A. and {Finethy}, T. and {Fox}, R.~T. and {Garrison}, L.~H. and {Gibbons}, S.~L.~J. and {Goldstein}, D.~A. and {Gommers}, R. and {Greco}, J.~P. and {Greenfield}, P. and {Groener}, A.~M. and {Grollier}, F. and {Hagen}, A. and {Hirst}, P. and {Homeier}, D. and {Horton}, A.~J. and {Hosseinzadeh}, G. and {Hu}, L. and {Hunkeler}, J.~S. and {Ivezi{\'c}}, {\v{Z}}. and {Jain}, A. and {Jenness}, T. and {Kanarek}, G. and {Kendrew}, S. and {Kern}, N.~S. and {Kerzendorf}, W.~E. and {Khvalko}, A. and {King}, J. and {Kirkby}, D. and {Kulkarni}, A.~M. and {Kumar}, A. and {Lee}, A. and {Lenz}, D. and {Littlefair}, S.~P. and {Ma}, Z. and {Macleod}, D.~M. and {Mastropietro}, M. and {McCully}, C. and {Montagnac}, S. and {Morris}, B.~M. and {Mueller}, M. and {Mumford}, S.~J. and {Muna}, D. and {Murphy}, N.~A. and {Nelson}, S. and {Nguyen}, G.~H. and {Ninan}, J.~P. and {N{\"o}the}, M. and {Ogaz}, S. and {Oh}, S. and {Parejko}, J.~K. and {Parley}, N. and {Pascual}, S. and {Patil}, R. and {Patil}, A.~A. and {Plunkett}, A.~L. and {Prochaska}, J.~X. and {Rastogi}, T. and {Reddy Janga}, V. and {Sabater}, J. and {Sakurikar}, P. and {Seifert}, M. and {Sherbert}, L.~E. and {Sherwood-Taylor}, H. and {Shih}, A.~Y. and {Sick}, J. and {Silbiger}, M.~T. and {Singanamalla}, S. and {Singer}, L.~P. and {Sladen}, P.~H. and {Sooley}, K.~A. and {Sornarajah}, S. and {Streicher}, O. and {Teuben}, P. and {Thomas}, S.~W. and {Tremblay}, G.~R. and {Turner}, J.~E.~H. and {Terr{\'o}n}, V. and {van Kerkwijk}, M.~H. and {de la Vega}, A. and {Watkins}, L.~L. and {Weaver}, B.~A. and {Whitmore}, J.~B. and {Woillez}, J. and {Zabalza}, V. and {Astropy Contributors}},
        title = "{The Astropy Project: Building an Open-science Project and Status of the v2.0 Core Package}",
      journal = {\aj},
         year = 2018,
        month = sep,
       volume = {156},
       number = {3},
          eid = {123},
        pages = {123},
          doi = {10.3847/1538-3881/aabc4f},
archivePrefix = {arXiv},
       eprint = {1801.02634},
 primaryClass = {astro-ph.IM},
       adsurl = {https://ui.adsabs.harvard.edu/abs/2018AJ....156..123A}
}

@ARTICLE{2022ApJ...935..167A,
       author = {{Astropy Collaboration} and {Price-Whelan}, Adrian M. and {Lim}, Pey Lian and {Earl}, Nicholas and {Starkman}, Nathaniel and {Bradley}, Larry and {Shupe}, David L. and {Patil}, Aarya A. and {Corrales}, Lia and {Brasseur}, C.~E. and {N{\"o}the}, Maximilian and {Donath}, Axel and {Tollerud}, Erik and {Morris}, Brett M. and {Ginsburg}, Adam and {Vaher}, Eero and {Weaver}, Benjamin A. and {Tocknell}, James and {Jamieson}, William and {van Kerkwijk}, Marten H. and {Robitaille}, Thomas P. and {Merry}, Bruce and {Bachetti}, Matteo and {G{\"u}nther}, H. Moritz and {Aldcroft}, Thomas L. and {Alvarado-Montes}, Jaime A. and {Archibald}, Anne M. and {B{\'o}di}, Attila and {Bapat}, Shreyas and {Barentsen}, Geert and {Baz{\'a}n}, Juanjo and {Biswas}, Manish and {Boquien}, M{\'e}d{\'e}ric and {Burke}, D.~J. and {Cara}, Daria and {Cara}, Mihai and {Conroy}, Kyle E. and {Conseil}, Simon and {Craig}, Matthew W. and {Cross}, Robert M. and {Cruz}, Kelle L. and {D'Eugenio}, Francesco and {Dencheva}, Nadia and {Devillepoix}, Hadrien A.~R. and {Dietrich}, J{\"o}rg P. and {Eigenbrot}, Arthur Davis and {Erben}, Thomas and {Ferreira}, Leonardo and {Foreman-Mackey}, Daniel and {Fox}, Ryan and {Freij}, Nabil and {Garg}, Suyog and {Geda}, Robel and {Glattly}, Lauren and {Gondhalekar}, Yash and {Gordon}, Karl D. and {Grant}, David and {Greenfield}, Perry and {Groener}, Austen M. and {Guest}, Steve and {Gurovich}, Sebastian and {Handberg}, Rasmus and {Hart}, Akeem and {Hatfield-Dodds}, Zac and {Homeier}, Derek and {Hosseinzadeh}, Griffin and {Jenness}, Tim and {Jones}, Craig K. and {Joseph}, Prajwel and {Kalmbach}, J. Bryce and {Karamehmetoglu}, Emir and {Ka{\l}uszy{\'n}ski}, Miko{\l}aj and {Kelley}, Michael S.~P. and {Kern}, Nicholas and {Kerzendorf}, Wolfgang E. and {Koch}, Eric W. and {Kulumani}, Shankar and {Lee}, Antony and {Ly}, Chun and {Ma}, Zhiyuan and {MacBride}, Conor and {Maljaars}, Jakob M. and {Muna}, Demitri and {Murphy}, N.~A. and {Norman}, Henrik and {O'Steen}, Richard and {Oman}, Kyle A. and {Pacifici}, Camilla and {Pascual}, Sergio and {Pascual-Granado}, J. and {Patil}, Rohit R. and {Perren}, Gabriel I. and {Pickering}, Timothy E. and {Rastogi}, Tanuj and {Roulston}, Benjamin R. and {Ryan}, Daniel F. and {Rykoff}, Eli S. and {Sabater}, Jose and {Sakurikar}, Parikshit and {Salgado}, Jes{\'u}s and {Sanghi}, Aniket and {Saunders}, Nicholas and {Savchenko}, Volodymyr and {Schwardt}, Ludwig and {Seifert-Eckert}, Michael and {Shih}, Albert Y. and {Jain}, Anany Shrey and {Shukla}, Gyanendra and {Sick}, Jonathan and {Simpson}, Chris and {Singanamalla}, Sudheesh and {Singer}, Leo P. and {Singhal}, Jaladh and {Sinha}, Manodeep and {Sip{\H{o}}cz}, Brigitta M. and {Spitler}, Lee R. and {Stansby}, David and {Streicher}, Ole and {{\v{S}}umak}, Jani and {Swinbank}, John D. and {Taranu}, Dan S. and {Tewary}, Nikita and {Tremblay}, Grant R. and {de Val-Borro}, Miguel and {Van Kooten}, Samuel J. and {Vasovi{\'c}}, Zlatan and {Verma}, Shresth and {de Miranda Cardoso}, Jos{\'e} Vin{\'\i}cius and {Williams}, Peter K.~G. and {Wilson}, Tom J. and {Winkel}, Benjamin and {Wood-Vasey}, W.~M. and {Xue}, Rui and {Yoachim}, Peter and {Zhang}, Chen and {Zonca}, Andrea and {Astropy Project Contributors}},
        title = "{The Astropy Project: Sustaining and Growing a Community-oriented Open-source Project and the Latest Major Release (v5.0) of the Core Package}",
      journal = {\apj},
         year = 2022,
        month = aug,
       volume = {935},
       number = {2},
          eid = {167},
        pages = {167},
          doi = {10.3847/1538-4357/ac7c74},
archivePrefix = {arXiv},
       eprint = {2206.14220},
 primaryClass = {astro-ph.IM},
       adsurl = {https://ui.adsabs.harvard.edu/abs/2022ApJ...935..167A}
}

@ARTICLE{2019MNRAS.488.3143B,
       author = {{Behroozi}, Peter and {Wechsler}, Risa H. and {Hearin}, Andrew P. and {Conroy}, Charlie},
        title = "{UNIVERSEMACHINE: The correlation between galaxy growth and dark matter halo assembly from z = 0-10}",
      journal = {\mnras},
         year = 2019,
        month = sep,
       volume = {488},
       number = {3},
        pages = {3143-3194},
          doi = {10.1093/mnras/stz1182},
archivePrefix = {arXiv},
       eprint = {1806.07893},
 primaryClass = {astro-ph.GA},
       adsurl = {https://ui.adsabs.harvard.edu/abs/2019MNRAS.488.3143B}
}

@ARTICLE{2025A&A...695A..20S,
       author = {{Shuntov}, M. and {Ilbert}, O. and {Toft}, S. and {Arango-Toro}, R.~C. and {Akins}, H.~B. and {Casey}, C.~M. and {Franco}, M. and {Harish}, S. and {Kartaltepe}, J.~S. and {Koekemoer}, A.~M. and {McCracken}, H.~J. and {Paquereau}, L. and {Laigle}, C. and {Bethermin}, M. and {Dubois}, Y. and {Drakos}, N.~E. and {Faisst}, A. and {Gozaliasl}, G. and {Gillman}, S. and {Hayward}, C.~C. and {Hirschmann}, M. and {Huertas-Company}, M. and {Jespersen}, C.~K. and {Jin}, S. and {Kokorev}, V. and {Lambrides}, E. and {Le Borgne}, D. and {Liu}, D. and {Magdis}, G. and {Massey}, R. and {McPartland}, C.~J.~R. and {Mercier}, W. and {McCleary}, J.~E. and {McKinney}, J. and {Oesch}, P.~A. and {Renzini}, A. and {Rhodes}, J.~D. and {Rich}, R.~M. and {Robertson}, B.~E. and {Sanders}, D. and {Trebitsch}, M. and {Tresse}, L. and {Valentino}, F. and {Vijayan}, A.~P. and {Weaver}, J.~R. and {Weibel}, A. and {Wilkins}, S.~M. and {Yang}, L.},
        title = "{COSMOS-Web: Stellar mass assembly in relation to dark matter halos across 0.2 < z < 12 of cosmic history}",
      journal = {\aap},
         year = 2025,
        month = mar,
       volume = {695},
          eid = {A20},
        pages = {A20},
          doi = {10.1051/0004-6361/202452570},
archivePrefix = {arXiv},
       eprint = {2410.08290},
 primaryClass = {astro-ph.GA},
       adsurl = {https://ui.adsabs.harvard.edu/abs/2025A&A...695A..20S}
}

@ARTICLE{2013ApJ...770...57B,
       author = {{Behroozi}, Peter S. and {Wechsler}, Risa H. and {Conroy}, Charlie},
        title = "{The Average Star Formation Histories of Galaxies in Dark Matter Halos from z = 0-8}",
      journal = {\apj},
         year = 2013,
        month = jun,
       volume = {770},
       number = {1},
          eid = {57},
        pages = {57},
          doi = {10.1088/0004-637X/770/1/57},
archivePrefix = {arXiv},
       eprint = {1207.6105},
 primaryClass = {astro-ph.CO},
       adsurl = {https://ui.adsabs.harvard.edu/abs/2013ApJ...770...57B}
}

@ARTICLE{2022A&A...664A..61S,
       author = {{Shuntov}, M. and {McCracken}, H.~J. and {Gavazzi}, R. and {Laigle}, C. and {Weaver}, J.~R. and {Davidzon}, I. and {Ilbert}, O. and {Kauffmann}, O.~B. and {Faisst}, A. and {Dubois}, Y. and {Koekemoer}, A.~M. and {Moneti}, A. and {Milvang-Jensen}, B. and {Mobasher}, B. and {Sanders}, D.~B. and {Toft}, S.},
        title = "{COSMOS2020: Cosmic evolution of the stellar-to-halo mass relation for central and satellite galaxies up to z {\ensuremath{\sim}} 5}",
      journal = {\aap},
         year = 2022,
        month = aug,
       volume = {664},
          eid = {A61},
        pages = {A61},
          doi = {10.1051/0004-6361/202243136},
archivePrefix = {arXiv},
       eprint = {2203.10895},
 primaryClass = {astro-ph.GA},
       adsurl = {https://ui.adsabs.harvard.edu/abs/2022A&A...664A..61S}
}

@ARTICLE{2022A&A...667L...3L,
       author = {{Laporte}, N. and {Zitrin}, A. and {Dole}, H. and {Roberts-Borsani}, G. and {Furtak}, L.~J. and {Witten}, C.},
        title = "{A lensed protocluster candidate at z = 7.66 identified in JWST observations of the galaxy cluster SMACS0723{\ensuremath{-}}7327}",
      journal = {\aap},
         year = 2022,
        month = nov,
       volume = {667},
          eid = {L3},
        pages = {L3},
          doi = {10.1051/0004-6361/202244719},
archivePrefix = {arXiv},
       eprint = {2208.04930},
 primaryClass = {astro-ph.GA},
       adsurl = {https://ui.adsabs.harvard.edu/abs/2022A&A...667L...3L}
}

@ARTICLE{2023MNRAS.521..497M,
       author = {{Mason}, Charlotte A. and {Trenti}, Michele and {Treu}, Tommaso},
        title = "{The brightest galaxies at cosmic dawn}",
      journal = {\mnras},
         year = 2023,
        month = may,
       volume = {521},
       number = {1},
        pages = {497-503},
          doi = {10.1093/mnras/stad035},
archivePrefix = {arXiv},
       eprint = {2207.14808},
 primaryClass = {astro-ph.GA},
       adsurl = {https://ui.adsabs.harvard.edu/abs/2023MNRAS.521..497M}
}

@ARTICLE{2025ApJ...988L..19J,
       author = {{Jespersen}, Christian Kragh and {Carnall}, Adam C. and {Lovell}, Christopher C.},
        title = "{Explaining Ultramassive Quiescent Galaxies at 3 < z < 5 in the Context of Their Environments}",
      journal = {\apjl},
         year = 2025,
        month = jul,
       volume = {988},
       number = {1},
          eid = {L19},
        pages = {L19},
          doi = {10.3847/2041-8213/adeb7c},
archivePrefix = {arXiv},
       eprint = {2507.05340},
 primaryClass = {astro-ph.GA},
       adsurl = {https://ui.adsabs.harvard.edu/abs/2025ApJ...988L..19J}
}

@ARTICLE{2012ApJ...745..179W,
       author = {{Whitaker}, Katherine E. and {Kriek}, Mariska and {van Dokkum}, Pieter G. and {Bezanson}, Rachel and {Brammer}, Gabriel and {Franx}, Marijn and {Labb{\'e}}, Ivo},
        title = "{A Large Population of Massive Compact Post-starburst Galaxies at z > 1: Implications for the Size Evolution and Quenching Mechanism of Quiescent Galaxies}",
      journal = {\apj},
         year = 2012,
        month = feb,
       volume = {745},
       number = {2},
          eid = {179},
        pages = {179},
          doi = {10.1088/0004-637X/745/2/179},
archivePrefix = {arXiv},
       eprint = {1112.0313},
 primaryClass = {astro-ph.CO},
       adsurl = {https://ui.adsabs.harvard.edu/abs/2012ApJ...745..179W}
}

@ARTICLE{2015ApJ...811L..12W,
       author = {{Whitaker}, Katherine E. and {Franx}, Marijn and {Bezanson}, Rachel and {Brammer}, Gabriel B. and {van Dokkum}, Pieter G. and {Kriek}, Mariska T. and {Labb{\'e}}, Ivo and {Leja}, Joel and {Momcheva}, Ivelina G. and {Nelson}, Erica J. and {Rigby}, Jane R. and {Rix}, Hans-Walter and {Skelton}, Rosalind E. and {van der Wel}, Arjen and {Wuyts}, Stijn},
        title = "{Galaxy Structure as a Driver of the Star Formation Sequence Slope and Scatter}",
      journal = {\apjl},
         year = 2015,
        month = sep,
       volume = {811},
       number = {1},
          eid = {L12},
        pages = {L12},
          doi = {10.1088/2041-8205/811/1/L12},
archivePrefix = {arXiv},
       eprint = {1508.04771},
 primaryClass = {astro-ph.GA},
       adsurl = {https://ui.adsabs.harvard.edu/abs/2015ApJ...811L..12W}
}

@ARTICLE{2013ApJ...779..127C,
       author = {{Chiang}, Yi-Kuan and {Overzier}, Roderik and {Gebhardt}, Karl},
        title = "{Ancient Light from Young Cosmic Cities: Physical and Observational Signatures of Galaxy Proto-clusters}",
      journal = {\apj},
         year = 2013,
        month = dec,
       volume = {779},
       number = {2},
          eid = {127},
        pages = {127},
          doi = {10.1088/0004-637X/779/2/127},
archivePrefix = {arXiv},
       eprint = {1310.2938},
 primaryClass = {astro-ph.CO},
       adsurl = {https://ui.adsabs.harvard.edu/abs/2013ApJ...779..127C}
}

@article{schuster1985incorporating,
  title={Incorporating support constraints into nonparametric estimators of densities},
  author={Schuster, Eugene F},
  journal={Communications in Statistics-Theory and methods},
  volume={14},
  number={5},
  pages={1123--1136},
  year={1985},
  publisher={Taylor \& Francis}
}

@article{jones1993simple,
  title={Simple boundary correction for kernel density estimation},
  author={Jones, M Chris},
  journal={Statistics and computing},
  volume={3},
  number={3},
  pages={135--146},
  year={1993},
  publisher={Springer}
}

@article{diggle1985kernel,
  title={A kernel method for smoothing point process data},
  author={Diggle, Peter},
  journal={Journal of the Royal Statistical Society: Series C (Applied Statistics)},
  volume={34},
  number={2},
  pages={138--147},
  year={1985},
  publisher={Wiley Online Library}
}

@article{marron1994transformations,
  title={Transformations to reduce boundary bias in kernel density estimation},
  author={Marron, James Stephen and Ruppert, David},
  journal={Journal of the Royal Statistical Society: Series B (Methodological)},
  volume={56},
  number={4},
  pages={653--671},
  year={1994},
  publisher={Wiley Online Library}
}

@article{muller1991smooth,
  title={Smooth optimum kernel estimators near endpoints},
  author={M{\"u}ller, Hans-Georg},
  journal={Biometrika},
  volume={78},
  number={3},
  pages={521--530},
  year={1991},
  publisher={Oxford University Press}
}

@article{chen1999beta,
  title={Beta kernel estimators for density functions},
  author={Chen, Song Xi},
  journal={Computational Statistics \& Data Analysis},
  volume={31},
  number={2},
  pages={131--145},
  year={1999},
  publisher={Elsevier}
}

@article{chen2000probability,
  title={Probability density function estimation using gamma kernels},
  author={Chen, Song Xi},
  journal={Annals of the institute of statistical mathematics},
  volume={52},
  number={3},
  pages={471--480},
  year={2000},
  publisher={Springer}
}

@article{cheng1997boundary,
  title={Boundary aware estimators of integrated density derivative products},
  author={Cheng, Ming-Yen},
  journal={Journal of the Royal Statistical Society: Series B (Statistical Methodology)},
  volume={59},
  number={1},
  pages={191--203},
  year={1997},
  publisher={Wiley Online Library}
}

@article{cowling1996pseudodata,
  title={On pseudodata methods for removing boundary effects in kernel density estimation},
  author={Cowling, Ann and Hall, Peter},
  journal={Journal of the Royal Statistical Society Series B: Statistical Methodology},
  volume={58},
  number={3},
  pages={551--563},
  year={1996},
  publisher={Oxford University Press}
}

@ARTICLE{2014ApJS..214...24S,
       author = {{Skelton}, Rosalind E. and {Whitaker}, Katherine E. and {Momcheva}, Ivelina G. and {Brammer}, Gabriel B. and {van Dokkum}, Pieter G. and {Labb{\'e}}, Ivo and {Franx}, Marijn and {van der Wel}, Arjen and {Bezanson}, Rachel and {Da Cunha}, Elisabete and {Fumagalli}, Mattia and {F{\"o}rster Schreiber}, Natascha and {Kriek}, Mariska and {Leja}, Joel and {Lundgren}, Britt F. and {Magee}, Daniel and {Marchesini}, Danilo and {Maseda}, Michael V. and {Nelson}, Erica J. and {Oesch}, Pascal and {Pacifici}, Camilla and {Patel}, Shannon G. and {Price}, Sedona and {Rix}, Hans-Walter and {Tal}, Tomer and {Wake}, David A. and {Wuyts}, Stijn},
        title = "{3D-HST WFC3-selected Photometric Catalogs in the Five CANDELS/3D-HST Fields: Photometry, Photometric Redshifts, and Stellar Masses}",
      journal = {\apjs},
         year = 2014,
        month = oct,
       volume = {214},
       number = {2},
          eid = {24},
        pages = {24},
          doi = {10.1088/0067-0049/214/2/24},
archivePrefix = {arXiv},
       eprint = {1403.3689},
 primaryClass = {astro-ph.GA},
       adsurl = {https://ui.adsabs.harvard.edu/abs/2014ApJS..214...24S}
}

@ARTICLE{2012ApJS..200...13B,
       author = {{Brammer}, Gabriel B. and {van Dokkum}, Pieter G. and {Franx}, Marijn and {Fumagalli}, Mattia and {Patel}, Shannon and {Rix}, Hans-Walter and {Skelton}, Rosalind E. and {Kriek}, Mariska and {Nelson}, Erica and {Schmidt}, Kasper B. and {Bezanson}, Rachel and {da Cunha}, Elisabete and {Erb}, Dawn K. and {Fan}, Xiaohui and {F{\"o}rster Schreiber}, Natascha and {Illingworth}, Garth D. and {Labb{\'e}}, Ivo and {Leja}, Joel and {Lundgren}, Britt and {Magee}, Dan and {Marchesini}, Danilo and {McCarthy}, Patrick and {Momcheva}, Ivelina and {Muzzin}, Adam and {Quadri}, Ryan and {Steidel}, Charles C. and {Tal}, Tomer and {Wake}, David and {Whitaker}, Katherine E. and {Williams}, Anna},
        title = "{3D-HST: A Wide-field Grism Spectroscopic Survey with the Hubble Space Telescope}",
      journal = {\apjs},
         year = 2012,
        month = jun,
       volume = {200},
       number = {2},
          eid = {13},
        pages = {13},
          doi = {10.1088/0067-0049/200/2/13},
archivePrefix = {arXiv},
       eprint = {1204.2829},
 primaryClass = {astro-ph.CO},
       adsurl = {https://ui.adsabs.harvard.edu/abs/2012ApJS..200...13B}
}

@ARTICLE{2025arXiv250300120K,
       author = {{Khostovan}, Ali Ahmad and {Kartaltepe}, Jeyhan S. and {Salvato}, Mara and {Ilbert}, Olivier and {Casey}, Caitlin M. and {Algera}, Hiddo and {Antwi-Danso}, Jacqueline and {Battisti}, Andrew and {Brinch}, Malte and {Brusa}, Marcella and {Calabro}, Antonello and {Capak}, Peter L. and {Chartab}, Nima and {Cooper}, Olivia R. and {Cox}, Isa G. and {Darvish}, Behnam and {Drakos}, Nicole E. and {Faisst}, Andreas L. and {George}, Matthew R. and {Gozaliasl}, Ghassem and {Harish}, Santosh and {Hasinger}, Gunther and {Hatamnia}, Hossein and {Iovino}, Angela and {Jin}, Shuowen and {Kashino}, Daichi and {Koekemoer}, Anton M. and {Laishram}, Ronaldo and {Lee}, Khee-Gan and {Lertprasertpong}, Jitrapon and {Lilly}, Simon J. and {Masters}, Daniel C. and {Mobasher}, Bahram and {Nagao}, Tohru and {Onodera}, Masato and {Peng}, Yingjie and {Sanders}, David B. and {Sanders}, Ryan L. and {Sattari}, Zahra and {Scoville}, Nick and {Shah}, Ekta A. and {Silverman}, John D. and {Suzuki}, Nao and {Tanaka}, Masayuki and {Toft}, Sune and {Trakhtenbrot}, Benny and {Trump}, Jonathan R. and {Vaccari}, Mattia and {Valentino}, Francesco and {Vanderhoof}, Brittany N. and {Weaver}, John R. and {Yun}, Min S. and {Zavala}, Jorge A.},
        title = "{COSMOS Spectroscopic Redshift Compilation (First Data Release): 165k Redshifts Encompassing Two Decades of Spectroscopy}",
      journal = {arXiv e-prints},
         year = 2025,
        month = feb,
          eid = {arXiv:2503.00120},
        pages = {arXiv:2503.00120},
          doi = {10.48550/arXiv.2503.00120},
archivePrefix = {arXiv},
       eprint = {2503.00120},
 primaryClass = {astro-ph.GA},
       adsurl = {https://ui.adsabs.harvard.edu/abs/2025arXiv250300120K}
}

@ARTICLE{2008A&A...478...83V,
       author = {{Vanzella}, E. and {Cristiani}, S. and {Dickinson}, M. and {Giavalisco}, M. and {Kuntschner}, H. and {Haase}, J. and {Nonino}, M. and {Rosati}, P. and {Cesarsky}, C. and {Ferguson}, H.~C. and {Fosbury}, R.~A.~E. and {Grazian}, A. and {Moustakas}, L.~A. and {Rettura}, A. and {Popesso}, P. and {Renzini}, A. and {Stern}, D. and {GOODS Team}},
        title = "{The great observatories origins deep survey. VLT/FORS2 spectroscopy in the GOODS-South field: Part III}",
      journal = {\aap},
         year = 2008,
        month = jan,
       volume = {478},
       number = {1},
        pages = {83-92},
          doi = {10.1051/0004-6361:20078332},
archivePrefix = {arXiv},
       eprint = {0711.0850},
 primaryClass = {astro-ph},
       adsurl = {https://ui.adsabs.harvard.edu/abs/2008A&A...478...83V}
}

@ARTICLE{2010A&A...512A..12B,
       author = {{Balestra}, I. and {Mainieri}, V. and {Popesso}, P. and {Dickinson}, M. and {Nonino}, M. and {Rosati}, P. and {Teimoorinia}, H. and {Vanzella}, E. and {Cristiani}, S. and {Cesarsky}, C. and {Fosbury}, R.~A.~E. and {Kuntschner}, H. and {Rettura}, A.},
        title = "{The Great Observatories Origins Deep Survey. VLT/VIMOS spectroscopy in the GOODS-south field: Part II}",
      journal = {\aap},
         year = 2010,
        month = mar,
       volume = {512},
          eid = {A12},
        pages = {A12},
          doi = {10.1051/0004-6361/200913626},
archivePrefix = {arXiv},
       eprint = {1001.1115},
 primaryClass = {astro-ph.CO},
       adsurl = {https://ui.adsabs.harvard.edu/abs/2010A&A...512A..12B}
}

@ARTICLE{2021A&A...647A.150G,
       author = {{Garilli}, B. and {McLure}, R. and {Pentericci}, L. and {Franzetti}, P. and {Gargiulo}, A. and {Carnall}, A. and {Cucciati}, O. and {Iovino}, A. and {Amorin}, R. and {Bolzonella}, M. and {Bongiorno}, A. and {Castellano}, M. and {Cimatti}, A. and {Cirasuolo}, M. and {Cullen}, F. and {Dunlop}, J. and {Elbaz}, D. and {Finkelstein}, S. and {Fontana}, A. and {Fontanot}, F. and {Fumana}, M. and {Guaita}, L. and {Hartley}, W. and {Jarvis}, M. and {Juneau}, S. and {Maccagni}, D. and {McLeod}, D. and {Nandra}, K. and {Pompei}, E. and {Pozzetti}, L. and {Scodeggio}, M. and {Talia}, M. and {Calabr{\`o}}, A. and {Cresci}, G. and {Fynbo}, J.~P.~U. and {Hathi}, N.~P. and {Hibon}, P. and {Koekemoer}, A.~M. and {Magliocchetti}, M. and {Salvato}, M. and {Vietri}, G. and {Zamorani}, G. and {Almaini}, O. and {Balestra}, I. and {Bardelli}, S. and {Begley}, R. and {Brammer}, G. and {Bell}, E.~F. and {Bowler}, R.~A.~A. and {Brusa}, M. and {Buitrago}, F. and {Caputi}, C. and {Cassata}, P. and {Charlot}, S. and {Citro}, A. and {Cristiani}, S. and {Curtis-Lake}, E. and {Dickinson}, M. and {Fazio}, G. and {Ferguson}, H.~C. and {Fiore}, F. and {Franco}, M. and {Georgakakis}, A. and {Giavalisco}, M. and {Grazian}, A. and {Hamadouche}, M. and {Jung}, I. and {Kim}, S. and {Khusanova}, Y. and {Le F{\`e}vre}, O. and {Longhetti}, M. and {Lotz}, J. and {Mannucci}, F. and {Maltby}, D. and {Matsuoka}, K. and {Mendez-Hernandez}, H. and {Mendez-Abreu}, J. and {Mignoli}, M. and {Moresco}, M. and {Nonino}, M. and {Pannella}, M. and {Papovich}, C. and {Popesso}, P. and {Roberts-Borsani}, G. and {Rosario}, D.~J. and {Saldana-Lopez}, A. and {Santini}, P. and {Saxena}, A. and {Schaerer}, D. and {Schreiber}, C. and {Stark}, D. and {Tasca}, L.~A.~M. and {Thomas}, R. and {Vanzella}, E. and {Wild}, V. and {Williams}, C. and {Zucca}, E.},
        title = "{The VANDELS ESO public spectroscopic survey. Final data release of 2087 spectra and spectroscopic measurements}",
      journal = {\aap},
         year = 2021,
        month = mar,
       volume = {647},
          eid = {A150},
        pages = {A150},
          doi = {10.1051/0004-6361/202040059},
archivePrefix = {arXiv},
       eprint = {2101.07645},
 primaryClass = {astro-ph.GA},
       adsurl = {https://ui.adsabs.harvard.edu/abs/2021A&A...647A.150G}
}

@ARTICLE{2011ApJS..197...35G,
       author = {{Grogin}, Norman A. and {Kocevski}, Dale D. and {Faber}, S.~M. and {Ferguson}, Henry C. and {Koekemoer}, Anton M. and {Riess}, Adam G. and {Acquaviva}, Viviana and {Alexander}, David M. and {Almaini}, Omar and {Ashby}, Matthew L.~N. and {Barden}, Marco and {Bell}, Eric F. and {Bournaud}, Fr{\'e}d{\'e}ric and {Brown}, Thomas M. and {Caputi}, Karina I. and {Casertano}, Stefano and {Cassata}, Paolo and {Castellano}, Marco and {Challis}, Peter and {Chary}, Ranga-Ram and {Cheung}, Edmond and {Cirasuolo}, Michele and {Conselice}, Christopher J. and {Roshan Cooray}, Asantha and {Croton}, Darren J. and {Daddi}, Emanuele and {Dahlen}, Tomas and {Dav{\'e}}, Romeel and {de Mello}, Du{\'\i}lia F. and {Dekel}, Avishai and {Dickinson}, Mark and {Dolch}, Timothy and {Donley}, Jennifer L. and {Dunlop}, James S. and {Dutton}, Aaron A. and {Elbaz}, David and {Fazio}, Giovanni G. and {Filippenko}, Alexei V. and {Finkelstein}, Steven L. and {Fontana}, Adriano and {Gardner}, Jonathan P. and {Garnavich}, Peter M. and {Gawiser}, Eric and {Giavalisco}, Mauro and {Grazian}, Andrea and {Guo}, Yicheng and {Hathi}, Nimish P. and {H{\"a}ussler}, Boris and {Hopkins}, Philip F. and {Huang}, Jia-Sheng and {Huang}, Kuang-Han and {Jha}, Saurabh W. and {Kartaltepe}, Jeyhan S. and {Kirshner}, Robert P. and {Koo}, David C. and {Lai}, Kamson and {Lee}, Kyoung-Soo and {Li}, Weidong and {Lotz}, Jennifer M. and {Lucas}, Ray A. and {Madau}, Piero and {McCarthy}, Patrick J. and {McGrath}, Elizabeth J. and {McIntosh}, Daniel H. and {McLure}, Ross J. and {Mobasher}, Bahram and {Moustakas}, Leonidas A. and {Mozena}, Mark and {Nandra}, Kirpal and {Newman}, Jeffrey A. and {Niemi}, Sami-Matias and {Noeske}, Kai G. and {Papovich}, Casey J. and {Pentericci}, Laura and {Pope}, Alexandra and {Primack}, Joel R. and {Rajan}, Abhijith and {Ravindranath}, Swara and {Reddy}, Naveen A. and {Renzini}, Alvio and {Rix}, Hans-Walter and {Robaina}, Aday R. and {Rodney}, Steven A. and {Rosario}, David J. and {Rosati}, Piero and {Salimbeni}, Sara and {Scarlata}, Claudia and {Siana}, Brian and {Simard}, Luc and {Smidt}, Joseph and {Somerville}, Rachel S. and {Spinrad}, Hyron and {Straughn}, Amber N. and {Strolger}, Louis-Gregory and {Telford}, Olivia and {Teplitz}, Harry I. and {Trump}, Jonathan R. and {van der Wel}, Arjen and {Villforth}, Carolin and {Wechsler}, Risa H. and {Weiner}, Benjamin J. and {Wiklind}, Tommy and {Wild}, Vivienne and {Wilson}, Grant and {Wuyts}, Stijn and {Yan}, Hao-Jing and {Yun}, Min S.},
        title = "{CANDELS: The Cosmic Assembly Near-infrared Deep Extragalactic Legacy Survey}",
      journal = {\apjs},
         year = 2011,
        month = dec,
       volume = {197},
       number = {2},
          eid = {35},
        pages = {35},
          doi = {10.1088/0067-0049/197/2/35},
archivePrefix = {arXiv},
       eprint = {1105.3753},
 primaryClass = {astro-ph.CO},
       adsurl = {https://ui.adsabs.harvard.edu/abs/2011ApJS..197...35G}
}

@ARTICLE{2011ApJS..197...36K,
       author = {{Koekemoer}, Anton M. and {Faber}, S.~M. and {Ferguson}, Henry C. and {Grogin}, Norman A. and {Kocevski}, Dale D. and {Koo}, David C. and {Lai}, Kamson and {Lotz}, Jennifer M. and {Lucas}, Ray A. and {McGrath}, Elizabeth J. and {Ogaz}, Sara and {Rajan}, Abhijith and {Riess}, Adam G. and {Rodney}, Steve A. and {Strolger}, Louis and {Casertano}, Stefano and {Castellano}, Marco and {Dahlen}, Tomas and {Dickinson}, Mark and {Dolch}, Timothy and {Fontana}, Adriano and {Giavalisco}, Mauro and {Grazian}, Andrea and {Guo}, Yicheng and {Hathi}, Nimish P. and {Huang}, Kuang-Han and {van der Wel}, Arjen and {Yan}, Hao-Jing and {Acquaviva}, Viviana and {Alexander}, David M. and {Almaini}, Omar and {Ashby}, Matthew L.~N. and {Barden}, Marco and {Bell}, Eric F. and {Bournaud}, Fr{\'e}d{\'e}ric and {Brown}, Thomas M. and {Caputi}, Karina I. and {Cassata}, Paolo and {Challis}, Peter J. and {Chary}, Ranga-Ram and {Cheung}, Edmond and {Cirasuolo}, Michele and {Conselice}, Christopher J. and {Roshan Cooray}, Asantha and {Croton}, Darren J. and {Daddi}, Emanuele and {Dav{\'e}}, Romeel and {de Mello}, Duilia F. and {de Ravel}, Loic and {Dekel}, Avishai and {Donley}, Jennifer L. and {Dunlop}, James S. and {Dutton}, Aaron A. and {Elbaz}, David and {Fazio}, Giovanni G. and {Filippenko}, Alexei V. and {Finkelstein}, Steven L. and {Frazer}, Chris and {Gardner}, Jonathan P. and {Garnavich}, Peter M. and {Gawiser}, Eric and {Gruetzbauch}, Ruth and {Hartley}, Will G. and {H{\"a}ussler}, Boris and {Herrington}, Jessica and {Hopkins}, Philip F. and {Huang}, Jia-Sheng and {Jha}, Saurabh W. and {Johnson}, Andrew and {Kartaltepe}, Jeyhan S. and {Khostovan}, Ali A. and {Kirshner}, Robert P. and {Lani}, Caterina and {Lee}, Kyoung-Soo and {Li}, Weidong and {Madau}, Piero and {McCarthy}, Patrick J. and {McIntosh}, Daniel H. and {McLure}, Ross J. and {McPartland}, Conor and {Mobasher}, Bahram and {Moreira}, Heidi and {Mortlock}, Alice and {Moustakas}, Leonidas A. and {Mozena}, Mark and {Nandra}, Kirpal and {Newman}, Jeffrey A. and {Nielsen}, Jennifer L. and {Niemi}, Sami and {Noeske}, Kai G. and {Papovich}, Casey J. and {Pentericci}, Laura and {Pope}, Alexandra and {Primack}, Joel R. and {Ravindranath}, Swara and {Reddy}, Naveen A. and {Renzini}, Alvio and {Rix}, Hans-Walter and {Robaina}, Aday R. and {Rosario}, David J. and {Rosati}, Piero and {Salimbeni}, Sara and {Scarlata}, Claudia and {Siana}, Brian and {Simard}, Luc and {Smidt}, Joseph and {Snyder}, Diana and {Somerville}, Rachel S. and {Spinrad}, Hyron and {Straughn}, Amber N. and {Telford}, Olivia and {Teplitz}, Harry I. and {Trump}, Jonathan R. and {Vargas}, Carlos and {Villforth}, Carolin and {Wagner}, Cory R. and {Wandro}, Pat and {Wechsler}, Risa H. and {Weiner}, Benjamin J. and {Wiklind}, Tommy and {Wild}, Vivienne and {Wilson}, Grant and {Wuyts}, Stijn and {Yun}, Min S.},
        title = "{CANDELS: The Cosmic Assembly Near-infrared Deep Extragalactic Legacy Survey{\textemdash}The Hubble Space Telescope Observations, Imaging Data Products, and Mosaics}",
      journal = {\apjs},
         year = 2011,
        month = dec,
       volume = {197},
       number = {2},
          eid = {36},
        pages = {36},
          doi = {10.1088/0067-0049/197/2/36},
archivePrefix = {arXiv},
       eprint = {1105.3754},
 primaryClass = {astro-ph.CO},
       adsurl = {https://ui.adsabs.harvard.edu/abs/2011ApJS..197...36K}
}

@ARTICLE{2014ApJ...788...28V,
       author = {{van der Wel}, A. and {Franx}, M. and {van Dokkum}, P.~G. and {Skelton}, R.~E. and {Momcheva}, I.~G. and {Whitaker}, K.~E. and {Brammer}, G.~B. and {Bell}, E.~F. and {Rix}, H. -W. and {Wuyts}, S. and {Ferguson}, H.~C. and {Holden}, B.~P. and {Barro}, G. and {Koekemoer}, A.~M. and {Chang}, Yu-Yen and {McGrath}, E.~J. and {H{\"a}ussler}, B. and {Dekel}, A. and {Behroozi}, P. and {Fumagalli}, M. and {Leja}, J. and {Lundgren}, B.~F. and {Maseda}, M.~V. and {Nelson}, E.~J. and {Wake}, D.~A. and {Patel}, S.~G. and {Labb{\'e}}, I. and {Faber}, S.~M. and {Grogin}, N.~A. and {Kocevski}, D.~D.},
        title = "{3D-HST+CANDELS: The Evolution of the Galaxy Size-Mass Distribution since z = 3}",
      journal = {\apj},
         year = 2014,
        month = jun,
       volume = {788},
       number = {1},
          eid = {28},
        pages = {28},
          doi = {10.1088/0004-637X/788/1/28},
archivePrefix = {arXiv},
       eprint = {1404.2844},
 primaryClass = {astro-ph.GA},
       adsurl = {https://ui.adsabs.harvard.edu/abs/2014ApJ...788...28V}
}

@MISC{2021jwst.prop.1837D,
       author = {{Dunlop}, James S. and {Abraham}, Roberto G. and {Ashby}, Matthew L.~N. and {Bagley}, Micaela and {Best}, Philip N. and {Bongiorno}, Angela and {Bouwens}, Rychard and {Bowler}, Rebecca A.~A. and {Brammer}, Gabriel and {Bremer}, Malcolm and {Calabro'}, Antonello and {Carnall}, Adam and {Castellano}, Marco and {Cirasuolo}, Michele and {Conselice}, Christopher and {Cullen}, Fergus and {Dave}, Romeel and {Dayal}, Pratika and {Dekel}, Avishai and {Dickinson}, Mark and {Duncan}, Kenneth James and {Elbaz}, David and {Ellis}, Richard S. and {Ferguson}, Harry C. and {Ferrara}, Andrea and {Finkelstein}, Steven L. and {Fontana}, Adriano and {Furlanetto}, Steven and {Fynbo}, Johan P.~U. and {Gallerani}, Simona and {Gardner}, Jonathan P. and {Giavalisco}, Mauro and {Grazian}, Andrea and {Grogin}, Norman and {Harikane}, Yuichi and {Hopkins}, Philip F. and {Ilbert}, Olivier and {Illingworth}, Garth D. and {Juneau}, Stephanie and {Jung}, Intae and {Kartaltepe}, Jeyhan and {Kassin}, Susan and {Kauffmann}, Olivier Benjamin and {Khochfar}, Sadegh and {Kirkpatrick}, Allison and {Kocevski}, Dale D. and {Koekemoer}, Anton M. and {Labbe}, Ivo and {Laporte}, Nicolas and {Larson}, Rebecca L. and {Lucas}, Ray A. and {Magee}, Daniel K. and {Mason}, Charlotte and {McCracken}, Henry Joy and {McLeod}, Derek and {McLure}, Ross and {Merlin}, Emiliano and {Mesinger}, Andrei and {Milvang-Jensen}, Bo and {Newman}, Jeffrey Allen and {Oesch}, Pascal and {Ouchi}, Masami and {Pacifici}, Camilla and {Papovich}, Casey and {Peacock}, John and {Peeples}, Molly and {Pentericci}, Laura and {Perez-Gonzalez}, Pablo G. and {Pirzkal}, Norbert and {Pope}, Alexandra and {Pye}, John P. and {Reddy}, Naveen A. and {Robertson}, Brant and {Salvato}, Mara and {Santini}, Paola and {Schaerer}, Daniel and {Shapley}, Alice E. and {Simons}, Raymond and {Smit}, Renske and {Smith}, Britton D. and {Snyder}, Greg and {Somerville}, Rachel S. and {Stanway}, Elizabeth R. and {Stefanon}, Mauro and {Tasca}, Lidia and {Tikkanen}, Tuomo and {Tresse}, Laurence and {Trump}, Jonathan R. and {Whitaker}, Katherine E. and {Wilkins}, Stephen Matthew and {Wright}, Gillian and {Wyithe}, J. Stuart B. and {van Dokkum}, Pieter and {van der Werf}, Paul},
        title = "{PRIMER: Public Release IMaging for Extragalactic Research}",
 howpublished = {JWST Proposal. Cycle 1, ID. \#1837},
         year = 2021,
        month = mar,
        pages = {1837},
       adsurl = {https://ui.adsabs.harvard.edu/abs/2021jwst.prop.1837D}
}

@ARTICLE{2016ApJ...826..114T,
       author = {{Toshikawa}, Jun and {Kashikawa}, Nobunari and {Overzier}, Roderik and {Malkan}, Matthew A. and {Furusawa}, Hisanori and {Ishikawa}, Shogo and {Onoue}, Masafusa and {Ota}, Kazuaki and {Tanaka}, Masayuki and {Niino}, Yuu and {Uchiyama}, Hisakazu},
        title = "{A Systematic Survey of Protoclusters at z \raisebox{-0.5ex}\textasciitilde 3-6 in the CFHTLS Deep Fields}",
      journal = {\apj},
         year = 2016,
        month = aug,
       volume = {826},
       number = {2},
          eid = {114},
        pages = {114},
          doi = {10.3847/0004-637X/826/2/114},
archivePrefix = {arXiv},
       eprint = {1605.01439},
 primaryClass = {astro-ph.GA},
       adsurl = {https://ui.adsabs.harvard.edu/abs/2016ApJ...826..114T}
}

@ARTICLE{2008A&A...478..299M,
       author = {{Meneux}, B. and {Guzzo}, L. and {Garilli}, B. and {Le F{\`e}vre}, O. and {Pollo}, A. and {Blaizot}, J. and {De Lucia}, G. and {Bolzonella}, M. and {Lamareille}, F. and {Pozzetti}, L. and {Cappi}, A. and {Iovino}, A. and {Marinoni}, C. and {McCracken}, H.~J. and {de la Torre}, S. and {Bottini}, D. and {Le Brun}, V. and {Maccagni}, D. and {Picat}, J.~P. and {Scaramella}, R. and {Scodeggio}, M. and {Tresse}, L. and {Vettolani}, G. and {Zanichelli}, A. and {Abbas}, U. and {Adami}, C. and {Arnouts}, S. and {Bardelli}, S. and {Bongiorno}, A. and {Charlot}, S. and {Ciliegi}, P. and {Contini}, T. and {Cucciati}, O. and {Foucaud}, S. and {Franzetti}, P. and {Gavignaud}, I. and {Ilbert}, O. and {Marano}, B. and {Mazure}, A. and {Merighi}, R. and {Paltani}, S. and {Pell{\`o}}, R. and {Radovich}, M. and {Vergani}, D. and {Zamorani}, G. and {Zucca}, E.},
        title = "{The VIMOS-VLT Deep Survey (VVDS). The dependence of clustering on galaxy stellar mass at z \raisebox{-0.5ex}\textasciitilde 1}",
      journal = {\aap},
         year = 2008,
        month = feb,
       volume = {478},
       number = {2},
        pages = {299-310},
          doi = {10.1051/0004-6361:20078182},
archivePrefix = {arXiv},
       eprint = {0706.4371},
 primaryClass = {astro-ph},
       adsurl = {https://ui.adsabs.harvard.edu/abs/2008A&A...478..299M}
}

@ARTICLE{2011ApJ...728...46W,
       author = {{Wake}, David A. and {Whitaker}, Katherine E. and {Labb{\'e}}, Ivo and {van Dokkum}, Pieter G. and {Franx}, Marijn and {Quadri}, Ryan and {Brammer}, Gabriel and {Kriek}, Mariska and {Lundgren}, Britt F. and {Marchesini}, Danilo and {Muzzin}, Adam},
        title = "{Galaxy Clustering in the NEWFIRM Medium Band Survey: The Relationship Between Stellar Mass and Dark Matter Halo Mass at 1 < z < 2}",
      journal = {\apj},
         year = 2011,
        month = feb,
       volume = {728},
       number = {1},
          eid = {46},
        pages = {46},
          doi = {10.1088/0004-637X/728/1/46},
archivePrefix = {arXiv},
       eprint = {1012.1317},
 primaryClass = {astro-ph.CO},
       adsurl = {https://ui.adsabs.harvard.edu/abs/2011ApJ...728...46W}
}

@ARTICLE{2014A&A...568A..24B,
       author = {{Bielby}, R.~M. and {Gonzalez-Perez}, V. and {McCracken}, H.~J. and {Ilbert}, O. and {Daddi}, E. and {Le F{\`e}vre}, O. and {Hudelot}, P. and {Kneib}, J. -P. and {Mellier}, Y. and {Willott}, C.},
        title = "{The WIRCam Deep Survey. II. Mass selected clustering}",
      journal = {\aap},
         year = 2014,
        month = aug,
       volume = {568},
          eid = {A24},
        pages = {A24},
          doi = {10.1051/0004-6361/201322814},
archivePrefix = {arXiv},
       eprint = {1310.2172},
 primaryClass = {astro-ph.CO},
       adsurl = {https://ui.adsabs.harvard.edu/abs/2014A&A...568A..24B}
}

@ARTICLE{2018MNRAS.481.4885Q,
       author = {{Qiu}, Yisheng and {Wyithe}, J. Stuart B. and {Oesch}, Pascal A. and {Mutch}, Simon J. and {Qin}, Yuxiang and {Labb{\'e}}, Ivo and {Bouwens}, Rychard J. and {Stefanon}, Mauro and {Illingworth}, Garth D.},
        title = "{Dependence of galaxy clustering on UV luminosity and stellar mass at z {\ensuremath{\sim}} 4-7}",
      journal = {\mnras},
         year = 2018,
        month = dec,
       volume = {481},
       number = {4},
        pages = {4885-4894},
          doi = {10.1093/mnras/sty2633},
archivePrefix = {arXiv},
       eprint = {1809.10161},
 primaryClass = {astro-ph.GA},
       adsurl = {https://ui.adsabs.harvard.edu/abs/2018MNRAS.481.4885Q}
}

@ARTICLE{2018A&A...612A..42D,
       author = {{Durkalec}, A. and {Le F{\`e}vre}, O. and {Pollo}, A. and {Zamorani}, G. and {Lemaux}, B.~C. and {Garilli}, B. and {Bardelli}, S. and {Hathi}, N. and {Koekemoer}, A. and {Pforr}, J. and {Zucca}, E.},
        title = "{The VIMOS Ultra Deep Survey. Luminosity and stellar mass dependence of galaxy clustering at z   3}",
      journal = {\aap},
         year = 2018,
        month = apr,
       volume = {612},
          eid = {A42},
        pages = {A42},
          doi = {10.1051/0004-6361/201730734},
archivePrefix = {arXiv},
       eprint = {1703.02049},
 primaryClass = {astro-ph.GA},
       adsurl = {https://ui.adsabs.harvard.edu/abs/2018A&A...612A..42D}
}

@ARTICLE{2018PASJ...70S..11H,
       author = {{Harikane}, Yuichi and {Ouchi}, Masami and {Ono}, Yoshiaki and {Saito}, Shun and {Behroozi}, Peter and {More}, Surhud and {Shimasaku}, Kazuhiro and {Toshikawa}, Jun and {Lin}, Yen-Ting and {Akiyama}, Masayuki and {Coupon}, Jean and {Komiyama}, Yutaka and {Konno}, Akira and {Lin}, Sheng-Chieh and {Miyazaki}, Satoshi and {Nishizawa}, Atsushi J. and {Shibuya}, Takatoshi and {Silverman}, John},
        title = "{GOLDRUSH. II. Clustering of galaxies at z {\ensuremath{\sim}} 4-6 revealed with the half-million dropouts over the 100 deg$^{2}$ area corresponding to 1 Gpc$^{3}$}",
      journal = {\pasj},
         year = 2018,
        month = jan,
       volume = {70},
          eid = {S11},
        pages = {S11},
          doi = {10.1093/pasj/psx097},
archivePrefix = {arXiv},
       eprint = {1704.06535},
 primaryClass = {astro-ph.GA},
       adsurl = {https://ui.adsabs.harvard.edu/abs/2018PASJ...70S..11H}
}

@ARTICLE{2010A&A...523A..13P,
       author = {{Pozzetti}, L. and {Bolzonella}, M. and {Zucca}, E. and {Zamorani}, G. and {Lilly}, S. and {Renzini}, A. and {Moresco}, M. and {Mignoli}, M. and {Cassata}, P. and {Tasca}, L. and {Lamareille}, F. and {Maier}, C. and {Meneux}, B. and {Halliday}, C. and {Oesch}, P. and {Vergani}, D. and {Caputi}, K. and {Kova{\v{c}}}, K. and {Cimatti}, A. and {Cucciati}, O. and {Iovino}, A. and {Peng}, Y. and {Carollo}, M. and {Contini}, T. and {Kneib}, J. -P. and {Le F{\'e}vre}, O. and {Mainieri}, V. and {Scodeggio}, M. and {Bardelli}, S. and {Bongiorno}, A. and {Coppa}, G. and {de la Torre}, S. and {de Ravel}, L. and {Franzetti}, P. and {Garilli}, B. and {Kampczyk}, P. and {Knobel}, C. and {Le Borgne}, J. -F. and {Le Brun}, V. and {Pell{\`o}}, R. and {Perez Montero}, E. and {Ricciardelli}, E. and {Silverman}, J.~D. and {Tanaka}, M. and {Tresse}, L. and {Abbas}, U. and {Bottini}, D. and {Cappi}, A. and {Guzzo}, L. and {Koekemoer}, A.~M. and {Leauthaud}, A. and {Maccagni}, D. and {Marinoni}, C. and {McCracken}, H.~J. and {Memeo}, P. and {Porciani}, C. and {Scaramella}, R. and {Scarlata}, C. and {Scoville}, N.},
        title = "{zCOSMOS - 10k-bright spectroscopic sample. The bimodality in the galaxy stellar mass function: exploring its evolution with redshift}",
      journal = {\aap},
         year = 2010,
        month = nov,
       volume = {523},
          eid = {A13},
        pages = {A13},
          doi = {10.1051/0004-6361/200913020},
archivePrefix = {arXiv},
       eprint = {0907.5416},
 primaryClass = {astro-ph.CO},
       adsurl = {https://ui.adsabs.harvard.edu/abs/2010A&A...523A..13P}
}

@ARTICLE{2006MNRAS.368....2D,
       author = {{Dekel}, Avishai and {Birnboim}, Yuval},
        title = "{Galaxy bimodality due to cold flows and shock heating}",
      journal = {\mnras},
         year = 2006,
        month = may,
       volume = {368},
       number = {1},
        pages = {2-20},
          doi = {10.1111/j.1365-2966.2006.10145.x},
archivePrefix = {arXiv},
       eprint = {astro-ph/0412300},
 primaryClass = {astro-ph},
       adsurl = {https://ui.adsabs.harvard.edu/abs/2006MNRAS.368....2D}
}

@ARTICLE{2003PASP..115..763C,
       author = {{Chabrier}, Gilles},
        title = "{Galactic Stellar and Substellar Initial Mass Function}",
      journal = {\pasp},
         year = 2003,
        month = jul,
       volume = {115},
       number = {809},
        pages = {763-795},
          doi = {10.1086/376392},
archivePrefix = {arXiv},
       eprint = {astro-ph/0304382},
 primaryClass = {astro-ph},
       adsurl = {https://ui.adsabs.harvard.edu/abs/2003PASP..115..763C}
}

@ARTICLE{2003MNRAS.344.1000B,
       author = {{Bruzual}, G. and {Charlot}, S.},
        title = "{Stellar population synthesis at the resolution of 2003}",
      journal = {\mnras},
         year = 2003,
        month = oct,
       volume = {344},
       number = {4},
        pages = {1000-1028},
          doi = {10.1046/j.1365-8711.2003.06897.x},
archivePrefix = {arXiv},
       eprint = {astro-ph/0309134},
 primaryClass = {astro-ph},
       adsurl = {https://ui.adsabs.harvard.edu/abs/2003MNRAS.344.1000B}
}

@ARTICLE{2000ApJ...533..682C,
       author = {{Calzetti}, Daniela and {Armus}, Lee and {Bohlin}, Ralph C. and {Kinney}, Anne L. and {Koornneef}, Jan and {Storchi-Bergmann}, Thaisa},
        title = "{The Dust Content and Opacity of Actively Star-forming Galaxies}",
      journal = {\apj},
         year = 2000,
        month = apr,
       volume = {533},
       number = {2},
        pages = {682-695},
          doi = {10.1086/308692},
archivePrefix = {arXiv},
       eprint = {astro-ph/9911459},
 primaryClass = {astro-ph},
       adsurl = {https://ui.adsabs.harvard.edu/abs/2000ApJ...533..682C}
}

@ARTICLE{2024ApJ...961...39S,
       author = {{Shi}, Ke and {Malavasi}, Nicola and {Toshikawa}, Jun and {Zheng}, Xianzhong},
        title = "{Nature versus Nurture: Revisiting the Environmental Impact on Star Formation Activities of Galaxies}",
      journal = {\apj},
         year = 2024,
        month = jan,
       volume = {961},
       number = {1},
          eid = {39},
        pages = {39},
          doi = {10.3847/1538-4357/ad11d7},
archivePrefix = {arXiv},
       eprint = {2311.18427},
 primaryClass = {astro-ph.GA},
       adsurl = {https://ui.adsabs.harvard.edu/abs/2024ApJ...961...39S}
}

@ARTICLE{2019A&A...626A...7T,
       author = {{Tarr{\'\i}o}, P. and {Melin}, J.-B. and {Arnaud}, M.},
        title = "{ComPRASS: a Combined Planck-RASS catalogue of X-ray-SZ clusters}",
      journal = {\aap},
         year = 2019,
        month = jun,
       volume = {626},
          eid = {A7},
        pages = {A7},
          doi = {10.1051/0004-6361/201834979},
archivePrefix = {arXiv},
       eprint = {1901.00873},
 primaryClass = {astro-ph.CO},
       adsurl = {https://ui.adsabs.harvard.edu/abs/2019A&A...626A...7T}
}

@ARTICLE{2019ApJ...878...55B,
       author = {{Bocquet}, S. and {Dietrich}, J.~P. and {Schrabback}, T. and {Bleem}, L.~E. and {Klein}, M. and {Allen}, S.~W. and {Applegate}, D.~E. and {Ashby}, M.~L.~N. and {Bautz}, M. and {Bayliss}, M. and {Benson}, B.~A. and {Brodwin}, M. and {Bulbul}, E. and {Canning}, R.~E.~A. and {Capasso}, R. and {Carlstrom}, J.~E. and {Chang}, C.~L. and {Chiu}, I. and {Cho}, H.-M. and {Clocchiatti}, A. and {Crawford}, T.~M. and {Crites}, A.~T. and {de Haan}, T. and {Desai}, S. and {Dobbs}, M.~A. and {Foley}, R.~J. and {Forman}, W.~R. and {Garmire}, G.~P. and {George}, E.~M. and {Gladders}, M.~D. and {Gonzalez}, A.~H. and {Grandis}, S. and {Gupta}, N. and {Halverson}, N.~W. and {Hlavacek-Larrondo}, J. and {Hoekstra}, H. and {Holder}, G.~P. and {Holzapfel}, W.~L. and {Hou}, Z. and {Hrubes}, J.~D. and {Huang}, N. and {Jones}, C. and {Khullar}, G. and {Knox}, L. and {Kraft}, R. and {Lee}, A.~T. and {von der Linden}, A. and {Luong-Van}, D. and {Mantz}, A. and {Marrone}, D.~P. and {McDonald}, M. and {McMahon}, J.~J. and {Meyer}, S.~S. and {Mocanu}, L.~M. and {Mohr}, J.~J. and {Morris}, R.~G. and {Padin}, S. and {Patil}, S. and {Pryke}, C. and {Rapetti}, D. and {Reichardt}, C.~L. and {Rest}, A. and {Ruhl}, J.~E. and {Saliwanchik}, B.~R. and {Saro}, A. and {Sayre}, J.~T. and {Schaffer}, K.~K. and {Shirokoff}, E. and {Stalder}, B. and {Stanford}, S.~A. and {Staniszewski}, Z. and {Stark}, A.~A. and {Story}, K.~T. and {Strazzullo}, V. and {Stubbs}, C.~W. and {Vanderlinde}, K. and {Vieira}, J.~D. and {Vikhlinin}, A. and {Williamson}, R. and {Zenteno}, A.},
        title = "{Cluster Cosmology Constraints from the 2500 deg$^{2}$ SPT-SZ Survey: Inclusion of Weak Gravitational Lensing Data from Magellan and the Hubble Space Telescope}",
      journal = {\apj},
         year = 2019,
        month = jun,
       volume = {878},
       number = {1},
          eid = {55},
        pages = {55},
          doi = {10.3847/1538-4357/ab1f10},
archivePrefix = {arXiv},
       eprint = {1812.01679},
 primaryClass = {astro-ph.CO},
       adsurl = {https://ui.adsabs.harvard.edu/abs/2019ApJ...878...55B}
}

@ARTICLE{2025arXiv250721459A,
       author = {{ACTDESHSC Collaboration} and {Aguena}, M. and {Aiola}, S. and {Allam}, S. and {Andrade-Oliveira}, F. and {Bacon}, D. and {Bahcall}, N. and {Battaglia}, N. and {Battistelli}, E.~S. and {Bocquet}, S. and {Bolliet}, B. and {Bond}, J.~R. and {Brooks}, D. and {Calabrese}, E. and {Carretero}, J. and {Choi}, S.~K. and {da Costa}, L.~N. and {Costanzi}, M. and {Coulton}, W. and {Davis}, T.~M. and {Desai}, S. and {Devlin}, M.~J. and {Dicker}, S. and {Doel}, P. and {Duivenvoorden}, A.~J. and {Dunkley}, J. and {Ferraro}, S. and {Flaugher}, B. and {Frieman}, J. and {Gallardo}, P.~A. and {Gatti}, M. and {Gaztanaga}, E. and {Gill}, A.~S. and {Golec}, J.~E. and {Gruen}, D. and {Gruendl}, R.~A. and {Halpern}, M. and {Hasselfield}, M. and {Hill}, J.~C. and {Hilton}, M. and {Hincks}, A.~D. and {Hinton}, S.~R. and {Hollowood}, D.~L. and {Honscheid}, K. and {Hubmayr}, J. and {Huffenberger}, K.~M. and {Hughes}, J.~P. and {James}, D.~J. and {Klein}, M. and {Knowles}, K. and {Koopman}, B.~J. and {Kosowsky}, A. and {Lahav}, O. and {Lee}, E. and {Lin}, Y. and {Lokken}, M. and {Madhavacheril}, M.~S. and {Plazas Malag{\'o}n}, A.~A. and {Marrewijk}, J. v. and {Marshall}, J.~L. and {McMahon}, J. and {Mena-Fern{\'a}ndez}, J. and {Miquel}, R. and {Miyatake}, H. and {Mohr}, J.~J. and {Moodley}, K. and {Mroczkowski}, T. and {Naess}, S. and {Nati}, F. and {Nicola}, A. and {Niemack}, M.~D. and {Ogando}, R.~L.~C. and {Oguri}, M. and {Orlowski-Scherer}, J. and {Page}, L.~A. and {Partridge}, B. and {da Silva Pereira}, M.~E. and {Porredon}, A. and {Qu}, F.~J. and {Ragavan}, D.~C. and {Ried Guachalla}, B. and {Romer}, A.~K. and {Carnero Rosell}, A. and {Rykoff}, E.~S. and {Samuroff}, S. and {Sanchez}, E. and {Sevilla-Noarbe}, I. and {Sierra}, C. and {Sif{\'o}n}, C. and {Smith}, M. and {Staggs}, S.~T. and {Suchyta}, E. and {Swanson}, M.~E.~C. and {Tucker}, D.~L. and {Vargas}, C. and {Vavagiakis}, E.~M. and {De Vicente}, J. and {Weaverdyck}, N. and {Weller}, J. and {Wollack}, E.~J. and {Zubeldia}, I.},
        title = "{The Atacama Cosmology Telescope: DR6 Sunyaev-Zel'dovich Selected Galaxy Clusters Catalog}",
      journal = {arXiv e-prints},
         year = 2025,
        month = jul,
          eid = {arXiv:2507.21459},
        pages = {arXiv:2507.21459},
          doi = {10.48550/arXiv.2507.21459},
archivePrefix = {arXiv},
       eprint = {2507.21459},
 primaryClass = {astro-ph.CO},
       adsurl = {https://ui.adsabs.harvard.edu/abs/2025arXiv250721459A}
}

@ARTICLE{2021MNRAS.503L...1K,
       author = {{Koyama}, Yusei and {Polletta}, Maria del Carmen and {Tanaka}, Ichi and {Kodama}, Tadayuki and {Dole}, Herv{\'e} and {Soucail}, Genevi{\`e}ve and {Frye}, Brenda and {Lehnert}, Matthew and {Scodeggio}, Marco},
        title = "{A Planck-selected dusty proto-cluster at z = 2.16 associated with a strong overdensity of massive H{\ensuremath{\alpha}}-emitting galaxies}",
      journal = {\mnras},
         year = 2021,
        month = may,
       volume = {503},
       number = {1},
        pages = {L1-L5},
          doi = {10.1093/mnrasl/slab013},
archivePrefix = {arXiv},
       eprint = {2008.13614},
 primaryClass = {astro-ph.GA},
       adsurl = {https://ui.adsabs.harvard.edu/abs/2021MNRAS.503L...1K}
}

@ARTICLE{2021A&A...654A.121P,
       author = {{Polletta}, M. and {Soucail}, G. and {Dole}, H. and {Lehnert}, M.~D. and {Pointecouteau}, E. and {Vietri}, G. and {Scodeggio}, M. and {Montier}, L. and {Koyama}, Y. and {Lagache}, G. and {Frye}, B.~L. and {Cusano}, F. and {Fumana}, M.},
        title = "{Spectroscopic observations of PHz G237.01+42.50: A galaxy protocluster at z = 2.16 in the Cosmos field}",
      journal = {\aap},
         year = 2021,
        month = oct,
       volume = {654},
          eid = {A121},
        pages = {A121},
          doi = {10.1051/0004-6361/202140612},
archivePrefix = {arXiv},
       eprint = {2109.04396},
 primaryClass = {astro-ph.GA},
       adsurl = {https://ui.adsabs.harvard.edu/abs/2021A&A...654A.121P}
}

@ARTICLE{2023Natur.615..809D,
       author = {{Di Mascolo}, Luca and {Saro}, Alexandro and {Mroczkowski}, Tony and {Borgani}, Stefano and {Churazov}, Eugene and {Rasia}, Elena and {Tozzi}, Paolo and {Dannerbauer}, Helmut and {Basu}, Kaustuv and {Carilli}, Christopher L. and {Ginolfi}, Michele and {Miley}, George and {Nonino}, Mario and {Pannella}, Maurilio and {Pentericci}, Laura and {Rizzo}, Francesca},
        title = "{Forming intracluster gas in a galaxy protocluster at a redshift of 2.16}",
      journal = {\nat},
         year = 2023,
        month = mar,
       volume = {615},
       number = {7954},
        pages = {809-812},
          doi = {10.1038/s41586-023-05761-x},
archivePrefix = {arXiv},
       eprint = {2303.16226},
 primaryClass = {astro-ph.CO},
       adsurl = {https://ui.adsabs.harvard.edu/abs/2023Natur.615..809D}
}

@ARTICLE{2015ApJ...808L..33C,
       author = {{Casey}, C.~M. and {Cooray}, A. and {Capak}, P. and {Fu}, H. and {Kovac}, K. and {Lilly}, S. and {Sanders}, D.~B. and {Scoville}, N.~Z. and {Treister}, E.},
        title = "{A Massive, Distant Proto-cluster at z = 2.47 Caught in a Phase of Rapid Formation?}",
      journal = {\apjl},
         year = 2015,
        month = aug,
       volume = {808},
       number = {2},
          eid = {L33},
        pages = {L33},
          doi = {10.1088/2041-8205/808/2/L33},
archivePrefix = {arXiv},
       eprint = {1506.01715},
 primaryClass = {astro-ph.GA},
       adsurl = {https://ui.adsabs.harvard.edu/abs/2015ApJ...808L..33C}
}

@ARTICLE{2016ApJ...828...56W,
       author = {{Wang}, Tao and {Elbaz}, David and {Daddi}, Emanuele and {Finoguenov}, Alexis and {Liu}, Daizhong and {Schreiber}, Corentin and {Mart{\'\i}n}, Sergio and {Strazzullo}, Veronica and {Valentino}, Francesco and {van der Burg}, Remco and {Zanella}, Anita and {Ciesla}, Laure and {Gobat}, Raphael and {Le Brun}, Amandine and {Pannella}, Maurilio and {Sargent}, Mark and {Shu}, Xinwen and {Tan}, Qinghua and {Cappelluti}, Nico and {Li}, Yanxia},
        title = "{Discovery of a Galaxy Cluster with a Violently Starbursting Core at z = 2.506}",
      journal = {\apj},
         year = 2016,
        month = sep,
       volume = {828},
       number = {1},
          eid = {56},
        pages = {56},
          doi = {10.3847/0004-637X/828/1/56},
archivePrefix = {arXiv},
       eprint = {1604.07404},
 primaryClass = {astro-ph.GA},
       adsurl = {https://ui.adsabs.harvard.edu/abs/2016ApJ...828...56W}
}

@ARTICLE{2007A&A...461..823V,
       author = {{Venemans}, B.~P. and {R{\"o}ttgering}, H.~J.~A. and {Miley}, G.~K. and {van Breugel}, W.~J.~M. and {de Breuck}, C. and {Kurk}, J.~D. and {Pentericci}, L. and {Stanford}, S.~A. and {Overzier}, R.~A. and {Croft}, S. and {Ford}, H.},
        title = "{Protoclusters associated with z > 2 radio galaxies . I. Characteristics of high redshift protoclusters}",
      journal = {\aap},
         year = 2007,
        month = jan,
       volume = {461},
       number = {3},
        pages = {823-845},
          doi = {10.1051/0004-6361:20053941},
archivePrefix = {arXiv},
       eprint = {astro-ph/0610567},
 primaryClass = {astro-ph},
       adsurl = {https://ui.adsabs.harvard.edu/abs/2007A&A...461..823V}
}

@ARTICLE{2014A&A...570A..16C,
       author = {{Cucciati}, O. and {Zamorani}, G. and {Lemaux}, B.~C. and {Bardelli}, S. and {Cimatti}, A. and {Le F{\`e}vre}, O. and {Cassata}, P. and {Garilli}, B. and {Le Brun}, V. and {Maccagni}, D. and {Pentericci}, L. and {Tasca}, L.~A.~M. and {Thomas}, R. and {Vanzella}, E. and {Zucca}, E. and {Amorin}, R. and {Capak}, P. and {Cassar{\`a}}, L.~P. and {Castellano}, M. and {Cuby}, J.~G. and {de la Torre}, S. and {Durkalec}, A. and {Fontana}, A. and {Giavalisco}, M. and {Grazian}, A. and {Hathi}, N.~P. and {Ilbert}, O. and {Moreau}, C. and {Paltani}, S. and {Ribeiro}, B. and {Salvato}, M. and {Schaerer}, D. and {Scodeggio}, M. and {Sommariva}, V. and {Talia}, M. and {Taniguchi}, Y. and {Tresse}, L. and {Vergani}, D. and {Wang}, P.~W. and {Charlot}, S. and {Contini}, T. and {Fotopoulou}, S. and {L{\'o}pez-Sanjuan}, C. and {Mellier}, Y. and {Scoville}, N.},
        title = "{Discovery of a rich proto-cluster at z = 2.9 and associated diffuse cold gas in the VIMOS Ultra-Deep Survey (VUDS)}",
      journal = {\aap},
         year = 2014,
        month = oct,
       volume = {570},
          eid = {A16},
        pages = {A16},
          doi = {10.1051/0004-6361/201423811},
archivePrefix = {arXiv},
       eprint = {1403.3691},
 primaryClass = {astro-ph.CO},
       adsurl = {https://ui.adsabs.harvard.edu/abs/2014A&A...570A..16C}
}

@ARTICLE{2018ApJ...856...72O,
       author = {{Oteo}, I. and {Ivison}, R.~J. and {Dunne}, L. and {Manilla-Robles}, A. and {Maddox}, S. and {Lewis}, A.~J.~R. and {de Zotti}, G. and {Bremer}, M. and {Clements}, D.~L. and {Cooray}, A. and {Dannerbauer}, H. and {Eales}, S. and {Greenslade}, J. and {Omont}, A. and {Perez{\textendash}Fourn{\'o}n}, I. and {Riechers}, D. and {Scott}, D. and {van der Werf}, P. and {Weiss}, A. and {Zhang}, Z.-Y.},
        title = "{An Extreme Protocluster of Luminous Dusty Starbursts in the Early Universe}",
      journal = {\apj},
         year = 2018,
        month = mar,
       volume = {856},
       number = {1},
          eid = {72},
        pages = {72},
          doi = {10.3847/1538-4357/aaa1f1},
archivePrefix = {arXiv},
       eprint = {1709.02809},
 primaryClass = {astro-ph.GA},
       adsurl = {https://ui.adsabs.harvard.edu/abs/2018ApJ...856...72O}
}

@ARTICLE{2018Natur.556..469M,
       author = {{Miller}, T.~B. and {Chapman}, S.~C. and {Aravena}, M. and {Ashby}, M.~L.~N. and {Hayward}, C.~C. and {Vieira}, J.~D. and {Wei{\ss}}, A. and {Babul}, A. and {B{\'e}thermin}, M. and {Bradford}, C.~M. and {Brodwin}, M. and {Carlstrom}, J.~E. and {Chen}, Chian-Chou and {Cunningham}, D.~J.~M. and {De Breuck}, C. and {Gonzalez}, A.~H. and {Greve}, T.~R. and {Harnett}, J. and {Hezaveh}, Y. and {Lacaille}, K. and {Litke}, K.~C. and {Ma}, J. and {Malkan}, M. and {Marrone}, D.~P. and {Morningstar}, W. and {Murphy}, E.~J. and {Narayanan}, D. and {Pass}, E. and {Perry}, R. and {Phadke}, K.~A. and {Rennehan}, D. and {Rotermund}, K.~M. and {Simpson}, J. and {Spilker}, J.~S. and {Sreevani}, J. and {Stark}, A.~A. and {Strandet}, M.~L. and {Strom}, A.~L.},
        title = "{A massive core for a cluster of galaxies at a redshift of 4.3}",
      journal = {\nat},
         year = 2018,
        month = apr,
       volume = {556},
       number = {7702},
        pages = {469-472},
          doi = {10.1038/s41586-018-0025-2},
archivePrefix = {arXiv},
       eprint = {1804.09231},
 primaryClass = {astro-ph.GA},
       adsurl = {https://ui.adsabs.harvard.edu/abs/2018Natur.556..469M}
}

@ARTICLE{2014ApJ...792...15T,
       author = {{Toshikawa}, Jun and {Kashikawa}, Nobunari and {Overzier}, Roderik and {Shibuya}, Takatoshi and {Ishikawa}, Shogo and {Ota}, Kazuaki and {Shimasaku}, Kazuhiro and {Tanaka}, Masayuki and {Hayashi}, Masao and {Niino}, Yuu and {Onoue}, Masafusa},
        title = "{A First Site of Galaxy Cluster Formation: Complete Spectroscopy of a Protocluster at z = 6.01}",
      journal = {\apj},
         year = 2014,
        month = sep,
       volume = {792},
       number = {1},
          eid = {15},
        pages = {15},
          doi = {10.1088/0004-637X/792/1/15},
archivePrefix = {arXiv},
       eprint = {1407.1851},
 primaryClass = {astro-ph.GA},
       adsurl = {https://ui.adsabs.harvard.edu/abs/2014ApJ...792...15T}
}

@ARTICLE{2023ApJ...953...53S,
       author = {{Sun}, Fengwu and {Egami}, Eiichi and {Pirzkal}, Nor and {Rieke}, Marcia and {Baum}, Stefi and {Boyer}, Martha and {Boyett}, Kristan and {Bunker}, Andrew J. and {Cameron}, Alex J. and {Curti}, Mirko and {Eisenstein}, Daniel J. and {Gennaro}, Mario and {Greene}, Thomas P. and {Jaffe}, Daniel and {Kelly}, Doug and {Koekemoer}, Anton M. and {Kumari}, Nimisha and {Maiolino}, Roberto and {Maseda}, Michael and {Perna}, Michele and {Rest}, Armin and {Robertson}, Brant E. and {Schlawin}, Everett and {Smit}, Renske and {Stansberry}, John and {Sunnquist}, Ben and {Tacchella}, Sandro and {Williams}, Christina C. and {Willmer}, Christopher N.~A.},
        title = "{First Sample of H{\ensuremath{\alpha}}+[O III]{\ensuremath{\lambda}}5007 Line Emitters at z > 6 Through JWST/NIRCam Slitless Spectroscopy: Physical Properties and Line-luminosity Functions}",
      journal = {\apj},
         year = 2023,
        month = aug,
       volume = {953},
       number = {1},
          eid = {53},
        pages = {53},
          doi = {10.3847/1538-4357/acd53c},
archivePrefix = {arXiv},
       eprint = {2209.03374},
 primaryClass = {astro-ph.GA},
       adsurl = {https://ui.adsabs.harvard.edu/abs/2023ApJ...953...53S}
}

@ARTICLE{2011Natur.470..233C,
       author = {{Capak}, Peter L. and {Riechers}, Dominik and {Scoville}, Nick Z. and {Carilli}, Chris and {Cox}, Pierre and {Neri}, Roberto and {Robertson}, Brant and {Salvato}, Mara and {Schinnerer}, Eva and {Yan}, Lin and {Wilson}, Grant W. and {Yun}, Min and {Civano}, Francesca and {Elvis}, Martin and {Karim}, Alexander and {Mobasher}, Bahram and {Staguhn}, Johannes G.},
        title = "{A massive protocluster of galaxies at a redshift of z\raisebox{-0.5ex}\textasciitilde5.3}",
      journal = {\nat},
         year = 2011,
        month = feb,
       volume = {470},
       number = {7333},
        pages = {233-235},
          doi = {10.1038/nature09681},
archivePrefix = {arXiv},
       eprint = {1101.3586},
 primaryClass = {astro-ph.CO},
       adsurl = {https://ui.adsabs.harvard.edu/abs/2011Natur.470..233C}
}

@ARTICLE{2024ApJ...962..124H,
       author = {{Helton}, Jakob M. and {Sun}, Fengwu and {Woodrum}, Charity and {Hainline}, Kevin N. and {Willmer}, Christopher N.~A. and {Rieke}, George H. and {Rieke}, Marcia J. and {Tacchella}, Sandro and {Robertson}, Brant and {Johnson}, Benjamin D. and {Alberts}, Stacey and {Eisenstein}, Daniel J. and {Hausen}, Ryan and {Bonaventura}, Nina R. and {Bunker}, Andrew and {Charlot}, Stephane and {Curti}, Mirko and {Curtis-Lake}, Emma and {Looser}, Tobias J. and {Maiolino}, Roberto and {Willott}, Chris and {Witstok}, Joris and {Boyett}, Kristan and {Chen}, Zuyi and {Egami}, Eiichi and {Endsley}, Ryan and {Hviding}, Raphael E. and {Jaffe}, Daniel T. and {Ji}, Zhiyuan and {Lyu}, Jianwei and {Sandles}, Lester},
        title = "{The JWST Advanced Deep Extragalactic Survey: Discovery of an Extreme Galaxy Overdensity at z = 5.4 with JWST/NIRCam in GOODS-S}",
      journal = {\apj},
         year = 2024,
        month = feb,
       volume = {962},
       number = {2},
          eid = {124},
        pages = {124},
          doi = {10.3847/1538-4357/ad0da7},
archivePrefix = {arXiv},
       eprint = {2302.10217},
 primaryClass = {astro-ph.GA},
       adsurl = {https://ui.adsabs.harvard.edu/abs/2024ApJ...962..124H}
}

@ARTICLE{2019ApJ...883..142H,
       author = {{Harikane}, Yuichi and {Ouchi}, Masami and {Ono}, Yoshiaki and {Fujimoto}, Seiji and {Donevski}, Darko and {Shibuya}, Takatoshi and {Faisst}, Andreas L. and {Goto}, Tomotsugu and {Hatsukade}, Bunyo and {Kashikawa}, Nobunari and {Kohno}, Kotaro and {Hashimoto}, Takuya and {Higuchi}, Ryo and {Inoue}, Akio K. and {Lin}, Yen-Ting and {Martin}, Crystal L. and {Overzier}, Roderik and {Smail}, Ian and {Toshikawa}, Jun and {Umehata}, Hideki and {Ao}, Yiping and {Chapman}, Scott and {Clements}, David L. and {Im}, Myungshin and {Jing}, Yipeng and {Kawaguchi}, Toshihiro and {Lee}, Chien-Hsiu and {Lee}, Minju M. and {Lin}, Lihwai and {Matsuoka}, Yoshiki and {Marinello}, Murilo and {Nagao}, Tohru and {Onodera}, Masato and {Toft}, Sune and {Wang}, Wei-Hao},
        title = "{SILVERRUSH. VIII. Spectroscopic Identifications of Early Large-scale Structures with Protoclusters over 200 Mpc at z {\ensuremath{\sim}} 6-7: Strong Associations of Dusty Star-forming Galaxies}",
      journal = {\apj},
         year = 2019,
        month = oct,
       volume = {883},
       number = {2},
          eid = {142},
        pages = {142},
          doi = {10.3847/1538-4357/ab2cd5},
archivePrefix = {arXiv},
       eprint = {1902.09555},
 primaryClass = {astro-ph.GA},
       adsurl = {https://ui.adsabs.harvard.edu/abs/2019ApJ...883..142H}
}

@ARTICLE{2019ApJ...877...51C,
       author = {{Chanchaiworawit}, Krittapas and {Guzm{\'a}n}, Rafael and {Salvador-Sol{\'e}}, Eduard and {Rodr{\'\i}guez Espinosa}, Jose Miguel and {Calvi}, Rosa and {Manrique}, Alberto and {Gallego}, Jesus and {Herrero}, Artemio and {Mar{\'\i}n-Franch}, Antonio and {Mas-Hesse}, Jose Miguel},
        title = "{Physical Properties of a Coma-analog Protocluster at z = 6.5}",
      journal = {\apj},
         year = 2019,
        month = may,
       volume = {877},
       number = {1},
          eid = {51},
        pages = {51},
          doi = {10.3847/1538-4357/ab1a34},
       adsurl = {https://ui.adsabs.harvard.edu/abs/2019ApJ...877...51C}
}

@ARTICLE{2024A&A...688A.146A,
       author = {{Arribas}, Santiago and {Perna}, Michele and {Rodr{\'\i}guez Del Pino}, Bruno and {Lamperti}, Isabella and {D'Eugenio}, Francesco and {P{\'e}rez-Gonz{\'a}lez}, Pablo G. and {Jones}, Gareth C. and {Crespo G{\'o}mez}, Alejandro and {Curti}, Mirko and {Lim}, Seunghwan and {{\'A}lvarez-M{\'a}rquez}, Javier and {Bunker}, Andrew J. and {Carniani}, Stefano and {Charlot}, St{\'e}phane and {Jakobsen}, Peter and {Maiolino}, Roberto and {{\"U}bler}, Hannah and {Willott}, Chris J. and {B{\"o}ker}, Torsten and {Chevallard}, Jacopo and {Circosta}, Chiara and {Cresci}, Giovanni and {Kumari}, Nimisha and {Parlanti}, Eleonora and {Scholtz}, Jan and {Venturi}, Giacomo and {Witstok}, Joris},
        title = "{GA-NIFS: The core of an extremely massive protocluster at the epoch of reionisation probed with JWST/NIRSpec}",
      journal = {\aap},
         year = 2024,
        month = aug,
       volume = {688},
          eid = {A146},
        pages = {A146},
          doi = {10.1051/0004-6361/202348824},
archivePrefix = {arXiv},
       eprint = {2312.00899},
 primaryClass = {astro-ph.GA},
       adsurl = {https://ui.adsabs.harvard.edu/abs/2024A&A...688A.146A}
}

@ARTICLE{2023ApJ...947L..24M,
       author = {{Morishita}, Takahiro and {Roberts-Borsani}, Guido and {Treu}, Tommaso and {Brammer}, Gabriel and {Mason}, Charlotte A. and {Trenti}, Michele and {Vulcani}, Benedetta and {Wang}, Xin and {Acebron}, Ana and {Bah{\'e}}, Yannick and {Bergamini}, Pietro and {Boyett}, Kristan and {Bradac}, Marusa and {Calabr{\`o}}, Antonello and {Castellano}, Marco and {Chen}, Wenlei and {De Lucia}, Gabriella and {Filippenko}, Alexei V. and {Fontana}, Adriano and {Glazebrook}, Karl and {Grillo}, Claudio and {Henry}, Alaina and {Jones}, Tucker and {Kelly}, Patrick L. and {Koekemoer}, Anton M. and {Leethochawalit}, Nicha and {Lu}, Ting-Yi and {Marchesini}, Danilo and {Mascia}, Sara and {Mercurio}, Amata and {Merlin}, Emiliano and {Metha}, Benjamin and {Nanayakkara}, Themiya and {Nonino}, Mario and {Paris}, Diego and {Pentericci}, Laura and {Rosati}, Piero and {Santini}, Paola and {Strait}, Victoria and {Vanzella}, Eros and {Windhorst}, Rogier A. and {Xie}, Lizhi},
        title = "{Early Results from GLASS-JWST. XIV. A Spectroscopically Confirmed Protocluster 650 Million Years after the Big Bang}",
      journal = {\apjl},
         year = 2023,
        month = apr,
       volume = {947},
       number = {2},
          eid = {L24},
        pages = {L24},
          doi = {10.3847/2041-8213/acb99e},
archivePrefix = {arXiv},
       eprint = {2211.09097},
 primaryClass = {astro-ph.GA},
       adsurl = {https://ui.adsabs.harvard.edu/abs/2023ApJ...947L..24M}
}

@ARTICLE{2025ApJ...982..153M,
       author = {{Morishita}, Takahiro and {Liu}, Zhaoran and {Stiavelli}, Massimo and {Treu}, Tommaso and {Trenti}, Michele and {Chartab}, Nima and {Roberts-Borsani}, Guido and {Vulcani}, Benedetta and {Bergamini}, Pietro and {Castellano}, Marco and {Grillo}, Claudio},
        title = "{Accelerated Emergence of Evolved Galaxies in Early Overdensities at z {\ensuremath{\sim}} 5.7}",
      journal = {\apj},
         year = 2025,
        month = apr,
       volume = {982},
       number = {2},
          eid = {153},
        pages = {153},
          doi = {10.3847/1538-4357/adb30f},
archivePrefix = {arXiv},
       eprint = {2408.10980},
 primaryClass = {astro-ph.GA},
       adsurl = {https://ui.adsabs.harvard.edu/abs/2025ApJ...982..153M}
}

@ARTICLE{2022ApJ...933....9L,
       author = {{Li}, Qingyang and {Yang}, Xiaohu and {Liu}, Chengze and {Jing}, Yipeng and {He}, Min and {Huang}, Jia-Sheng and {Dai}, Y. Sophia and {Sawicki}, Marcin and {Arnouts}, Stephane and {Gwyn}, Stephen and {Moutard}, Thibaud and {Mo}, H.~J. and {Wang}, Kai and {Katsianis}, Antonios and {Cui}, Weiguang and {Han}, Jiaxin and {Chiu}, I.-Non and {Gu}, Yizhou and {Xu}, Haojie},
        title = "{Groups and Protocluster Candidates in the CLAUDS and HSC-SSP Joint Deep Surveys}",
      journal = {\apj},
         year = 2022,
        month = jul,
       volume = {933},
       number = {1},
          eid = {9},
        pages = {9},
          doi = {10.3847/1538-4357/ac6e69},
archivePrefix = {arXiv},
       eprint = {2205.05517},
 primaryClass = {astro-ph.CO},
       adsurl = {https://ui.adsabs.harvard.edu/abs/2022ApJ...933....9L}
}

@ARTICLE{2007ApJ...671..153Y,
       author = {{Yang}, Xiaohu and {Mo}, H.~J. and {van den Bosch}, Frank C. and {Pasquali}, Anna and {Li}, Cheng and {Barden}, Marco},
        title = "{Galaxy Groups in the SDSS DR4. I. The Catalog and Basic Properties}",
      journal = {\apj},
         year = 2007,
        month = dec,
       volume = {671},
       number = {1},
        pages = {153-170},
          doi = {10.1086/522027},
archivePrefix = {arXiv},
       eprint = {0707.4640},
 primaryClass = {astro-ph},
       adsurl = {https://ui.adsabs.harvard.edu/abs/2007ApJ...671..153Y}
}

@article{daddi2022bending,
  title={The bending of the star-forming main sequence traces the cold-to hot-accretion transition mass over 0< z< 4},
  author={Daddi, Emanuele and Delvecchio, Ivan and Dimauro, Paola and Magnelli, Benjamin and Gomez-Guijarro, Carlos and Coogan, Rosemary and Elbaz, David and Kalita, Boris S and Le Bail, Aurelien and Rich, R Michael and others},
  journal={Astronomy \& Astrophysics},
  volume={661},
  pages={L7},
  year={2022},
  publisher={EDP Sciences}
}

@article{mandelker2020instability,
  title={Instability of supersonic cold streams feeding galaxies--IV. Survival of radiatively cooling streams},
  author={Mandelker, Nir and Nagai, Daisuke and Aung, Han and Dekel, Avishai and Birnboim, Yuval and van den Bosch, Frank C},
  journal={Monthly Notices of the Royal Astronomical Society},
  volume={494},
  number={2},
  pages={2641--2663},
  year={2020},
  publisher={Oxford University Press}
}

@ARTICLE{2016A&ARv..24...14O,
       author = {{Overzier}, Roderik A.},
        title = "{The realm of the galaxy protoclusters. A review}",
      journal = {\aapr},
         year = 2016,
        month = nov,
       volume = {24},
       number = {1},
          eid = {14},
        pages = {14},
          doi = {10.1007/s00159-016-0100-3},
archivePrefix = {arXiv},
       eprint = {1610.05201},
 primaryClass = {astro-ph.GA},
       adsurl = {https://ui.adsabs.harvard.edu/abs/2016A&ARv..24...14O}
}

@article{wilson1927probable,
  title={Probable inference, the law of succession, and statistical inference},
  author={Wilson, Edwin B},
  journal={Journal of the American Statistical Association},
  volume={22},
  number={158},
  pages={209--212},
  year={1927},
  publisher={Taylor \& Francis}
}

@article{kendall1938new,
  title={A new measure of rank correlation},
  author={Kendall, Maurice G},
  journal={Biometrika},
  volume={30},
  number={1-2},
  pages={81--93},
  year={1938},
  publisher={Oxford University Press}
}

@article{curran2014monte,
  title={Monte Carlo error analyses of Spearman's rank test},
  author={Curran, Peter A},
  journal={arXiv preprint arXiv:1411.3816},
  year={2014}
}

@article{weaver2023cosmos2020,
  title={COSMOS2020: The galaxy stellar mass function-The assembly and star formation cessation of galaxies at 0.2< z≤ 7.5},
  author={Weaver, JR and Davidzon, I and Toft, S and Ilbert, O and McCracken, HJ and Gould, KML and Jespersen, CK and Steinhardt, C and Lagos, CDP and Capak, PL and others},
  journal={Astronomy \& Astrophysics},
  volume={677},
  pages={A184},
  year={2023},
  publisher={EDP Sciences}
}

@article{muzzin2013evolution,
  title={The evolution of the stellar mass functions of star-forming and quiescent galaxies to z= 4 from the COSMOS/UltraVISTA survey},
  author={Muzzin, Adam and Marchesini, Danilo and Stefanon, Mauro and Franx, Marijn and McCracken, Henry J and Milvang-Jensen, Bo and Dunlop, James S and Fynbo, JPU and Brammer, Gabriel and Labb{\'e}, Ivo and others},
  journal={The Astrophysical Journal},
  volume={777},
  number={1},
  pages={18},
  year={2013},
  publisher={IOP Publishing}
}

@article{ilbert2013mass,
  title={Mass assembly in quiescent and star-forming galaxies since z≃ 4 from UltraVISTA},
  author={Ilbert, Olivier and McCracken, Henry J and Le F{\`e}vre, Olivier and Capak, Peter and Dunlop, James and Karim, Alexander and Renzini, MA and Caputi, Karina and Boissier, Samuel and Arnouts, St{\'e}phane and others},
  journal={Astronomy \& Astrophysics},
  volume={556},
  pages={A55},
  year={2013},
  publisher={EDP Sciences}
}

@article{edward2024stellar,
  title={The stellar mass function of quiescent galaxies in 2< z< 2.5 protoclusters},
  author={Edward, Adit H and Balogh, Michael L and Bahe, Yannick M and Cooper, Michael C and Hatch, Nina A and Marchioni, Justin and Muzzin, Adam and Noble, Allison and Rudnick, Gregory H and Vulcani, Benedetta and others},
  journal={Monthly Notices of the Royal Astronomical Society},
  volume={527},
  number={3},
  pages={8598--8617},
  year={2024},
  publisher={Oxford University Press}
}

@article{daddi2022evidence,
  title={Evidence for Cold-stream to Hot-accretion Transition as Traced by Ly$\alpha$ Emission from Groups and Clusters at 2< z< 3.3},
  author={Daddi, E and Rich, R Michael and Valentino, F and Jin, S and Delvecchio, I and Liu, D and Strazzullo, V and Neill, J and Gobat, R and Finoguenov, A and others},
  journal={The Astrophysical Journal Letters},
  volume={926},
  number={2},
  pages={L21},
  year={2022},
  publisher={IOP Publishing}
}

@article{van2017environmental,
  title={The environmental dependence of gas accretion on to galaxies: quenching satellites through starvation},
  author={van de Voort, Freeke and Bah{\'e}, Yannick M and Bower, Richard G and Correa, Camila A and Crain, Robert A and Schaye, Joop and Theuns, Tom},
  journal={Monthly Notices of the Royal Astronomical Society},
  volume={466},
  number={3},
  pages={3460--3471},
  year={2017},
  publisher={Oxford University Press}
}

@article{toni2025cosmos,
  title={COSMOS-Web galaxy groups: Evolution of red sequence and quiescent galaxy fraction},
  author={Toni, Greta and Maturi, Matteo and Castignani, Gianluca and Moscardini, Lauro and Gozaliasl, Ghassem and Finoguenov, Alexis and Taamoli, Sina and Akins, B Hollis and Arango-Toro, C Rafael and Casey, M Caitlin and others},
  journal={arXiv preprint arXiv:2509.08040},
  year={2025}
}

@article{shi2024nature,
  title={Nature versus Nurture: Revisiting the Environmental Impact on Star Formation Activities of Galaxies},
  author={Shi, Ke and Malavasi, Nicola and Toshikawa, Jun and Zheng, Xianzhong},
  journal={The Astrophysical Journal},
  volume={961},
  number={1},
  pages={39},
  year={2024},
  publisher={IOP Publishing}
}

@article{hatamnia2025large,
  title={Large-Scale Structure in COSMOS-Web: Tracing Galaxy Evolution in the Cosmic Web up to $ z$\backslash$sim 7$ with the Largest JWST Survey},
  author={Hatamnia, Hossein and Mobasher, Bahram and Taamoli, Sina and Kartaltepe, Jeyhan S and Casey, Caitlin M and Akins, Hollis B and Brinch, Malte and Chartab, Nima and Drakos, Nicole E and Faisst, Andreas L and others},
  journal={arXiv preprint arXiv:2511.10727},
  year={2025}
}

@article{darvish2016effects,
  title={The effects of the local environment and stellar mass on galaxy quenching to z~ 3},
  author={Darvish, Behnam and Mobasher, Bahram and Sobral, David and Rettura, Alessandro and Scoville, Nick and Faisst, Andreas and Capak, Peter},
  journal={The Astrophysical Journal},
  volume={825},
  number={2},
  pages={113},
  year={2016},
  publisher={IOP Publishing}
}

@article{lemaux2022vimos,
  title={The VIMOS Ultra Deep Survey: The reversal of the star-formation rate- density relation at 2< z< 5},
  author={Lemaux, BC and Cucciati, O and Le F{\`e}vre, O and Zamorani, G and Lubin, LM and Hathi, N and Ilbert, O and Pelliccia, D and Amor{\'\i}n, R and Bardelli, S and others},
  journal={Astronomy \& Astrophysics},
  volume={662},
  pages={A33},
  year={2022},
  publisher={EDP Sciences}
}

@article{mcgee2014overconsumption,
  title={Overconsumption, outflows and the quenching of satellite galaxies},
  author={McGee, Sean L and Bower, Richard G and Balogh, Michael L},
  journal={Monthly Notices of the Royal Astronomical Society: Letters},
  volume={442},
  number={1},
  pages={L105--L109},
  year={2014},
  publisher={Oxford University Press}
}

@article{balogh2016evidence,
  title={Evidence for a change in the dominant satellite galaxy quenching mechanism at z= 1},
  author={Balogh, Michael L and McGee, Sean L and Mok, Angus and Muzzin, Adam and van der Burg, Remco FJ and Bower, Richard G and Finoguenov, Alexis and Hoekstra, Henk and Lidman, Chris and Mulchaey, John S and others},
  journal={Monthly Notices of the Royal Astronomical Society},
  volume={456},
  number={4},
  pages={4364--4376},
  year={2016},
  publisher={Oxford University Press}
}

@article{shibuya2025galaxy,
  title={Galaxy morphologies revealed with Subaru HSC and super-resolution techniques. II. Environmental dependence of galaxy mergers at z~ 2--5},
  author={Shibuya, Takatoshi and Ito, Yohito and Asai, Kenta and Kirihara, Takanobu and Fujimoto, Seiji and Toba, Yoshiki and Miura, Noriaki and Umayahara, Takuya and Iwadate, Kenji and Ali, Sadman S and others},
  journal={Publications of the Astronomical Society of Japan},
  volume={77},
  number={1},
  pages={21--45},
  year={2025},
  publisher={Oxford University Press}
}

@article{mcconachie2025galaxies,
  title={Where Galaxies Go to Die: The Environments of Massive Quiescent Galaxies at $3< z< 5$},
  author={McConachie, Ian and de Graaff, Anna and Maseda, Michael V and Leja, Joel and Zhang, Yunchong and Setton, David J and Bezanson, Rachel and Boogaard, Leindert A and Brammer, Gabriel and Cleri, Nikko J and others},
  journal={arXiv preprint arXiv:2510.25024},
  year={2025}
}

@article{huvsko2023buildup,
  title={The buildup of galaxies and their spheroids: The contributions of mergers, disc instabilities, and star formation},
  author={Hu{\v{s}}ko, Filip and Lacey, Cedric G and Baugh, Carlton M},
  journal={Monthly Notices of the Royal Astronomical Society},
  volume={518},
  number={4},
  pages={5323--5339},
  year={2023},
  publisher={Oxford University Press}
}

@ARTICLE{1991ApJ...379...52W,
       author = {{White}, Simon D.~M. and {Frenk}, Carlos S.},
        title = "{Galaxy Formation through Hierarchical Clustering}",
      journal = {\apj},
         year = 1991,
        month = sep,
       volume = {379},
        pages = {52},
          doi = {10.1086/170483},
       adsurl = {https://ui.adsabs.harvard.edu/abs/1991ApJ...379...52W}
}

@article{springel2005simulations,
  title={Simulations of the formation, evolution and clustering of galaxies and quasars},
  author={Springel, Volker and White, Simon DM and Jenkins, Adrian and Frenk, Carlos S and Yoshida, Naoki and Gao, Liang and Navarro, Julio and Thacker, Robert and Croton, Darren and Helly, John and others},
  journal={nature},
  volume={435},
  number={7042},
  pages={629--636},
  year={2005},
  publisher={Nature Publishing Group UK London}
}

@article{cole2000hierarchical,
  title={Hierarchical galaxy formation},
  author={Cole, Shaun and Lacey, Cedric G and Baugh, Carlton M and Frenk, Carlos S},
  journal={Monthly Notices of the Royal Astronomical Society},
  volume={319},
  number={1},
  pages={168--204},
  year={2000},
  publisher={Blackwell Science Ltd Oxford, UK}
}

@article{dressler1980galaxy,
  title={Galaxy morphology in rich clusters-Implications for the formation and evolution of galaxies},
  author={Dressler, Alan},
  journal={Astrophysical Journal, Part 1, vol. 236, Mar. 1, 1980, p. 351-365.},
  volume={236},
  pages={351--365},
  year={1980}
}

@article{butcher1978evolution,
  title={The evolution of galaxies in clusters. I-ISIT photometry of C1 0024+ 1654 and 3C 295},
  author={Butcher, Harvey and Oemler Jr, Augustus},
  journal={Astrophysical Journal, Part 1, vol. 219, Jan. 1, 1978, p. 18-30.},
  volume={219},
  pages={18--30},
  year={1978}
}

@article{moore1998morphological,
  title={Morphological transformation from galaxy harassment},
  author={Moore, Ben and Lake, George and Katz, Neal},
  journal={The Astrophysical Journal},
  volume={495},
  number={1},
  pages={139},
  year={1998},
  publisher={IOP Publishing}
}

@article{moore1996galaxy,
  title={Galaxy harassment and the evolution of clusters of galaxies},
  author={Moore, Ben and Katz, Neal and Lake, George and Dressler, Alan and Oemler, Augustus},
  journal={nature},
  volume={379},
  number={6566},
  pages={613--616},
  year={1996},
  publisher={Nature Publishing Group UK London}
}

@article{abadi1999ram,
  title={Ram pressure stripping of spiral galaxies in clusters},
  author={Abadi, Mario G and Moore, Ben and Bower, Richard G},
  journal={Monthly Notices of the Royal Astronomical Society},
  volume={308},
  number={4},
  pages={947--954},
  year={1999},
  publisher={Blackwell Science Ltd Oxford, UK}
}

@ARTICLE{okegun,
       author = {{Oke}, J.~B. and {Gunn}, J.~E.},
        title = "{Secondary standard stars for absolute spectrophotometry.}",
      journal = {\apj},
         year = 1983,
        month = mar,
       volume = {266},
        pages = {713-717},
          doi = {10.1086/160817},
       adsurl = {https://ui.adsabs.harvard.edu/abs/1983ApJ...266..713O}
}

@article{adam2019euclid,
  title={Euclid preparation-III. Galaxy cluster detection in the wide photometric survey, performance and algorithm selection},
  author={Adam, R and Vannier, M and Maurogordato, S and Biviano, ANDREA and Adami, C and Ascaso, B and Bellagamba, F and Benoist, C and Cappi, Alberto and D{\'\i}az-S{\'a}nchez, A and others},
  journal={Astronomy \& Astrophysics},
  volume={627},
  pages={A23},
  year={2019},
  publisher={EDP sciences}
}

@article{gunn1972infall,
  title={On the infall of matter into clusters of galaxies and some effects on their evolution},
  author={Gunn, James E and Gott III, J Richard},
  journal={Astrophysical Journal, vol. 176, p. 1},
  volume={176},
  pages={1},
  year={1972}
}

@article{larson1980evolution,
  title={The evolution of disk galaxies and the origin of S0 galaxies},
  author={Larson, Richard B and Tinsley, Beatrice M and Caldwell, C Nelson},
  journal={Astrophysical Journal, Part 1, vol. 237, May 1, 1980, p. 692-707. Research supported by the Alfred P. Sloan Foundation},
  volume={237},
  pages={692--707},
  year={1980}
}

@article{muldrew2015protoclusters,
  title={What are protoclusters?--Defining high-redshift galaxy clusters and protoclusters},
  author={Muldrew, Stuart I and Hatch, Nina A and Cooke, Elizabeth A},
  journal={Monthly Notices of the Royal Astronomical Society},
  volume={452},
  number={3},
  pages={2528--2539},
  year={2015},
  publisher={Oxford University Press}
}

@article{davis2011virialization,
  title={Virialization of high-redshift dark matter haloes},
  author={Davis, Andrew J and D’Aloisio, Anson and Natarajan, Priyamvada},
  journal={Monthly Notices of the Royal Astronomical Society},
  volume={416},
  number={1},
  pages={242--247},
  year={2011},
  publisher={Blackwell Publishing Ltd Oxford, UK}
}

@article{naufal2024revealing,
  title={Revealing the quiescent galaxy population in the Spiderweb protocluster at z= 2.16 with deep HST/WFC3 slitless spectroscopy},
  author={Naufal, Abdurrahman and Koyama, Yusei and D’Eugenio, Chiara and Dannerbauer, Helmut and Shimakawa, Rhythm and P{\'e}rez-Mart{\'\i}nez, Jose Manuel and Kodama, Tadayuki and Zhang, Yuheng and Daikuhara, Kazuki},
  journal={The Astrophysical Journal},
  volume={977},
  number={1},
  pages={58},
  year={2024},
  publisher={IOP Publishing}
}

@article{shimakawa2015early,
  title={An early phase of environmental effects on galaxy properties unveiled by near-infrared spectroscopy of protocluster galaxies at z> 2},
  author={Shimakawa, Rhythm and Kodama, Tadayuki and Tadaki, Ken-ichi and Hayashi, Masao and Koyama, Yusei and Tanaka, Ichi},
  journal={Monthly Notices of the Royal Astronomical Society},
  volume={448},
  number={1},
  pages={666--680},
  year={2015},
  publisher={Oxford University Press}
}

@article{taamoli2024cosmos2020,
  title={COSMOS2020: Disentangling the Role of Mass and Environment in Star Formation Activity of Galaxies at 0.4< z< 4},
  author={Taamoli, Sina and Nezhad, Negin and Mobasher, Bahram and Manesh, Faezeh and Chartab, Nima and Weaver, John R and Capak, Peter L and Casey, Caitlin M and Gozaliasl, Ghassem and Heintz, Kasper E and others},
  journal={The Astrophysical Journal},
  volume={977},
  number={2},
  pages={263},
  year={2024},
  publisher={IOP Publishing}
}

@article{dekel2009cold,
  title={Cold streams in early massive hot haloes as the main mode of galaxy formation},
  author={Dekel, A and Birnboim, Y and Engel, G and Freundlich, J and Goerdt, T and Mumcuoglu, M and Neistein, E and Pichon, C and Teyssier, R and Zinger, E},
  journal={Nature},
  volume={457},
  number={7228},
  pages={451--454},
  year={2009},
  publisher={Nature Publishing Group UK London}
}

@article{sillassen2024noema,
  title={NOEMA formIng Cluster survEy (NICE): Characterizing eight massive galaxy groups at 1.5< z< 4 in the COSMOS field},
  author={Sillassen, Nikolaj B and Jin, Shuowen and Magdis, Georgios E and Daddi, Emanuele and Wang, Tao and Lu, Shiying and Sun, Hanwen and Arumugam, Vinod and Liu, Daizhong and Brinch, Malte and others},
  journal={Astronomy \& Astrophysics},
  volume={690},
  pages={A55},
  year={2024},
  publisher={EDP sciences}
}

@article{adachi2025enhanced,
  title={Enhanced gas-phase metallicities and suppressed outflows for galaxies in a rich cluster core at cosmic noon},
  author={Adachi, Kota and Kodama, Tadayuki and P{\'e}rez-Mart{\'\i}nez, Jose Manuel and Suzuki, Tomoko L and Onodera, Masato},
  journal={arXiv preprint arXiv:2506.01088},
  year={2025}
}

@article{perez2024enhanced,
  title={Enhanced star formation and metallicity deficit in the USS 1558- 003 forming protocluster at z= 2.53},
  author={P{\'e}rez-Mart{\'\i}nez, Jose Manuel and Kodama, Tadayuki and Koyama, Yusei and Shimakawa, Rhythm and Suzuki, Tomoko L and Daikuhara, Kazuki and Adachi, Kota and Onodera, Masato and Tanaka, Ichi},
  journal={Monthly Notices of the Royal Astronomical Society},
  volume={527},
  number={4},
  pages={10221--10238},
  year={2024},
  publisher={Oxford University Press}
}

@article{chartab2021mosdef,
  title={The MOSDEF Survey: environmental dependence of the gas-phase metallicity of galaxies at 1.4≤ z≤ 2.6},
  author={Chartab, Nima and Mobasher, Bahram and Shapley, Alice E and Shivaei, Irene and Sanders, Ryan L and Coil, Alison L and Kriek, Mariska and Reddy, Naveen A and Siana, Brian and Freeman, William R and others},
  journal={The Astrophysical Journal},
  volume={908},
  number={2},
  pages={120},
  year={2021},
  publisher={IOP Publishing}
}

@article{madau2014cosmic,
  title={Cosmic star-formation history},
  author={Madau, Piero and Dickinson, Mark},
  journal={Annual Review of Astronomy and Astrophysics},
  volume={52},
  number={1},
  pages={415--486},
  year={2014},
  publisher={Annual Reviews}
}

@article{cardelli1989relationship,
  title={The relationship between infrared, optical, and ultraviolet extinction},
  author={Cardelli, Jason A and Clayton, Geoffrey C and Mathis, John S},
  journal={Astrophysical Journal, Part 1 (ISSN 0004-637X), vol. 345, Oct. 1, 1989, p. 245-256.},
  volume={345},
  pages={245--256},
  year={1989}
}

@article{inoue2011rest,
  title={Rest-frame ultraviolet-to-optical spectral characteristics of extremely metal-poor and metal-free galaxies},
  author={Inoue, Akio K},
  journal={Monthly Notices of the Royal Astronomical Society},
  volume={415},
  number={3},
  pages={2920--2931},
  year={2011},
  publisher={Blackwell Publishing Ltd Oxford, UK}
}

@article{ferland1998cloudy,
  title={CLOUDY 90: numerical simulation of plasmas and their spectra},
  author={Ferland, GJ and Korista, KT and Verner, DA and Ferguson, JW and Kingdon, JB and Verner, EM},
  journal={Publications of the Astronomical Society of the Pacific},
  volume={110},
  number={749},
  pages={761},
  year={1998},
  publisher={IOP Publishing}
}

@article{ferland20132013,
  title={The 2013 release of cloudy},
  author={Ferland, Gary J and Porter, RL and Van Hoof, PAM and Williams, RJR and Abel, NP and Lykins, ML and Shaw, Gargi and Henney, William J and Stancil, PC},
  journal={Revista mexicana de astronom{\'\i}a y astrof{\'\i}sica},
  volume={49},
  number={1},
  pages={137--163},
  year={2013},
  publisher={Instituto de Astronom{\'\i}a}
}

@article{chiang2013ancient,
  title={Ancient light from young cosmic cities: physical and observational signatures of galaxy proto-clusters},
  author={Chiang, Yi-Kuan and Overzier, Roderik and Gebhardt, Karl},
  journal={The Astrophysical Journal},
  volume={779},
  number={2},
  pages={127},
  year={2013},
  publisher={IOP Publishing}
}

@article{cooper2008deep2,
  title={The DEEP2 Galaxy Redshift Survey: the role of galaxy environment in the cosmic star formation history},
  author={Cooper, Michael C and Newman, Jeffrey A and Weiner, Benjamin J and Yan, Renbin and Willmer, Christopher NA and Bundy, Kevin and Coil, Alison L and Conselice, Christopher J and Davis, Marc and Faber, SM and others},
  journal={Monthly Notices of the Royal Astronomical Society},
  volume={383},
  number={3},
  pages={1058--1078},
  year={2008},
  publisher={Blackwell Publishing Ltd Oxford, UK}
}

@article{elbaz2007reversal,
  title={The reversal of the star formation-density relation in the distant universe},
  author={Elbaz, David and Daddi, E and Le Borgne, D and Dickinson, M and Alexander, DM and Chary, R-R and Starck, J-L and Brandt, WN and Kitzbichler, M and MacDonald, E and others},
  journal={Astronomy \& Astrophysics},
  volume={468},
  number={1},
  pages={33--48},
  year={2007},
  publisher={EDP Sciences}
}
\bibliographystyle{aasjournal}

\end{document}